\documentclass[twocolumn,floats,floatfix,
               showpacs,prd,superscriptaddress,
               longbibliography,
               nofootinbib]{revtex4-2}
\usepackage[utf8]{inputenc}

\usepackage{graphicx,epsfig}
\usepackage{epstopdf}
\usepackage{amssymb,amsmath,amsthm,amsfonts}
\usepackage{bm}

\usepackage[caption=false]{subfig}
\usepackage[usenames,dvipsnames]{xcolor}
\usepackage{url}
\usepackage[inline]{enumitem}
\usepackage{dsfont}
\usepackage{float}
\usepackage{placeins}
\usepackage[T1]{fontenc}
\usepackage{mathrsfs}
\usepackage{anyfontsize}
\usepackage{xspace}
\usepackage{tensor}
\usepackage{microtype}
\usepackage{tikz}
\usetikzlibrary{positioning}
\usepackage{acronym}

\usepackage[linktocpage,breaklinks]{hyperref}
\hypersetup{colorlinks=true,
            citecolor=NavyBlue,
            linkcolor=NavyBlue,
            urlcolor=NavyBlue}

\def\canuda{\textsc{Canuda}}
\def\ETK{\textsc{Einstein Toolkit}}
\def\newacronym#1#2#3{\gdef#1{\gdef#1{#2\xspace}#3 (#2)\xspace}}
\newacronym{\amr}{AMR}{adaptive mesh refinement}
\newacronym{\bhs}{BHs}{black holes}
\newacronym{\bh}{BH}{Black hole}
\newacronym{\ns}{NS}{neutron star}
\newacronym{\bssn}{BSSN}{Baumgarte-Shapiro-Shibata-Nakamura}
\newacronym{\CAH}{CAH}{common apparent horizon}
\newacronym{\eft}{EFT}{effective field theory}
\newacronym{\eob}{EOB}{effective-one-body}
\newacronym{\gr}{GR}{general relativity}
\newacronym{\gw}{GW}{gravitational waves}
\newacronym{\hpc}{HPC}{High-performance computing}
\newacronym{\sGB}{sGB}{scalar Gauss-Bonnet}
\newacronym{\GB}{GB}{Gauss-Bonnet}
\newacronym{\ode}{ODE}{ordinary differential equation}
\newacronym{\pde}{PDE}{partial differential equation}
\newacronym{\snr}{SNR}{signal-to-noise ratio}
\newacronym{\qnm}{QNM}{quasinormal mode}
\newacronym{\PN}{PN}{post-Newtonian}

\def\dif{\textrm{d}}
\def\p{\partial}

\def\aGB{\alpha_{\rm GB}}
\def\RGB{\mathscr{G}}

\def\R4D{R}
\def\Kphi{K_{\rm \Phi}}
\def\rex{r_{\rm ex}}
\def\tmg{t_{\rm M}}

\def\rBL{\bar{r}}
\def\H{\mathcal{H}}
\def\Lie{\mathcal{L}}

\def\cM{\mathscr{M}}

\begin{document}

\title{
Spin-induced dynamical scalarization, de-scalarization and stealthness \\ in scalar-Gauss-Bonnet gravity during black hole coalescence}

\author{Matthew Elley}
\email{matthew.elley@kcl.ac.uk}
\affiliation{Department of Physics, King's College London, Strand, London, WC2R 2LS, United Kingdom}

\author{Hector O. Silva}
\email{hector.silva@aei.mpg.de}
\affiliation{Max Planck Institute for Gravitational Physics (Albert Einstein Institute), Am M\"uhlenberg 1, 14476 Potsdam, Germany}

\author{Helvi Witek}
\email{hwitek@illinois.edu}
\affiliation{Illinois Center for Advanced Studies of the Universe \& Department of Physics, University of Illinois at Urbana-Champaign, Urbana, Illinois 61801, USA}

\author{Nicol\'as Yunes}
\email{nyunes@illinois.edu}
\affiliation{Illinois Center for Advanced Studies of the Universe \& Department of Physics, University of Illinois at Urbana-Champaign, Urbana, Illinois 61801, USA}

\begin{abstract}
Particular couplings between a scalar field and the Gauss-Bonnet invariant lead to spontaneous scalarization of black holes. Here we continue our work on simulating this phenomenon in the context of binary black hole systems. We consider a negative coupling for which the black-hole spin plays a major role in the scalarization process. We find two main phenomena: (i) dynamical descalarization, in which initially scalarized black holes form an unscalarized remnant, and (ii) dynamical scalarization, whereby the late merger of initially unscalarized black holes can cause scalar hair to grow. An important consequence of the latter case is that modifications to the gravitational waveform due to the scalar field may only occur post-merger, as its presence is hidden during the entirety of the inspiral. However, with a sufficiently strong coupling, we find that scalarization can occur before the remnant has even formed. We close with a discussion of observational implications for gravitational-wave tests of general relativity.
\end{abstract}

\date{{\today}}

\maketitle

\section{Introduction}
\label{sec:introduction}
The detection of \gw produced by coalescing compact binaries by the LIGO-Virgo-Kagra Collaboration~\cite{LIGOScientific:2018mvr,LIGOScientific:2020ibl,LIGOScientific:2021djp}
have opened a new avenue to test \gr{} in its strong-field, nonlinear regime~\cite{Yunes:2013dva,Berti:2015itd,Berti:2018cxi,Berti:2018vdi,Yunes:2016jcc}.
In fact, the first three catalogs of observations have already been used to perform several null tests of \gr~\cite{Yunes:2016jcc,LIGOScientific:2016lio,LIGOScientific:2018dkp,LIGOScientific:2019fpa,Cardenas-Avendano:2019zxd,LIGOScientific:2020tif,Silva:2020acr,LIGOScientific:2021sio,Ghosh:2021mrv,Carullo:2021dui},
as well as theory-specific tests~\cite{Sennett:2019bpc,Mehta:2022pcn,Zhao:2019suc,Wong:2022wni,Zhao:2019suc,Nair:2019iur,Yamada:2019zrb,Perkins:2021mhb,Lyu:2022gdr,Mehta:2022pcn,Silva:2022srr}.
The latter have placed constraints on quadratic gravity theories~\cite{Nair:2019iur,Yamada:2019zrb,Perkins:2021mhb,Lyu:2022gdr,Wong:2022wni}.

In these theories, a scalar field couples to a curvature scalar, which is
quadratic in the Riemann tensor (see e.g.~Ref.~\cite{Yagi:2015oca} for an overview).
Well-known examples include coupling to the Pontryagin density or the \GB invariant.
The latter theories are often named \sGB gravity. They can emerge in the low-energy limit of string theory (see, for instance, Refs.~\cite{Metsaev:1987zx,Kanti:1995cp,Cano:2021rey}), as well as through a dimensional reduction of Lovelock gravity~\cite{Charmousis:2014mia}, and belong to the wider class of Horndeski gravity theories~\cite{Kobayashi:2011nu,Kobayashi:2019hrl}.

\bh solutions in this theory have long been known to have a nontrivial scalar field (i.e., a ``hair''), to which we can associate
a monopole scalar charge that depends on the \bh's mass and spin.
When the \bh{s} are found in a binary, their motion can lead to the emission of scalar dipole radiation, which in turn
modifies the system's orbital dynamics and the \gw signal with respect to \gr's
prediction. Such phenomenology has been explored with both \PN~\cite{Yagi:2011xp,Yagi:2012vf,Yagi:2013mbt,Shiralilou:2020gah,Shiralilou:2021mfl,Julie:2019sab,Julie:2022huo} and numerical relativity~\cite{Witek:2018dmd,Okounkova:2020rqw,East:2020hgw,East:2021bqk,Silva:2020omi,Doneva:2022byd} techniques.
The scalar field can also affect the post-merger signal, modifying the remnant \bh's ringdown~\cite{Pani:2009wy,Blazquez-Salcedo:2016enn,Blazquez-Salcedo:2020rhf,Blazquez-Salcedo:2020caw,Pierini:2021jxd,Bryant:2021xdh}.
In \sGB gravity, the presence of scalar hair depends on the functional form of the coupling between scalar field and the \GB invariant.

More specifically, if the functional form of the coupling always has a non-vanishing first derivative, such as for a linear or exponential coupling, \bh{s} are known
to invariably have scalar hair~\cite{Campbell:1991kz,Mignemi:1992nt,Kanti:1995vq,Torii:1996yi,Guo:2008hf,Yunes:2011we,Sotiriou:2013qea,Sotiriou:2014pfa,Benkel:2016kcq,Benkel:2016rlz,Antoniou:2017acq,Antoniou:2017hxj,Prabhu:2018aun,Saravani:2019xwx,R:2022cwe}.
Hence, the observation of \gw{s} from \bh binaries and
mixed \ns-\bh binaries have allowed us to constrain the length scale at which the
scalar-field-\GB interaction becomes relevant to less than approximately one kilometer~\cite{Nair:2019iur,Yamada:2019zrb,Perkins:2021mhb,Lyu:2022gdr}.

In contrast, if the first derivative of the coupling function vanishes for
some constant background scalar field,
both scalarized and unscalarized \bh solutions
can exist~\cite{Doneva:2017bvd,Silva:2017uqg}.
Depending on the length scale associated with the scalar-field-\GB
interaction, and the \bh's mass~\cite{Doneva:2017bvd,Silva:2017uqg,Macedo:2019sem} and spin~\cite{Cunha:2019dwb,Collodel:2019kkx,Dima:2020yac,Herdeiro:2020wei,Berti:2020kgk,Hod:2020jjy,Doneva:2020nbb,Hod:2022hfm},
the \bh solutions of \gr become unstable to scalar field perturbations, and the end-state of this instability is a \emph{scalarized} \bh~\cite{Ripley:2020vpk}.
This process is similar to spontaneous scalarization of \ns{s} in scalar-tensor gravity~\cite{Damour:1993hw,Damour:1996ke}.
The difference lies in the fact that for \ns{s} the scalar field is sourced by
matter, while for \bh{s}
the scalar is sourced by the spacetime curvature alone.
Thus, one could envision that the aforementioned \gw constraints (such as e.g.~\cite{Wong:2022wni}) can be avoided
if scalarization occurs \emph{right before merger, or possibly only after merger.}

Can such a scenario happen?
Here we continue our previous work~\cite{Silva:2020omi} and explore how the onset of scalarization plays out during binary \bh mergers.
As in our previous paper, we work in the decoupling approximation,
i.e., we evolve the scalar field on a time-depenedent \gr background.
In Ref.~\cite{Silva:2020omi}, we studied a variety of possible processes for head-on \bh collisions, as well as a quasi-circular inspiral-merger of equal mass non-spinning binaries
using a positive sign of the scalar-field-\GB coupling.
We demonstrated the existence of a process we coined \textit{dynamical descalarization}, whereby initially scalarized \bh{s} merged to form a larger remnant that descalarized
because its \GB curvature was too small to sustain the scalar hair.
The alternative, the dynamical scalarization of the remnant, was not possible because its larger mass (compared to the initial \bh{s}' masses) inevitably leads to a smaller \GB curvature near the horizon.

However, for a negative sign of the coupling, the scalar field instability happens
only for sufficiently rapidly-spinning \bh{s} (``spin-induced scalarization'')~\cite{Dima:2020yac,Hod:2020jjy,Herdeiro:2020wei,Berti:2020kgk,Doneva:2020nbb}.
This leads to the following questions:
\begin{enumerate*}[label={(\arabic*)}]
\item Does the formation of a highly spinning remnant cause spin-induced dynamical scalarization? If so, at what stage in the binary's evolution is the scalar hair excited?
\item Can the process of dynamical descalarization found in Ref.~\cite{Silva:2020omi} be generalized to the negative coupling case?
\end{enumerate*}
Here we address these questions with a new suite of binary \bh{} simulations
and negative sign of the coupling constant.

We find that indeed spin-induced descalarization and scalarization of the \bh remnant are both possible. The spin-induced descalarization of initially scalarized, spinning \bhs, extends and completes the work in Ref.~\cite{Silva:2020omi}. The spin-induced scalarization of the remnant is a new result. For values of the coupling constant close to the scalarization threshold,
the growth of the scalar field has
a large instability time-scale.
Therefore, scalarization only becomes significant significantly after the
remnant \bh's ringdown begins. We therefore now coin the term \textit{stealth dynamical scalarization}, whereby the scalar field remains hidden
throughout the full inspiral, merger and early ringdown evolution of the \bh binary and is thus unconstrainable with \gw observations.

In the remainder of this work we explain how we arrived at these conclusions.
In Sec.~\ref{sec:SGBSetup} we review both scalarization and descalarization of \bh{s}
in \sGB gravity.
Next, in Sec.~\ref{sec:Numerical_Method} we discuss our numerical methods and
our numerical relativity simulations designed to answer our previously stated questions.
In Sec.~\ref{sec:Results} we present our findings and we finish by discussing
some of their observational implications in Sec.~\ref{sec:discussions}.
We work with geometric units $G = 1 = c$.

\section{Scalar Gauss--Bonnet gravity}\label{sec:SGBSetup}
\subsection{Action and field equations}\label{ssec:action_field}

\sGB gravity modifies \gr via a nonminimal coupling between a real scalar field $\Phi$
and the \GB invariant $\RGB$, as
described by the action
\begin{align}
\label{eq:ActionSGB}
S =  \frac{1}{16\pi} \int \dif^{4} x \sqrt{-g} \left[
       \R4D
     - \frac{1}{2} \left( \nabla\Phi \right)^{2}
     + \frac{\aGB}{4} f(\Phi) \, \RGB
     \right]
\,,
\nonumber \\
\end{align}
where $R$ is the Ricci scalar, $g = {\rm det} (g_{\mu\nu})$ the metric determinant,
$(\nabla \Phi)^2 = g^{\mu\nu} \nabla_{\mu} \Phi \nabla_{\nu} \Phi$ the scalar
field kinetic term, and
\begin{equation}
\RGB = \R4D^{2}
- 4 \R4D_{\mu\nu} \R4D^{\mu\nu}
+ \R4D_{\mu\nu\rho\sigma} \R4D^{\mu\nu\rho\sigma}
\,,
\end{equation}
is the \GB invariant, where $\R4D_{\mu\nu\rho\sigma}$ and $\R4D_{\mu\nu}$
are the Riemann and Ricci tensor respectively.
The particular form of the theory is parametrized by the coupling function $f(\Phi)$ and
the coupling constant $\aGB$ with units of $[{\rm Length}]^{2}$.

As in our previous study~\cite{Silva:2020omi}, we work in the decoupling limit.
That is, we neglect the backreaction of the scalar field onto the spacetime metric:
the scalar field evolves
on a dynamical, vacuum background spacetime of \gr.
The action~\eqref{eq:ActionSGB} gives rise to the
field equation for $\Phi$
\begin{equation}
\label{eq:SGBEoMScalar}
\Box\Phi   =  - \tfrac{1}{4} \aGB f'(\Phi) \RGB
\,,
\end{equation}
where a prime denotes a derivative with respect to $\Phi$.
Since we work in the decoupling limit, the d'Alembertian and the \GB invariant are those of the time-dependent GR background.

The choice of the coupling function $f(\Phi)$ determines specific \sGB models.
As we already alluded to in Sec.~\ref{sec:introduction}, the models can be classified into two types depending on the properties of their \bh solutions.
We label models as \textit{type~I}
if the derivative of the coupling function $f'(\Phi) \neq 0$.
In this case, \bh solutions always have scalar hair~\cite{Campbell:1991kz,Mignemi:1992nt,Kanti:1995vq,Torii:1996yi,Guo:2008hf,Yunes:2011we,Sotiriou:2013qea,Sotiriou:2014pfa,Benkel:2016kcq,Benkel:2016rlz,Antoniou:2017acq,Antoniou:2017hxj,Prabhu:2018aun,Saravani:2019xwx,R:2022cwe}.
Examples of type~I models
include the dilatonic $f(\Phi) \propto \exp(\Phi)$~\cite{Mignemi:1992nt,Kanti:1995vq,Torii:1996yi,Guo:2008hf}
and shift-symmetric $f(\Phi) \propto \Phi$~\cite{Yunes:2011we,Sotiriou:2013qea,Sotiriou:2014pfa} coupling functions.
We label models as type~II
if the derivative of the coupling function $f'(\Phi_0) = 0$, for some
constant $\Phi_{0}$.
In this case, the theory admits the stationary vacuum \bh solutions of \gr, as
proved by the no-hair theorem of~\cite{Silva:2017uqg}, but also admits, when
the theorem is violated, scalarized \bh{s}.
Examples include quadratic $f(\Phi) \propto \Phi^2$~\cite{Silva:2017uqg} and Gaussian $f(\Phi) \propto \exp(\Phi^2)$~\cite{Doneva:2017bvd} coupling functions.
Here we consider type~II models only.

\subsection{Scalarization of isolated black holes}\label{ssec:SBH_overview}
In the second type of sGB model the onset of scalarization is found by linearizing Eq.~\eqref{eq:SGBEoMScalar} around the background \bh spacetime, i.e.,~$\Phi = \Phi_0 + \delta \Phi$, where $\Phi_0$ is a constant.
This results in the scalar-field evolution equation
\begin{equation}
\label{eq:lin_pert}
    \left( \Box - m^{2}_\text{eff} \right)\delta \Phi = 0 \,,
\end{equation}
with an effective mass squared
\begin{equation}
\label{eq:eff_mass}
   m^{2}_\text{eff} :=  -\tfrac{1}{4} \aGB f''(\Phi_{0}) \, \RGB \,,
\end{equation}
which can become tachyonically unstable; in other words, the \bh can scalarize if $m^{2}_\text{eff} < 0$~\cite{Doneva:2017bvd,Silva:2017uqg}.
This, however, is a necessary, but not sufficient condition for scalarization.
The scalarization threshold can be calculated by finding a bound state solution, i.e,
a time independent solution of Eq.~\eqref{eq:lin_pert} which is regular at the \bh
horizon and that vanishes at spatial infinity.
By imposing these boundary conditions on $\delta \Phi$, the calculation
of the scalarization threshold is reduced to a boundary value problem,
with the dimensionless ratio between $\alpha_{\rm GB}$ and the \bh's mass squared
playing the role of the eigenvalue. The smallest eigenvalue provides
the scalarization threshold for the ``fundamental'' (i.e., the nodeless solution) family of
scalarized \bh{s}, while the other eigenvalues determine the threshold for the
formation of ``excited states'' (i.e., solutions with one or more nodes).
We focus on the latter here.
See Fig.~1 in Ref.~\cite{Silva:2017uqg} or Sec.~4.3 of Ref.~\cite{Silva:2014fca}
for further details.
To be more concrete, here we consider a quadratic coupling function,
\begin{equation}
\label{eq:SGBCouplingFunction}
    f(\Phi) = \Phi^2 \,.
\end{equation}
The coupling strength is determined by the dimensionless constant\footnote{With respect to
the notation of Ref.~\cite{Silva:2020omi}, we are omitting the subscript ``2'' and fixing $\Bar\beta = 1$.}
\begin{equation}
\beta = \alpha_{\rm GB} / \cM^2
\,,
\label{eq:def_of_beta}
\end{equation}
where $\cM$ is the characteristic mass of the system.
The effective mass then becomes
\begin{equation}
m^2_{\rm eff} = - \tfrac{1}{2} \beta \, \cM^2 \, \RGB \,.
\label{eq:effective_mass_quad}
\end{equation}

If $\RGB$ is positive-definite in the \bh exterior, then the instability can only
happen for positive $\beta$.
However, if $\RGB$ is negative, at least in some regions outside the horizon, then the instability can also be triggered with a negative $\beta$.
For example, consider the Kerr metric, for which
the \GB invariant in Boyer-Lindquist coordinates $(t,\rBL,\theta,\varphi)$ is given by
\begin{equation}
\label{eq:GB_Kerr}
    \RGB_{\mathrm{Kerr}}=\frac{48 m^{2}}{\left(\rBL^{2}+\sigma^{2} \right)^{6}}
    \left(\rBL^{6}-15  \rBL^{4} \sigma^{2} +15 \rBL^{2}  \sigma^{4} -\sigma^{6} \right) \,,
\end{equation}
where $\sigma = a \cos{\theta}$ and $a = J/m$ is the angular momentum per unit
mass of the \bh.
When the dimensionless spin $\chi = a / m < 0.5$, $\RGB$ is positive everywhere outside
the event horizon and so scalarization can only take place if $\beta$ is positive. This also holds true in the limiting case of a Schwarzschild \bh.
However, for sufficiently rapidly rotating \bh{s} (i.e.,~those
with $\chi = a / m \geqslant 0.5$),
the \GB invariant can become negative in the exterior of the outer \bh horizon
in regions along the rotation axis~\cite{Cherubini:2002gen}.
Hence, spin can induce scalarization of \bh{s} if $\beta$ is negative
and $\chi \geqslant 0.5$~\cite{Dima:2020yac,Hod:2020jjy,Herdeiro:2020wei,Berti:2020kgk,Doneva:2020nbb,Hod:2022hfm}
and suppress it if $\beta$ is positive~\cite{Cunha:2019dwb,Collodel:2019kkx}.

One may note that scalarized solutions in quadratic \sGB gravity
with a positive coupling constant, $\beta>0$,
are unstable to radial perturbations~\cite{Blazquez-Salcedo:2018jnn}. Although this is true, such \bhs can be stabilized by including higher-order scalar terms in the coupling $f(\Phi)$~\cite{Minamitsuji:2018xde,Silva:2018qhn}, through
the addition of scalar field self-interactions while retaining the
quadratic form of $f(\Phi)$~\cite{Macedo:2019sem}, or
through the addition of a coupling of scalar field to the Ricci scalar~\cite{Antoniou:2021zoy,Antoniou:2022agj}.
Since we are investigating the onset of scalarization, it is unnecessary to include such terms
and so we focus only on the quadratic coupling case here.

%%%%%%%%%%%%%%%%%%%%%%%%%%%%%%%%%%%%%%%%%%%%%%%%%%%%%%%%%%%%%%%%%%%%%%%%%%%%%%
\subsection{Scalarization and Descalarization \\ in black hole binaries}\label{ssec:BBH_overview}

What could be the consequences of scalarization in \bh binaries?
To answer this question, in Ref.~\cite{Silva:2020omi} we performed the
first numerical relativity simulations of both head-on collisions and
quasi-circular inspirals of \bh{s} in quadratic \sGB gravity with a
positive coupling $\beta$.
We identified a new effect, that we named \textit{dynamical descalarization}, in which
initially non-spinning scalarized \bh{s} shed-off completely their scalar hair
after the merger.
This is a result of the comparatively weaker curvature generated near the horizon of the resulting larger remnant \bh.
Consequently, several possible dynamical processes were discovered for particular combinations of mass ratio and coupling strength, as illustrated in Fig.~1 of Ref.~\cite{Silva:2020omi}.
We can contrast this with similar simulations in type I theories in which
the remnant \bh always retains some scalar hair~\cite{Witek:2018dmd}.

Here we extend our previous work by considering negative coupling $\beta < 0$ values.
For this case the spins of the initial and/or remnant \bh{s} play a crucial role
in the development of the scalar field of the system due the possibility of
spin-induced scalarization.
Specifically, the formation of negative \GB regions
close to merger causes the remnant to scalarize, a process that
we call \textit{spin-induced dynamical scalarization}.
Additionally, we also demonstrate \textit{spin-induced dynamical descalarization} -- the spin analogue of the aforementioned dynamical descalarization mechanism -- as high-spinning binary components merge to produce a lower spin remnant that cannot support the instability.

\section{Simulating binary black holes in \sGB gravity -- Methods and setup}\label{sec:Numerical_Method}

\subsection{Time evolution formulation}\label{ssec:Time_Evolution_Problem}

We investigate the dynamics of the \sGB scalar field, determined by its equation of motion~\eqref{eq:SGBEoMScalar}, and sourced by a binary \bh background spacetime.
We perform a series of time evolution simulations in $3+1$ dimensions by adopting standard numerical relativity techniques; see e.g. Ref.~\cite{Alcubierre:2008}.
That is, we foliate the four-dimensional spacetime into three-dimensional spatial hypersurfaces $\Sigma_{t}$, parametrized by a time parameter $t$, with an induced spatial metric $\gamma_{ij}$. We introduce the timelike vector $n^{\mu}$ that is orthonormal to the hypersurface.
Then, the spacetime metric $g_{\mu\nu}$ can be decomposed as
\begin{align}
\label{eq:3p1metric}
\dif s^{2} & = g_{\mu\nu} \dif x^{\mu} \dif x^{\nu}
\\ &
    = - \left(\alpha^2 - \beta^{k}\beta_{k} \right) \dif t^2
      + 2 \gamma_{ij} \beta^{i} \dif t \dif x^{j}
      + \gamma_{ij} \dif x^{i} \dif x^{j}
\,,\nonumber
\end{align}
where $\alpha$ is the lapse function (not to be confused with the dimensional coupling constant $\alpha_{\rm GB}$) and $\beta^{i}$ is the shift vector
(not to be confused with the dimensionless coupling constant $\beta$).
Finally, we introduce the extrinsic curvature
$K_{ij}=-\frac{1}{2\alpha} \left(\p_{t} - \Lie_{\beta}\right) \gamma_{ij}$, where $\Lie_{\beta}$ is the Lie-derivative along the shift vector $\beta^{i}$.

To simulate the background \bh binary we write Einstein's equations as a Cauchy problem and adopt the
\bssn formulation~\cite{Shibata:1995we,Baumgarte:1998te}
together with the moving puncture gauge conditions~\cite{Campanelli:2005dd,Baker:2005vv}.
We prepare initial data describing a quasi-circular binary of two spinning \bh{s} with the Bowen-York approach~\cite{Bowen:1980yu,Brandt:1997tf}.

To evolve the scalar field $\Phi$ in this time-dependent GR background, we write its field equation~\eqref{eq:SGBEoMScalar} as a set of time evolution equations.
Therefore, we introduce the scalar field's momentum
$\Kphi{}=-\frac{1}{\alpha}\left(\p_{t}-\Lie_{\beta}\right)\Phi$
and we apply the spacetime decomposition to Eq.~\eqref{eq:SGBEoMScalar}.
This procedure gives the equations
\begin{subequations}
\label{eq:SF_EoM_Decomposed}
\begin{align}
   \left(\p_{t}-\Lie_{\beta}\right) \Phi    &= - \alpha \Kphi
   \,, \\
   \left(\p_{t}-\Lie_{\beta}\right) \Kphi   &= - D^{i} \alpha D_{i} \Phi
   \\ & \quad
        - \alpha \left( D^{i} D_{i} \Phi - K \Kphi
        + \tfrac{1}{4} \aGB f' \, \RGB  \right)
\,,\nonumber
\end{align}
\end{subequations}
where $D_{i}$, $\RGB$ and $K=\gamma^{ij}K_{ij}$ are
the covariant derivative with respect to the induced metric,
the four-dimensional \GB invariant and the trace of the extrinsic curvature
of the  background spacetime.

We initialize the scalar field to represent multiple scalarized \bh{s}. For simplicity, we neglect the scalar field's initial linear and angular momentum, because it relaxes to its equilibrium configuration within about $100M$ from the start of  the evolution, i.e., within approximately one orbit~\cite{Witek:2018dmd,Witek:2020uzz}.
Since the scalar field equation~\eqref{eq:SGBEoMScalar} is linear,
we can superpose the static bound-state solution anchored around an isolated \bh{.}
For $N$ \bh{s}, we then have
\begin{equation}
\label{eq:SGBScalarID}
\left.\Phi\right|_{t=0} = \sum^{N}_{a=1} \Phi_{(a)}
\,,\quad
\left.\Kphi\right|_{t=0} = 0
\,,
\end{equation}
where the subscript $(a)$ labels the $a$-th \bh{.}
The bound state of the \sGB scalar field around an isolated, non-spinning \bh with a coupling of the form~\eqref{eq:SGBCouplingFunction}
was obtained numerically in Ref.~\cite{Silva:2017uqg}.
We approximate this solution with
the fit
\begin{align}
\label{eq:SF_ID_Bound}
\left.\Phi_{(a)}\right|_{t=0} = \frac{m_{(a)} r_{(a)}}{\varrho_{(a)}^{2}}
    \left[ c_{1} + c_{2} \frac{m_{(a)} r_{(a)}}{\varrho_{(a)}^{2}} + c_{3} \frac{(m_{(a)} r_{(a)})^{2}}{\varrho_{(a)}^{4}} \right]
\,,
\nonumber \\
\end{align}
where $\varrho_{(a)} = m_{(a)} + 2 \, r_{(a)}$, $r_{(a)}$ is field point distance from the location of the $a$-th \bh{} in quasi-isotropic radial coordinates of the background spacetime, $m_{(a)}$ is the mass of the $a$-th \bh, and $c_1 = 3.68375$, $c_2 = 4.97242$,
$c_3 = 2.29938 \times 10^2$ are fitting constants, where we corrected a misprint in $c_3$ in Ref.~\cite{Silva:2020omi}.

\subsection{Code description}\label{ssec:Code_Description}
We performed the simulations with
\canuda~\cite{witek_helvi_2021_5520862},
our open-source numerical relativity code for fundamental physics~\cite{Okawa:2014nda,Zilhao:2015tya,Witek:2018dmd,Silva:2020omi}.
\canuda{} is fully compatible with the \ETK~\cite{steven_r_brandt_2021_5770803, Loffler:2011ay,Zilhao:2013hia},
a public numerical relativity software for computational astrophysics.
The \ETK{} is based on the {\textsc{Cactus}} computational toolkit~\cite{Goodale:2002a,Cactuscode:web} and uses the {\textsc{Carpet}} driver~\cite{Schnetter:2003rb,CarpetCode:web} to provide boxes-in-boxes \amr as well as MPI parallelization.
To evolve the field equations we employ the method-of-lines.
Spatial derivatives are typically realized by fourth-order finite differences
(with sixth order also being available)
and for the time integration we  use a fourth-order Runge-Kutta scheme.

The background spacetime, consisting of two spinning \bh{s} in a quasi-circular orbit, is initialized with the {\textsc{TwoPunctures}} spectral code~\cite{Ansorg:2004ds} that solves the constraint equations of \gr with the Bowen-York approach~\cite{Bowen:1980yu,Brandt:1997tf}.
We evolve Einstein's equations using \canuda{'s} modern version of the {\textsc{Lean}} thorn~\cite{Sperhake:2006cy} that implements the \bssn equations with the moving puncture gauge.
The \sGB scalar field evolution equations~\eqref{eq:SF_EoM_Decomposed} and its initial data~\eqref{eq:SF_ID_Bound} are implemented in \canuda{'s} arrangement {\textsc{Canuda\_EdGB\_dec}}. Details of the implementation are described in Refs.~\cite{Benkel:2016rlz,Witek:2018dmd,Silva:2020omi}.
To analyse the numerical data, we compute the Newman-Penrose scalar $\Psi_{4}$ as a measure for gravitational radiation and we extract the gravitational and scalar field multipoles on spheres of constant extraction radius $\rex$ using the \textsc{QuasiLocalMeasures} thorn~\cite{Dreyer:2002mx}.
We find the \bh{s'} apparent horizons and compute their properties with the \textsc{AHFinderDirect} thorn~\cite{Thornburg:1995cp,Thornburg:2003sf}.

\subsection{Setup of simulations}\label{ssec:SimulationSetup}
To investigate spin-induced dynamical scalarization or
descalarization in binary \bh mergers, we have performed
a series of simulations of equal-mass, quasi-circular inspirals
for the negative coupling case, $\beta<0$.
The initial \bh{s} have either zero spin or a spin (anti-)aligned with the orbital angular momentum.

\begin{figure}[t]
\includegraphics[width=\columnwidth]{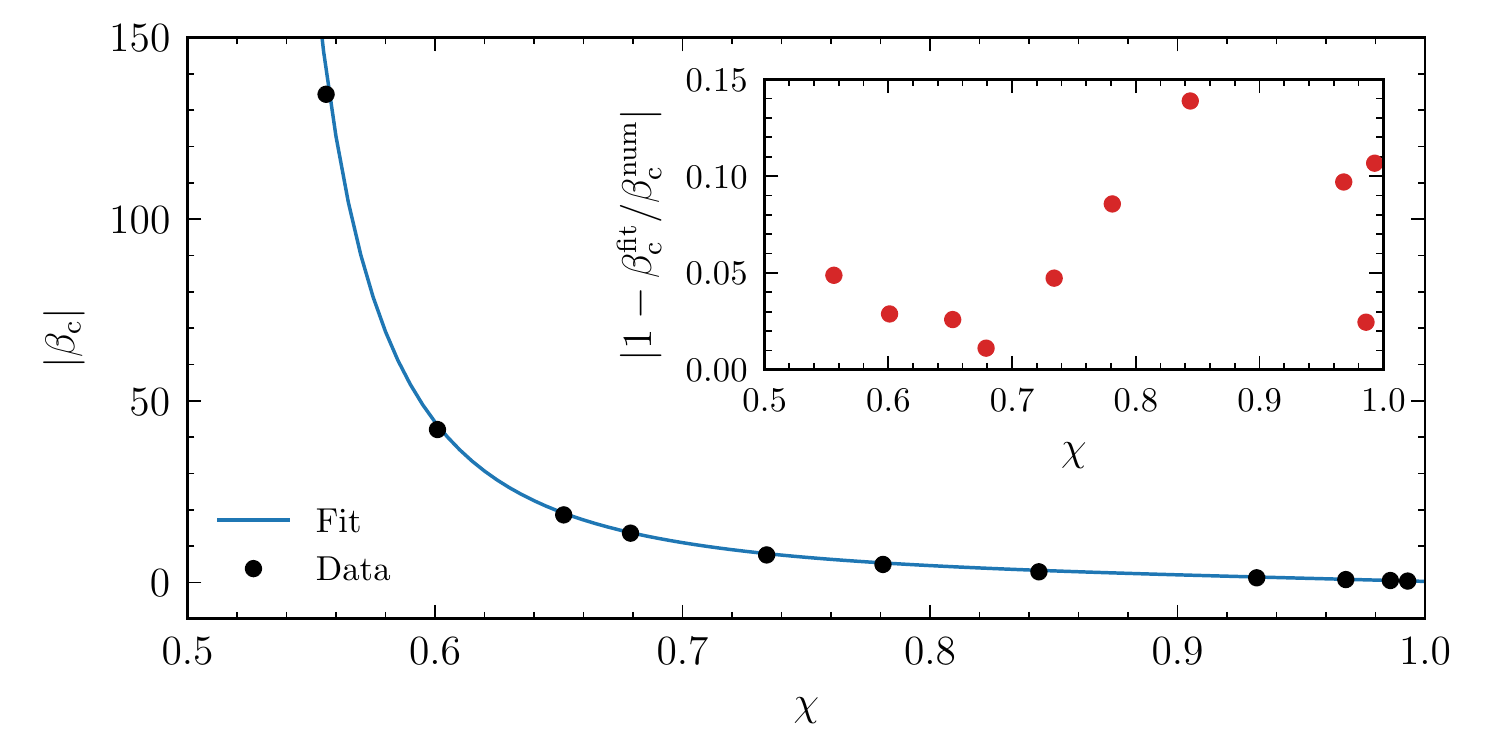}
\caption{Absolute value of the critical coupling, $\beta_{\rm c}$, for spin-induced scalarization of a single \bh{} as a function of the dimensionless spin $\chi$.
We show the  numerical data of Ref.~\cite{Herdeiro:2020wei} and the fitting formula~\eqref{eq:beta_spin_fit}.
The inset shows the
relative error between the fit and the data. We see that the error is less than $15\%$
in the range $0.5 \leqslant \chi < 1$ and
less than $5 \%$ for $\chi \lesssim 0.74$.}
\label{fig:beta_fit}
\end{figure}

To choose the values of the coupling constant $\beta$
in our simulations, we used the numerical
data found in Ref.~\cite{Herdeiro:2020wei} (cf.~Supplemental Material, Table I) to obtain a fitting formula that returns
the value of $\beta$ at the threshold for spin-induced scalarization as a function of the dimensionless spin $\chi$; we will refer to this threshold value as the \textit{critical value} of the dimensionless coupling constant.
The critical value for the coupling constant satisfies the scaling
\begin{equation}
    \beta_{\rm c}(m/M,\chi) = (m/M)^2 \, \beta_{\rm c}(1,\chi)
    \,,
    \label{eq:beta_crit_code}
\end{equation}
where $m$ is a place-holder for either the individual masses of the binary $m_{(a)}$ or the final remnant mass $m_f$, while $M = m_1 + m_2$ is the initial total mass of the binary.
The quantity $\beta_{\rm c}(1, \chi)$ is the critical value of the coupling that leads to scalarization for a \bh{} of mass $1 M$ and dimensionless spin $\chi$, namely
\begin{equation}
    \beta_{\rm c}(1,\chi) = - \frac{0.422}{(|\chi| - 1/2)^2} + 1.487 \, |\chi|^{7.551}\,,
    \label{eq:beta_spin_fit}
\end{equation}
where $\beta_{\rm c}(1,\chi)$ diverges as $|\chi|$ tends to $0.5$, in agreement with Ref.~\cite{Hod:2020jjy}. For instance, if we wish to scalarize the initial components of the binary, and if the mass ratio is unity, then $m_{(a)} = M/2$, and $\beta_{{\rm c}, \, (a)}(1/2,\chi_{(a)}) = (1/4) \,  \beta_{\rm c}(1,\chi_{(a)})$.
In Fig.~\ref{fig:beta_fit}, we show Eq.~\eqref{eq:beta_spin_fit}
and compare it against the numerical results of Ref.~\cite{Herdeiro:2020wei}.
We obtain
relative errors smaller than $15\%$
in the range $0.5 \leqslant \chi < 1$ and
less than $5 \%$ for $\chi \lesssim 0.74$.
We use Eq.~\eqref{eq:beta_crit_code} as reference to choose the
values of $\beta$ to probe scalarization of either one (or both) of the initial
binary components or of the remnant \bh.

\begin{table*}[t]
\begin{tabular}{c|ccccccc|cc}
\hline
\hline
Run            & $d/M$ & $\chi_{1} $ & $\chi_{2} $  & $\chi_{f} $ & $\beta$ & $\beta_{{\rm c},1}$ & $\beta_{{\rm c},f}$& process \\
\hline
\ref{item:setup_A} & $10$ & $0$   & $0$   & $0.68$ & $-14.30$ & -- & $-12.96$ & $\bar{s} + \bar{s} \rightarrow s_{\uparrow}$  \\
\ref{item:setup_B} & $10$ & $-0.6$ & $-0.6$ & $0.48$& $-11.00$ & $-10.55$ & -- & $s_{\downarrow} + s_{\downarrow} \rightarrow \bar{s}_{\uparrow}$\\
\hline
\hline
\end{tabular}
\caption{
Setup of the simulations of equal-mass, quasi-circular \bh binaries.
We show the initial separation $d/M$, the initial dimensionless spins $\chi_{1}$ and $\chi_{2}$ of each binary component, the dimensionless spin
$\chi_{f}$ of the remnant,
and the dimensionless coupling constant $\beta$ used in the simulations.
For reference, we also show the critical values
to scalarize the initial
($\beta_{\rm c,1}=\beta_{\rm c,2}$)
or final
($\beta_{{\rm c}, f}$)
\bh{s}, calculated using Eqs.~\eqref{eq:beta_crit_code} and~\eqref{eq:beta_spin_fit}.
The last column summarizes the process that unfolds during the simulation.
We use $\bar{s}$ and $s$ to denote unscalarized and scalarized states, respectively, and
the subscript $\uparrow$ ($\downarrow$) indicates spin aligned (anti-aligned) with the
orbital angular momentum, which is assumed to be $\uparrow$. See Fig.~\ref{fig:diagram_simulations} for additional details.
}
\label{tab:setups}
\end{table*}

\begin{figure}[t]
\subfloat[\ref{item:setup_A}]{\includegraphics[width=0.425\columnwidth]{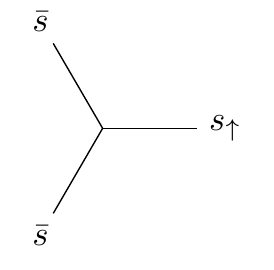}\label{fig:diagram_simulationsA}}
\qquad
\subfloat[\ref{item:setup_B}]{\includegraphics[width=0.425\columnwidth]{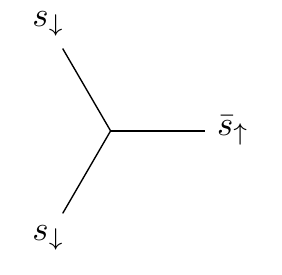}\label{fig:diagram_simulationsB}
}
\caption{
Binary \bh simulations, where $s$ ($\bar{s}$) stands for initial or final \bh states that are scalarized (unscalarized) and with spin along the positive ($\uparrow$) or negative ($\downarrow$) $z$-direction (i.e., aligned or anti-aligned with the orbital angular momentum, assuming the latter is $\uparrow$).
\bh states without an arrow are non-spinning.
Panel~\ref{fig:diagram_simulationsA}
illustrates a process of spin-induced dynamical scalarization: two initially unscalarized \bh{s} produce a spinning, scalarized remnant.
Panel~\ref{fig:diagram_simulationsB}
illustrates a process of dynamical descalarization: two initially rotating, scalarized \bh{s}
whose spin is anti-aligned with the orbital angular momentum merge into a rotating \bh with a smaller spin magnitude.
Consequently, the remnant descalarizes.
}
\label{fig:diagram_simulations}
\end{figure}

Here, we present two key simulations, listed in Table~\ref{tab:setups} and illustrated in Fig.~\ref{fig:diagram_simulations}, with the following setups:
\begin{enumerate}[label=\textbf{Setup \Alph*}, ref={Setup \Alph*}, wide, labelindent=0pt]
\item in Table~\ref{tab:setups} is designed to address our first question: does the formation of a highly spinning remnant cause spin-induced dynamical scalarization?
Here, we consider a binary of initially non-spinning, unscalarized \bh{s} that merges into a spinning,
scalarized remnant as illustrated in Fig.~\ref{fig:diagram_simulationsA}.
The \bh{s} complete $10$ orbits prior to their merger at $\tmg=927M$, as estimated from the peak
in the gravitational (quadrupole) waveform; see the bottom panel of Fig.~\ref{fig:bet-143_Phi00_Psi22_evol_rex100}.
When the coupling $\beta$ is negative, the squared effective mass~\eqref{eq:eff_mass} of the initial \bh{s} (with $\chi=0$) is positive definite everywhere outside their horizons, and so they are initially not scalarized.
The final \bh has a dimensionless spin of $\chi_{f}=0.68$ and mass $m_f \sim M$.
For a \bh with these parameters,
the critical coupling is $\beta_{{\rm c},f} \approx \beta_{\rm c}(1,0.68) \approx - 12.96$; cf. Eq.~\eqref{eq:beta_crit_code}.
In our simulation we chose $|\beta|>|\beta_{{\rm c},f}|$ such that the remnant \bh is indeed scalarized.
In this simulation, we initialize the scalar field according to Eq.~\eqref{eq:SF_ID_Bound}
around each binary component. The scalar field disperses early in the simulation,
leaving each \bh unscalarized and a negligible, but nonvanishing ambient scalar field in the numerical grid.
Notice that if we had set $\Phi\vert_{t=0} = 0$, there would be no scalar field dynamics [see Eq.~\eqref{eq:SGBEoMScalar}].
\label{item:setup_A}
\item in Table~\ref{tab:setups} is designed to address our second question:
is the dynamical descalarization found in Ref.~\cite{Silva:2020omi} a general phenomenon? Is there a spin-induced dynamical descalarization?
Here we consider a binary of initially rotating, scalarized \bh{s} with spins $\chi_{1}=\chi_{2}=-0.6$, anti-aligned with the orbital angular momentum as illustrated in
Fig.~\ref{fig:diagram_simulationsB}.
Each of the components of the binary has a mass $m_1 = m_2 = M/2$.
Inserting these parameters in Eq.~\eqref{eq:beta_crit_code}, we find $\beta_{{\rm c},1} = \beta_{{\rm c},2} = \beta_{\rm c}(1/2,-0.6) \approx -10.55$.
In our simulations, we set $|\beta| \gtrsim |\beta_{\rm c}(1/2,-0.6)|$ such that the initial \bh{s} are scalarized.
The initial \bh{s} merge into a final rotating \bh that has a spin aligned with the orbital angular momentum of the previously inspiralling system, with a spin magnitude $\chi_{f}=0.48$.
This value is below the threshold for spin-induced scalarization, and so the remnant \bh does not support scalar hair.
\label{item:setup_B}
\end{enumerate}

To show that our qualitative results are robust for a large variety of \bh spin parameters, we have performed a series of additional simulations listed in Table~\ref{tab:prod_list} of Appendix~\ref{app:all_simulations}.
All simulations presented in
Tables~\ref{tab:setups} and~\ref{tab:prod_list}
have the same grid setup: the numerical domain was composed of a Cartesian box-in-box \amr grid structure with seven refinement levels.
The outer boundary was located at $255.5M$.
We use a grid spacing of $\dif x=0.7M$ on the outermost refinement level to ensure a sufficiently high resolution in the wave zone. The region around the \bh{s} has a resolution of $\dif x=0.011M$.
To validate our code and estimate the numerical error of our simulations, we performed convergence tests for our most demanding simulation with $\chi_{1,2}=-0.6$, corresponding to Setup B in Table~\ref{tab:setups}.
The relative error in the gravitational quadrupole waveform is $\Delta\Psi_{4,22} / \Psi_{4,22}\leqslant 0.8\%$, while
the relative error of the scalar charge accumulates to
$\Delta\Phi_{00}/\Phi_{00}\leqslant30\%$ in the last orbits before merger; the latter is $\Delta\Phi_{00}/\Phi_{00}\leqslant 15\%$
in the merger and ringdown phase.
The large error in the scalar field, close to the \bh{s} merger, is a consequence of the exponential growth of the scalar field during inspiral. As our investigation is of a qualitative nature, this cumulative error is not a cause of concern for our results. However, a future quantitative analysis
would have to address this issue.
See Appendix~\ref{app:CodeVerification} for details.

\section{Results}\label{sec:Results}
\subsection{Spin-induced dynamical scalarization}
\label{sec:results_setupA}
Here we present key results obtained with simulation \ref{item:setup_A} (see~Sec.~\ref{ssec:SimulationSetup}), corresponding to Fig.~\ref{fig:diagram_simulationsA}.
In particular, we show that
an initially unscalarized \bh binary can indeed form
a hairy, rotating remnant.

\begin{figure}[t]
\includegraphics[width=0.95\columnwidth]{"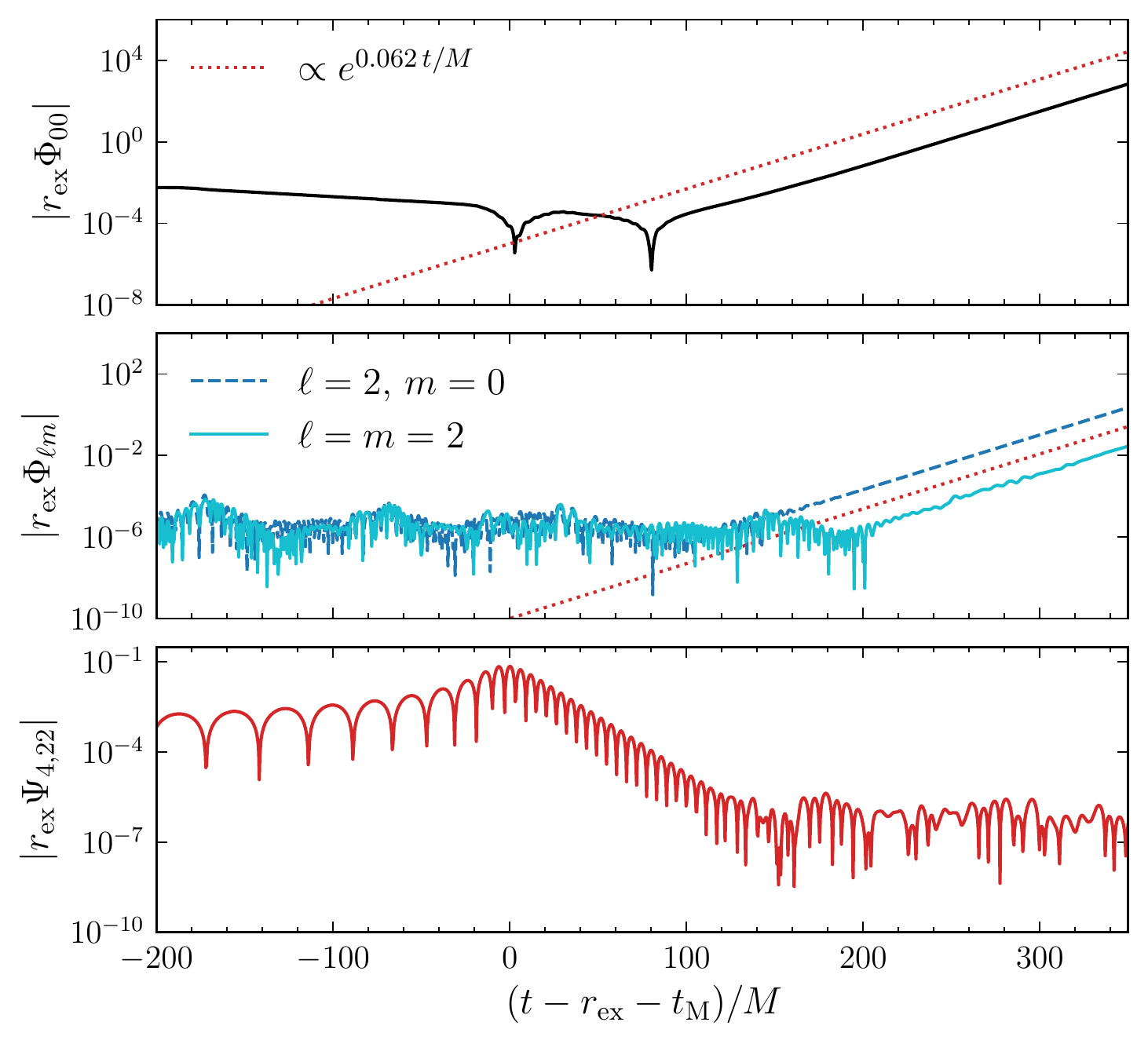"}
\caption{Evolution of the scalar field monopole (top panel), scalar field $\ell=2$ multipoles (middle panel) and the gravitational waveform of the background spacetime (bottom panel) for \ref{item:setup_A} in Table~\ref{tab:setups}.
We rescale the multipoles by the extraction radius $\rex=100M$, and shift them in time such that
$(t-\rex-\tmg)/M = 0$ indicates the time of merger,
determined by the peak of the gravitational waveform.
}
\label{fig:bet-143_Phi00_Psi22_evol_rex100}
\end{figure}

This process
is illustrated in the top panel of Fig.~\ref{fig:bet-143_Phi00_Psi22_evol_rex100},
where we present the time evolution of the scalar field's monopole charge, $\rex \Phi_{00}$, measured at $\rex=100M$, and shifted in time such that
$(t - \rex - \tmg) / M = 0$ indicates the time of merger.
The scalar field perturbation that is initially present in our simulations remains small during the entire inspiral.
See, for instance, the amplitudes $\rex \Phi_{\ell m}$ at $(t - \rex - \tmg) / M < 0$ which are of ${\cal{O}}(10^{-4})$ or ${\cal{O}}(10^{-6})$.
Yet, we see an exponential growth of the scalar charge,
$\rex \Phi_{00} \sim e^{\omega_{\rm I,00} t}$,
that exceeds the background fluctuations, approximately $100M$ after the merger.
We estimate the growth rate (for our choice of $\beta$) to be $M\omega_{\rm I,00}\sim0.062$ by fitting to the numerical data.
We show this with the dotted red line in the top and middle panels.

We find a similar behavior in the scalar field quadrupole, as shown in the middle panel of Fig.~\ref{fig:bet-143_Phi00_Psi22_evol_rex100}.
That is, both the axisymmetric $(\ell,m)=(2,0)$ and the $(\ell,m)=(2,2)$ multipoles are excited and grow exponentially with a rate of
$M\omega_{\rm I} \sim 0.062$.
For the form of the coupling function considered here, the rate appears to be independent of the $(\ell,m)$ multipole and is determined by the coupling constant $\beta$, as we further discuss later.
The quadrupole scalar field is absent in the initial data because we initialized the scalar field
with a spherically symmetric distribution around each of the \bh{s}.
Hence, the scalar field quadrupole we observe is caused by the ``stirring'' of the
ambient scalar field due to the dynamical binary \bh spacetime, which has a quadrupole moment.
These $\Phi_{2m}$ multipoles also become unstable eventually, but at a later time relative to the monopole, as is evident
by comparing the top and middle panels of Fig.~\ref{fig:bet-143_Phi00_Psi22_evol_rex100}.
The exponential growth of the $\Phi_{2m}$ multipoles is consistent with the findings in Refs.~\cite{Dima:2020yac,Doneva:2020nbb},
showing that higher-$\ell$ and $m\neq0$ scalar field multipoles
can also become unstable.

All of these results beg for the following questions: at what stage in the binary's evolution is the scalar field
instability induced?
Is it due to the orbital angular momentum at the late inspiral or is it due to the
angular momentum of the remnant \bh{?}
As we discussed in Sec.~\ref{ssec:SBH_overview},
a necessary (but not sufficient) condition for the tachyonic instability to occur
is for the \GB invariant to become negative outside the \bh horizon in the $\beta<0$ case; see~Eq.~\eqref{eq:effective_mass_quad}.
To address these questions, we inspect the behavior of the \GB invariant at different stages throughout the evolution.

\begin{figure}[t]
\includegraphics[width=\columnwidth]{"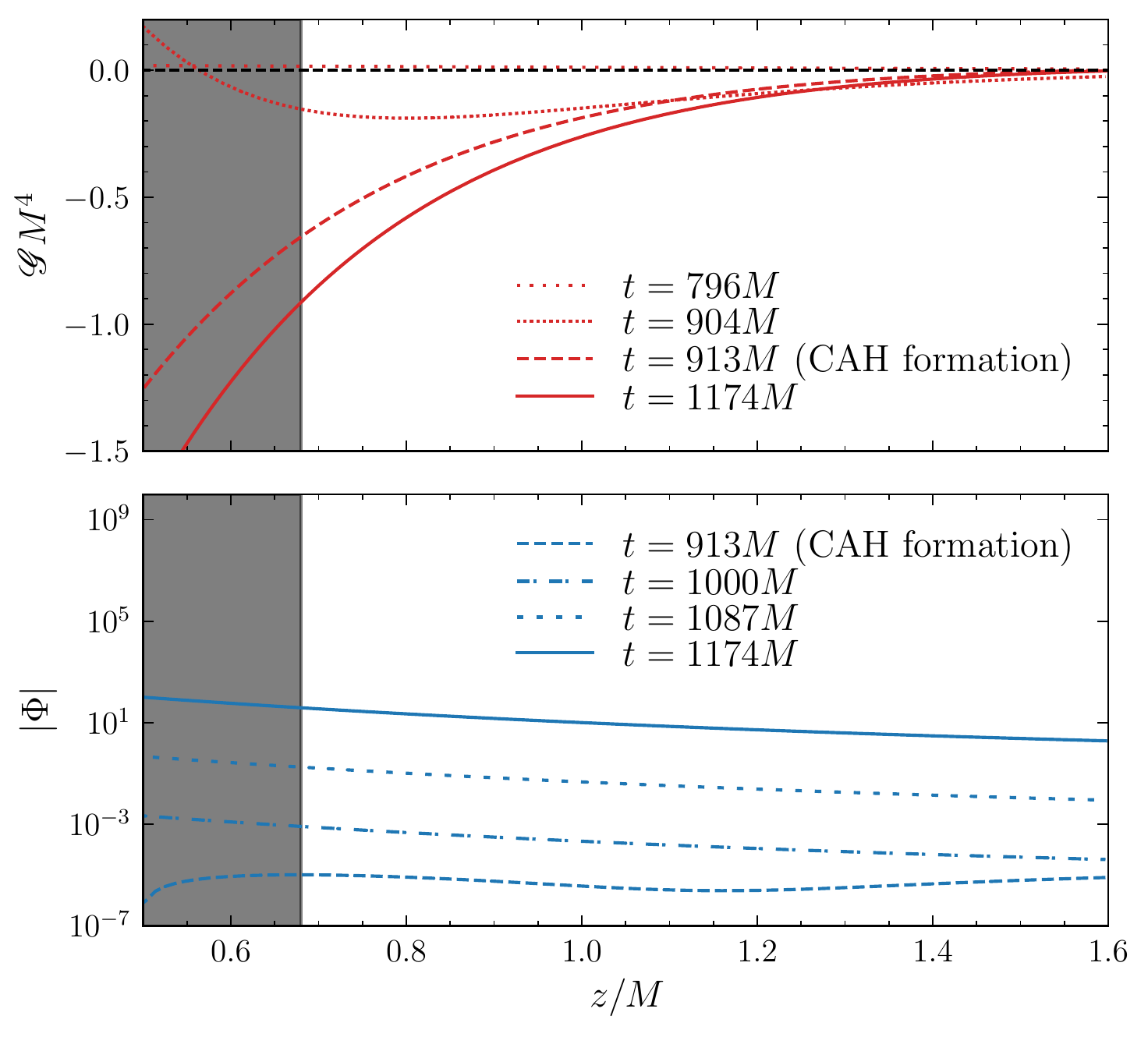"}
\caption{Profiles of the \GB invariant (top panel) and
of the scalar field (bottom panel), corresponding to \ref{item:setup_A} in Table~\ref{tab:setups}, along the $z$-axis in a close-up region near the \CAH{.}
The curves correspond to different times throughout the evolution.
The shaded region indicates the \CAH{,} shown $t=100M$ after its formation when the final \bh{} has relaxed to its stationary state.
The \GB invariant becomes negative during the \bh{s'} last orbit before merger, and settles to its profile around the final rotating \bh with dimensionless spin $\chi_f=0.68$.
In response, the scalar field becomes unstable.
}
\label{fig:SpatialProfilesSetupA}
\end{figure}

\begin{figure}[t]
\includegraphics[width=\columnwidth]{"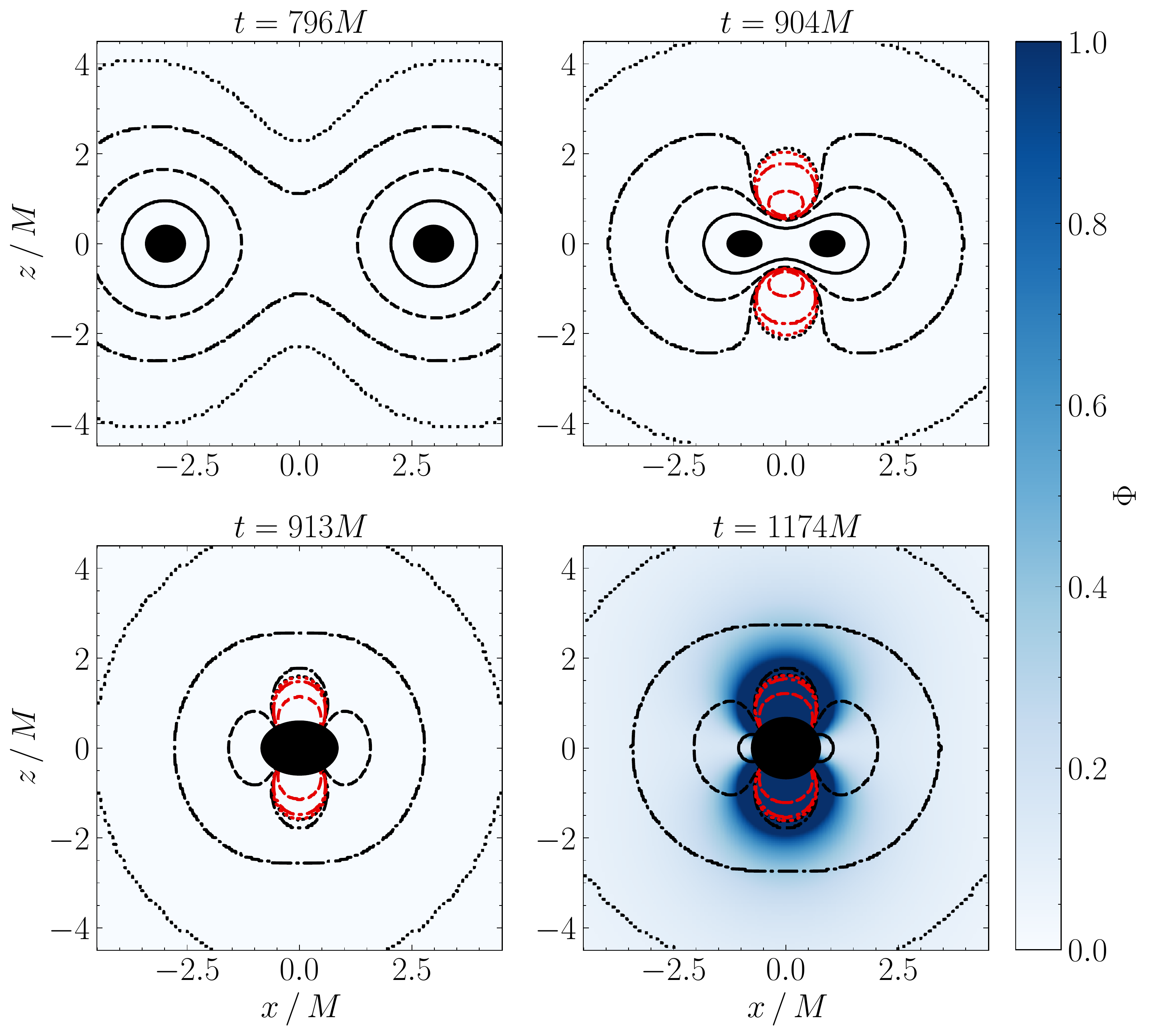"}
\caption{Snapshots of the scalar field, $\Phi$, and the \GB invariant in the $xz$-plane corresponding to \ref{item:setup_A} in Table~\ref{tab:setups}.
The color map indicates the amplitude of the scalar field.
The isocurvature contours of the \GB invariant correspond to
$|\RGB M^{4}|=1$ (solid line),
$|\RGB M^{4}|=10^{-1}$ (dashed line),
$|\RGB M^{4}|=10^{-2}$ (dot-dashed line),
$|\RGB M^{4}|=10^{-3}$ (dotted line),
Black (red) lines correspond to positive (negative) values of $\RGB$.
We show the inspiral (top left),
half an orbit before merger (top right),
formation of the first \CAH (bottom left)
and about $200M$ after the merger.
}
\label{fig:SnapshotsSetupA}
\end{figure}

In Fig.~\ref{fig:SpatialProfilesSetupA} we show a close-up of the \GB invariant's (top panel)
and the scalar field's (bottom panel) profiles along the $z$-axis, parallel to
the orbital angular momentum, at different
time snapshots throughout the evolution.
In Fig.~\ref{fig:SnapshotsSetupA} we show the \GB invariant $\RGB$ together with snapshots of the
scalar field $\Phi$
in the $xz$-plane, perpendicular to the orbital plane of the binary.
The snapshots correspond to time instants
during the inspiral (top left),
half an orbit before merger (top right),
at the formation of the \CAH (bottom left)
and about $200M$ after the merger (bottom right).
The color map represents the scalar field amplitude and is shared among all panels, while
the contours are isocurvature levels
$|\RGB M^4| = \{1, \, 10^{-1},\, 10^{-2},\, 10^{-3} \}$,
with positive (negative) values of $\RGB$ in black (red).
We also show the location of the individual \bh{s} using their apparent horizons,
represented as ellipses with center, semi-major and semi-minor axes given
by the centroid, maximum and minimum radial directions as obtained with the
\textsc{AHFinderDirect} thorn~\cite{Thornburg:1995cp,Thornburg:2003sf}.
We do not show the evolution of $\RGB$ in the equatorial plane because
we did not observe negative regions forming on this plane throughout the
entire simulation.

During the early inspiral, the \GB invariant is positive around the individual, non-spinning \bh{s}, and the scalar field remains small across the numerical grid as can be seen in the top left panel of Fig.~\ref{fig:SnapshotsSetupA}.
However, about half an orbit before merger,
we see the formation of
regions between the two \bh{s} where the \GB invariant is negative; see top right panel of Fig.~\ref{fig:SnapshotsSetupA}
and top panel of Fig.~\ref{fig:SpatialProfilesSetupA}, $t = 904M$ curve.
By the time $t = 904M$, the effective mass squared defined in Eq.~\eqref{eq:effective_mass_quad} has become negative and this, we re-emphasize, is a necessary, but not sufficient condition for
the tachyonic instability to occur.

As the \bh{s} merge and the system settles to a final, rotating \bh{,} the \GB invariant remains negative along the $z$-axis, which now coincides with the remnant \bh{'s} rotation axis.
This is illustrated in the bottom panels of Fig.~\ref{fig:SnapshotsSetupA}, which
correspond
to the instant of the formation of the \CAH (bottom left)
and to about $200M$ after the merger (bottom right).
In response, the scalar field grows exponentially
as can be seen in its profiles
shown in the bottom panel of Fig.~\ref{fig:SpatialProfilesSetupA} for different times after the \CAH has formed.
The scalar field assumes a predominantly dipolar spatial distribution along the \bh{'s} spin axis,
a consequence of the regions where the \GB invariant is negative.
We note that the scalar field continues to grow
instead of settling to a stationary bound state because the magnitude of the coupling is larger than the critical value for spin-induced scalarization for the final \bh with spin $\chi_f=0.68$; see~Table~\ref{tab:setups}.

\begin{figure}[t]
\includegraphics[width=\columnwidth]{"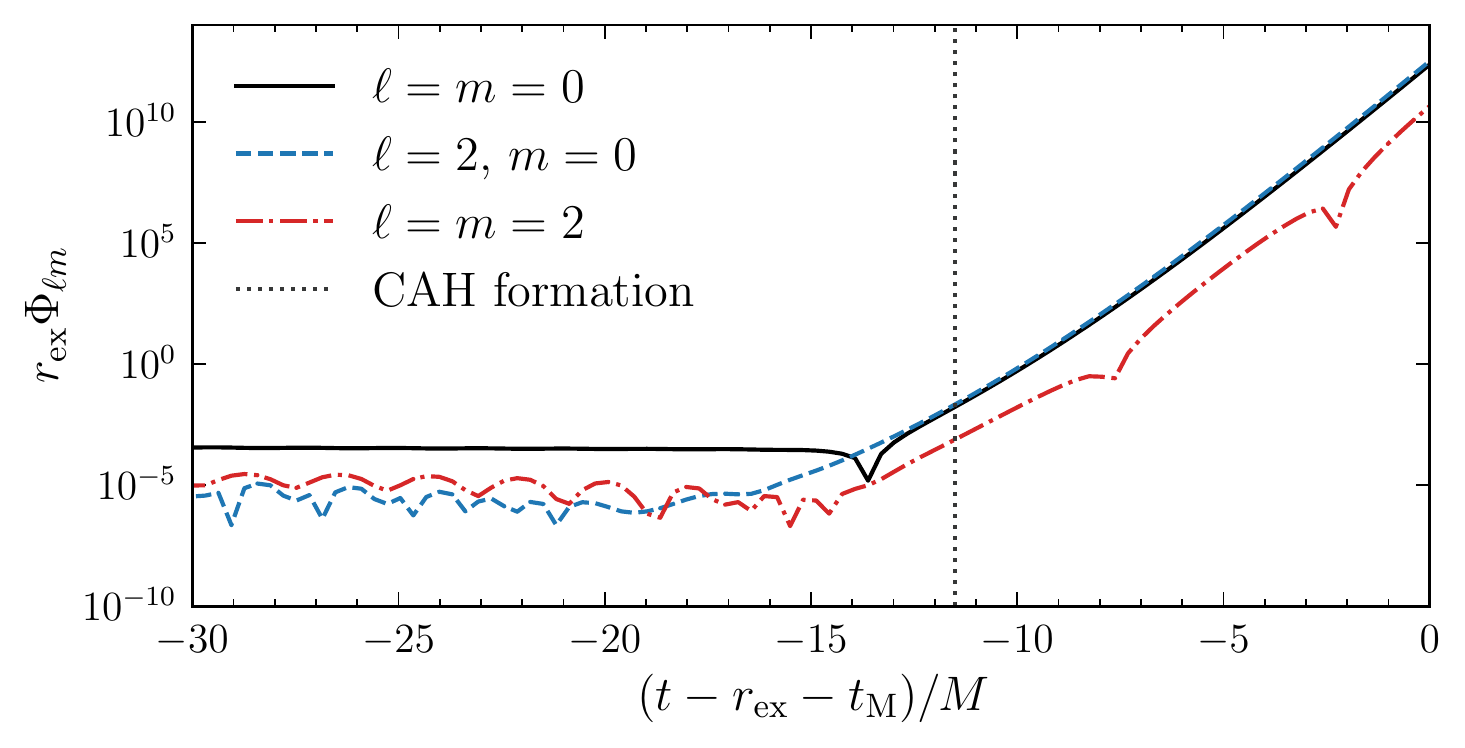"}
\caption{Evolution of the $\ell=m=0$ (solid line), $\ell=2$, $m=0$ (dashed line) and $\ell=m=2$ (dot-dashed line) scalar field multipoles
for the coupling $\beta=-10^3$; cf. Setup~A1 in Table~\ref{tab:prod_list}.
We rescale the multipoles by the extraction radius $\rex=50M$ and shift them such that
$(t-\rex-\tmg)/M=0$ indicates the time of merger
determined by the peak in the gravitational waveform.
For comparison we also show the formation of the \CAH (dotted line).
We observe that the scalar field grows exponentially about $20M$ prior to the merger.
}
\label{fig:d06_bet-1000_monopole}
\end{figure}

\begin{figure}[t]
\includegraphics[width=\columnwidth]{"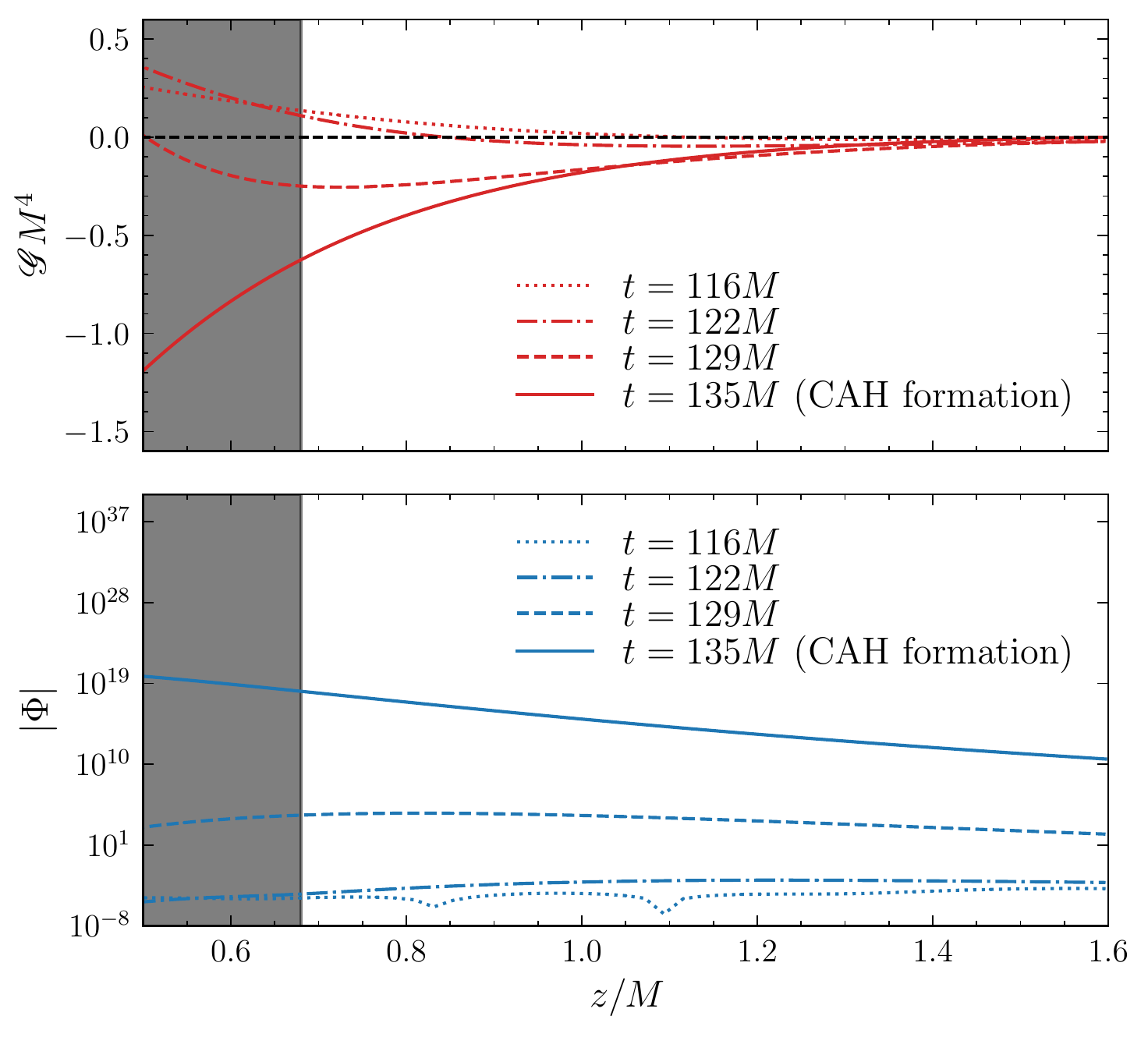"}
\caption{
Same as Fig.~\ref{fig:SpatialProfilesSetupA}, but for Setup~A1 in Table~\ref{tab:prod_list}.
We see that the \GB invariant (top panel) becomes negative
and triggers the excitation of the scalar field (bottom panel)
before the formation of the \CAH{,}
indicated by the gray region.
}
\label{fig:bet-1000_GB_Phi_profiles}
\end{figure}

To verify that the regions of negative \GB curvature
before the merger can induce the instability, we repeated the simulation of \ref{item:setup_A}
with a smaller initial \bh separation of $d=6M$
and a large-in-magnitude coupling constant $\beta=-10^{3}$;
see~Setup~A1 in Table~\ref{tab:prod_list}.
Although this choice of coupling, with $|\beta|\gg|\beta_{{\rm c},f}|=|\beta_{\rm c}(1,0.68)|$, may appear unphysical\footnote{Such a large value of $|\beta|$ may be unphysical because the phase space of nonlinear \bh solutions (i.e.,~including backreaction) has a band structure~\cite{Silva:2017uqg}: given a fixed value of $M$  there is a maximum value of $|\beta$| for which scalarized \bhs exist.
The domain of existence of scalarized \bh{s} depends on $f(\Phi)$, the \bh mass,
and its spin. Thus, if this $\beta$ is physical requires a careful, nonlinear analysis.
Here we focus only on the scalarization threshold.}
it has the desired effect of being able to cause the instability before the merger
and with a short time-scale; both effects are controlled by $|\beta|$.
This can be seen in Fig.~\ref{fig:d06_bet-1000_monopole}, where we show
the evolution of the scalar field multipoles,
and in Fig.~\ref{fig:bet-1000_GB_Phi_profiles}, where we show
the field's profile along the rotation axis.
Indeed, shortly after the \GB invariant becomes negative,
the scalar field grows exponentially and exceeds the magnitude of its background fluctuations
at about $t=20M$ before the \CAH{} is first found.

In summary, if $|\beta|$ is large enough, the \bh{s'}
late inspiral and merger may be affected by the \sGB scalar field.
However, for $|\beta|$-values near the scalarization threshold, the inspiral and merger of initially unscalarized \bh{} binaries, and their \gw emission, are identical to that of \gr and imprints of the \sGB scalar field only appear during the late ringdown.
Such effects may be very difficult (if not impossible) to detect, and this is what we refer to as \textit{stealth scalarization}.

\subsection{Spin-induced dynamical descalarization}
\label{sec:results_setupB}
In this section we present our key results obtained with simulation \ref{item:setup_B} in Table~\ref{tab:prod_list}
(see~Sec.~\ref{ssec:SimulationSetup}),
illustrated in Fig.~\ref{fig:diagram_simulationsB}.
The setup corresponds to two initially rotating, scalarized \bh{s}
(whose spin is anti-aligned with the orbital angular momentum)
that produce a unscalarized remnant
with a spin magnitude below the scalarization threshold
for any choice of the coupling constant.

\begin{figure}[t]
\includegraphics[width=\columnwidth]{"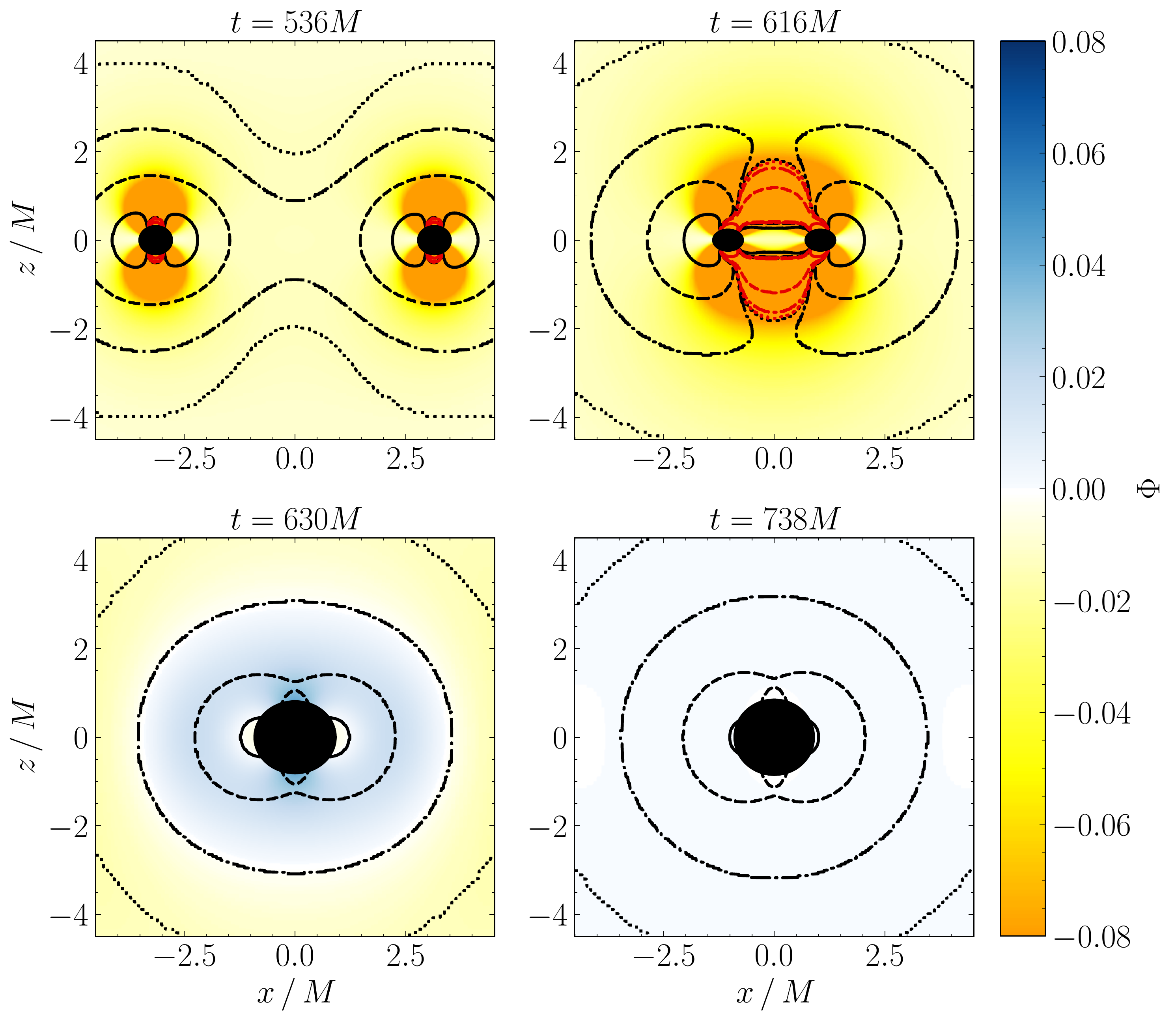"}
\caption{Snapshots of the scalar field, $\Phi$, and the \GB invariant, $\RGB$, in the $xz$-plane, corresponding to \ref{item:setup_B} in Table~\ref{tab:prod_list}.
The color map represents the amplitude of the scalar field.
The isocurvature contours indicate the magnitude of the \GB invariant with
$|\RGB M^{4}|=1$ (solid line),
$|\RGB M^{4}|=10^{-1}$ (dashed line),
$|\RGB M^{4}|=10^{-2}$ (dot-dashed line),
$|\RGB M^{4}|=10^{-3}$ (dotted line),
with positive (negative) values shown in black (red).
We show the inspiral (top left),
half an orbit before merger (top right),
$10M$ after the \CAH formation (bottom left)
and about $100M$ after the merger (bottom right).
}
\label{fig:SnapshotsSetupB}
\end{figure}

In Fig.~\ref{fig:SnapshotsSetupB}
we show snapshots of the scalar field and the \GB invariant in the $xz$-plane,
perpendicular to the binary's orbital plane,
during the inspiral (top left), half an orbit before the merger (top right),
at the merger (bottom left) and about $t=100M$ after the merger (bottom right).
We illustrate the location of the \bh{s} by their apparent horizons.
The color-coding represents the amplitude of the scalar field and is shared among all panels.
The contours represent the isocurvature lines
$|\RGB M^4| = \{1,\, 10^{-1},\, 10^{-2},\, 10^{-3} \}$,
with positive (negative) values shown in black (red).
The spin magnitude of the two inspiraling \bh{s} is sufficiently large to yield a \GB invariant that has negative regions outside the \bh{s'} horizon.
Combined with our choice of $|\beta|$, the \bh{s} sustain a scalar field bound state,
as shown in the top left panel of Fig.~\ref{fig:SnapshotsSetupB} and the \bh{s} carry
a scalar ``charge'' during the inspiral.
As the \bh{s} merge, they form a single, rotating \bh
which has a spin aligned with the orbital angular momentum and a magnitude of $\chi_f=0.48$.
For this spin magnitude, the \GB invariant is positive everywhere outside the \bh{'s} horizon, as shown in the bottom row of Fig.~\ref{fig:SnapshotsSetupB}.
As a consequence, the effective mass-squared becomes positive everywhere in the \bh{'s} exterior and
the scalar field bound states are no longer supported.
That is, the scalar field dissipates, and the \bh dynamically descalarizes, in agreement with
the no-hair theorem of Ref.~\cite{Silva:2017uqg}\footnote{
One might wonder if the final rotating \bh{} may become superradiantly unstable due to the presence of an effective mass for the scalar field $\Phi$. While the necessary conditions are satisfied~\cite{Shlapentokh-Rothman:2013ysa,Brito:2015oca,Moschidis:2016wew},
the instability
for a \bh{} of $\chi_{f}\lesssim0.5$
would evolve on e-folding timescales much longer than those studied here~\cite{Dolan:2007mj,Dolan:2012yt}; see Ref.~\cite{Dima:2020yac} for a comparison
against spin-induced scalarization.
Moreover, if backreaction of $\Phi$ onto the metric was included, the \bh{} mass and spin would decrease until the superradiance condition is saturated and the instability is turned off. Then, the scalar decays and the end-state is a \bh{} with no scalar field.
}.

\begin{figure}[t]
\includegraphics[width=0.91\columnwidth]{"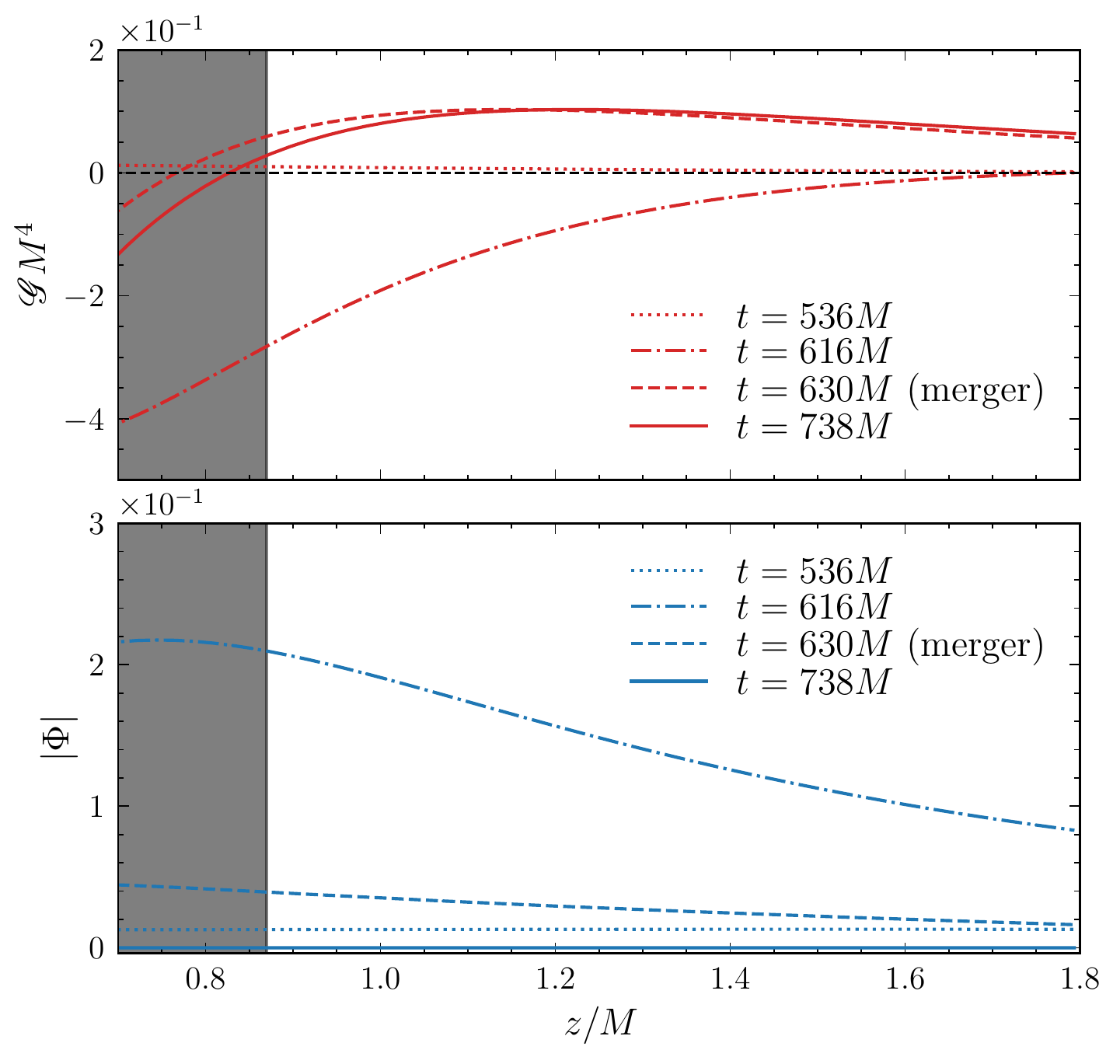"}
\caption{
Profiles of the \GB invariant (top panel) and of the scalar field (bottom panel) for \ref{item:setup_B} in Table~\ref{tab:prod_list} along the $z$-axis.
The lines correspond to different times during the evolution.
The shaded region indicates the \CAH{,} shown $100M$
after its formation.
The \GB invariant becomes positive outside the horizon when the \CAH{} is first formed.
Consequently, the scalar field magnitude decreases and the remnant \bh descalarizes.
}
\label{fig:SpatialProfilesSetupB}
\end{figure}

These phenomena can also be seen in Fig.~\ref{fig:SpatialProfilesSetupB}, where we show the profiles of the \GB invariant (top panel) and of the scalar field (bottom panel) along the $z$-axis
(parallel to orbital angular momentum) for several instants during the evolution.
The shaded region indicates the apparent horizon of the final \bh{.}
The \GB invariant remains negative outside the individual \bh{s}
during their (late) inspiral.
Only when the \CAH first forms,
does the \GB invariant become positive everywhere outside the remnant \bh{'s} horizon
At this point, the effective mass-squared becomes positive, the tachyonic instability that kept
each \bh scalarized switches off, and the scalar field dissipates as shown in the bottom panel of Fig.~\ref{fig:SpatialProfilesSetupB}.

\begin{figure}[t]
\includegraphics[width=\columnwidth]{"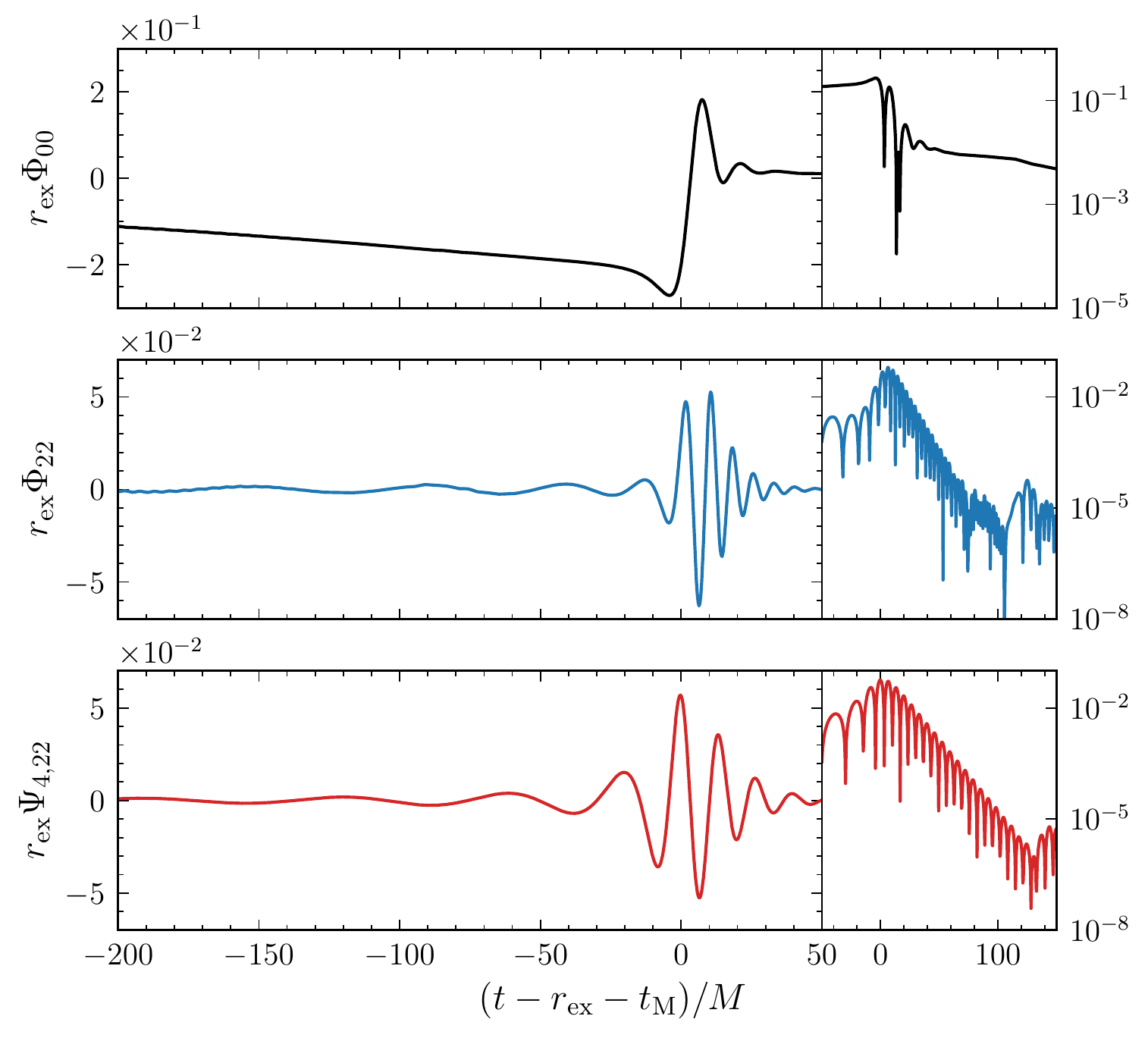"}
\caption{Evolution of the scalar field monopole (top panel)
and quadrupole (middle panel) and gravitational quadrupole (bottom panel)
for \ref{item:setup_B} in Table~\ref{tab:setups}.
The waveforms are rescaled by the extractions radius $\rex = 100 M$ and shifted in time such that $(t - \rex - \tmg)/M = 0$ at the merger.
In the insets we show the absolute values of
the multipoles, in logarithmic scale, during the merger and ringdown.
}
\label{fig:WaveformsSetupB}
\end{figure}

Does the presence of scalar charges during the inspiral produce scalar radiation?
The answer is affirmative as can be seen in Fig.~\ref{fig:WaveformsSetupB} where we show the time evolution of the scalar field monopole (top panel) and quadrupole (middle panel). For comparison, we also display the gravitational quadrupole waveform of the background spacetime (bottom panel).
The scalar field monopole quantifies the development of the combined scalar charge of the \bh{} binary
measured on spheres of radius $\rex=100M$,
i.e., enclosing the entire binary.
The total scalar charge remains approximately constant during the inspiral
as the coupling is close to its critical value.
Its magnitude increases about $10M$ before the merger which coincides with the formation of a joined region in which the \GB invariant is negative due to the proximity of the two \bh{s}
As the \bh{s} merge into a single rotating remnant with a spin below the threshold for the spin-induced scalarization,
the scalar charge decays as illustrated in the inset of Fig.~\ref{fig:WaveformsSetupB} (top panel).
Because the scalar charges anchored around each \bh{} follow the holes' orbital motion, they generate scalar radiation.
In general, one would expect the scalar dipole to dominate the signal, as is also the case for shift-symmetric \sGB gravity~\cite{Witek:2018dmd,Shiralilou:2020gah,Shiralilou:2021mfl}.
In the simulations shown here, however, the scalar dipole is suppressed due to the symmetry of the system (equal mass and spin of the companions), and the $\ell=m=2$ multipole dominates.

The scalar waveform is displayed in the middle panel of Fig.~\ref{fig:WaveformsSetupB} and shows the familiar chirp pattern:
its amplitude and frequency increase as the scalar charges inspiral (following the  inspiraling \bh{s} in the background),
and culminates in a peak as the \bhs merge.
The phase of the scalar field quadrupole clearly tracks its gravitational counterpart. Therefore, we deduce that the morphology (phase evolution) of the observed scalar quadrupole radiation is a result of the orbital dynamics of the system.
A sufficiently large magnitude of the coupling constant may lead to an additional scalarization of the $\ell=2$ mode, which would become manifest as an exponential growth of the signal superposed with the chirp.
This situation is analogous to the evolutions
with positive coupling shown in our previous work \cite{Silva:2020omi}.

After the merger, the scalar quadrupole exhibits a quasi-normal ringdown pattern, i.e., an exponentially damped sinusoid, shown in the inset of Fig.~\ref{fig:WaveformsSetupB} (middle panel).
Here, in contrast to Ref.~\cite{Silva:2020omi},  descalarization occurs due to the vanishing
of negative \GB regions outside the remnant \bh (because its final spin is $|\chi_f| < 0.5$),
rather than due to a reduction of positive curvature (because of an increase in mass).
We note that the scalar field rings down on similar timescales as the \gw signal shown in the bottom panel of Fig.~\ref{fig:WaveformsSetupB} for comparison.
Therefore, one might expect a modification to the \gw ringdown if backreaction
onto the spacetime is included.

\section{Discussion}
\label{sec:discussions}
In this paper, we continued our study of dynamical scalarization and descalarization in binary \bh{} mergers in \sGB gravity by extending our previous work~\cite{Silva:2020omi}. The latter focused on a positive coupling constant between the
scalar field and the \GB invariant, yielding dynamical descalarization in binary \bh{} mergers.
As a natural continuation, here we studied a negative coupling for which the \bh{s}' spins play a major role in determining the onset of scalarization. In particular, we have shown that
the merger remnant
can either dynamically scalarize or dynamically descalarize depending on its spin and mass.

Spin-induced dynamical scalarization occurs when the merger remnant grows a scalar charge during coalescence due to the large spin of the remnant. In cases like this, the initial binary components lack a charge because their spins are not large enough to support one~\cite{Dima:2020yac,Hod:2020jjy,Herdeiro:2020wei,Berti:2020kgk,Doneva:2020nbb,Hod:2022hfm}.
However, after the objects merge, the remnant \bh{} spins faster than either component, allowing for a charge to grow.
We found that it is possible for the scalar charge to grow as early as 1--2 orbits before a \CAH has formed if the coupling $|\beta|$ is extremely large.
This occurs because there are spacetime regions before merger (and near the poles of the future remnant) with a negative \GB invariant, and a sufficient large value of $|\beta|$ allows bound states to form fast enough.
We also found that if the coupling $|\beta|$ is close to the threshold, then scalarization occurs only in the late ringdown, because of the timescale required for the bound states to form.

Is such spin-induced scalarization detectable with current or future GW observatories?
For values of $|\beta|$ near the scalarization threshold the instability timescale
is large and the effects of the scalar field growth would only appear at times
much later than the merger and, more importantly, after the start of the ringdown.
Hence, the inspiral-merger-ringdown
of such a binary would be indistinguishable from one in \gr, and scalarization would be a hidden or ``stealth'' effect, i.e.,~the remnant \bh would acquire a charge, but its formation would not lead to an easily measurable effect.
For instance, during the \gw ringdown, which is dominated by the fundamental $(\ell,m) = (2,2)$ \qnm frequency, we know that at a spin of $\chi \approx 0.68$, the decay time
is approximately $\tau \approx 12.3 M$~\cite{Berti:2009kk}.
Hence, after $100 M$ from
the peak in the waveform,
the dominant mode has decayed by roughly $\exp(-t/\tau) \approx \exp(-100/12.3) \approx 10^{-4}$.
If the dominant \qnm frequency begins to be modified only after $100 M$, the GW has decayed so much that detecting this change or constraining it would be essentially impossible.

Is there no hope to detect such late times scalarization? Not necessarily. If we were to include the scalar field backreaction onto the spacetime, one could entertain the possibility that the late time growth of the scalar field (in particular of $\Phi_{22}$) and the subsequent readjustment of the spinning remnant \bh to its scalarized counterpart could result in a \textit{second} \gw signal.
Confirming this possibility and, if confirmed, characterizing such a \gw signal is left for future work.

Spin-induced dynamical descalarization occurs when the merger remnant loses its scalar charge during coalescence due to the low spin of the remnant. In cases like this, the initial binary components are spinning fast enough that each of them has a scalar charge and the remnant descalarizes if it has spin $\chi_f \leqslant 0.5$.
Here we demonstrated this effect in a example in which the initial binary components have their spin angular momenta anti-aligned with the orbital angular momentum.
The merger produces
a remnant \bh with $\chi_{f} = 0.48$, for which no scalar field bound states are supported and the field
is radiated away
shortly ($\sim 10M$) after the \CAH formation.

Is such spin-induced descalarization detectable with current or future GW observatories? For such descalarization to be detectable, one must first detect that the binary components were scalarized during the inspiral. Our simulations showed that the scalar charges lead to scalar quadrupole radiation because of the highly symmetric configurations (equal mass, equal spin magnitude) we chose to evolve. More realistic astrophysical configurations (with unequal masses and unequal spin magnitudes) forces the binary to emit scalar dipole radiation. Such emission of dipole or quadrupole radiation  accelerates the inspiral, and thus affect the GW phase at $-$1\PN and 0\PN respectively, as shown in shift-symmetric theories~\cite{Yagi:2011xp,Yagi:2012vf,Yagi:2013mbt,Shiralilou:2020gah,Shiralilou:2021mfl,Julie:2019sab}.
These effects in the inspiral are observable and can thus be constrained with current ground-based~\cite{Yunes:2016jcc,Nair:2019iur,Yamada:2019zrb,Perkins:2021mhb,Lyu:2022gdr} and future detectors~\cite{Carson:2019fxr,Perkins:2020tra} within the parameterized post-Einsteinian framework~\cite{Yunes:2009ke,Cornish:2011ys,Tahura:2018zuq,Perkins:2022fhr}, provided the binary is of sufficiently low mass such that enough of the inspiral is observed~\cite{Perkins:2020tra}.
In fact, a constraint of this type was recently obtained using the GW190814 event~\cite{LIGOScientific:2020zkf} in~\cite{Wong:2022wni}.

Let us then assume, for the sake of argument, that some future event reveals a scalar charge in the inspiraling binary components. Our results then indicate that
descalarization may be detectable, if there is enough signal-to-noise ratio in the merger and ringdown~\cite{Witek:2018dmd,Okounkova:2020rqw}.
This is because this process occurs at the same time and with the same timescales as the \gw merger and ringdown, see Fig.~\ref{fig:WaveformsSetupB}.
Future work could study the backreaction of the scalar field onto the metric to determine the magnitude of these modifications in the transient phase, without which one cannot assess detectability confidently.
Our results indicate that descalarization might be best probed with a full inspiral-merger-ringdown analysis of the \gw signal.

\acknowledgements
We thank
A.~C\'ardenas-Avenda\~{n}o,
A.~Dima,
and R.~Teixeira da Costa
for useful discussions.
H.~W. acknowledges financial support provided by NSF grants
No.~OAC-2004879 and No.~PHY-2110416, and Royal Society (UK) Research Grant RGF\textbackslash R1\textbackslash 180073.
NY acknowledges support from the Simons Foundation through award 896696.
M.~E acknowledges support from the Science and Technology Facilities Council (STFC).
% Clusters
This work made use of several computing infrastructures: the Extreme Science and Engineering Discovery Environment (XSEDE) Expanse through the allocation TG-PHY210114, which is supported by NSF Grant No. ACI-1548562; the Blue Waters sustained-petascale computing project which was supported by NSF Award No.~OCI-0725070 and No.~ACI-1238993, the State of Illinois and the National Geospatial Intelligence Agency
(Blue Waters is a joint effort of the University of Illinois at Urbana-Champaign and its National Center for Supercomputing Applications); the Illinois Campus Cluster, a computing
resource that is operated by the Illinois Campus Cluster Program
(ICCP) in conjunction with the National Center for Supercomputing Applications (NCSA) and which is supported by funds from the University of Illinois at Urbana-Champaign; the {\sc Minerva} cluster at the Max Planck Institute for Gravitational Physics; the Leibnitz Supercomputing Centre SuperMUC-NG under
PRACE Grant No. 2018194669; the J\"ulich
Supercomputing Center JUWELS HPC under PRACE
Grant No. 2020225359; COSMA7 in Durham
and Leicester DiAL HPC under DiRAC RAC13 Grant No.
ACTP238 and the Cambridge Data Driven CSD3 facility which is operated by the University of Cambridge
Research Computing on behalf of the STFC DiRAC
HPC Facility.
Our simulations were performed with
\canuda~\cite{witek_helvi_2021_5520862,Zilhao:2015tya,Witek:2018dmd}
and the \ETK~\cite{steven_r_brandt_2021_5770803, Loffler:2011ay,Zilhao:2013hia}.
Some of our calculations were performed with the {\sc Mathematica} packages
{\sc xPert}~\cite{Brizuela:2008ra} and {\sc
Invar}~\cite{Martin-Garcia:2007bqa,Martin-Garcia:2008yei}, part of the {\sc xAct/xTensor}
suite~\cite{Mart_n_Garc_a_2008,xAct}.
The figures in this work were produced with {\sc Matplotlib}~\cite{Hunter:2007}, {\sc kuibit}~\cite{kuibit} and {\sc TikZ-Feynman}~\cite{Ellis:2016jkw}.

\appendix
\section{Full suite of simulations}~\label{app:all_simulations}

\begin{table*}[t]
\begin{tabular}{c|ccccccc|cc}
\hline
\hline
Setup            & $d/M$ & $\chi_{1} $ & $\chi_{2} $  & $\chi_{f} $ & $\beta$ & $\beta_{{\rm c}, 1}$ & $\beta_{{\rm c}, f}$& process \\
\hline
A    & $10$ & $0$   & $0$   & $0.68$ & $-14.30$ & -- & $-12.96$ & $\bar{s} + \bar{s} \rightarrow s_{\uparrow}$  \\
A1   & $6$ & $0$ & $0$ & $0.68$& $-1000$ & -- & $-12.96$ & $\bar{s}+\bar{s} \rightarrow s_{\uparrow}$ \\
A2   & $10$ & $0.6$ & $0.6$ & $0.85$ & $-2.9$   & $-10.55$ & $-3.01$ & $\bar{s}_{\uparrow} + \bar{s}_{\uparrow} \rightarrow s_{\uparrow}$ \\
A3   & $10$ & $0.6$ & $0.6$ & $0.85$ & $-12.0$ & $-10.55$ & $-3.01$ & $s_{\uparrow} + s_{\uparrow} \rightarrow s_{\uparrow}$ \\
A4   & $10$ & $0.0$ & $0.6$ & $0.77$ & $-12.0$ & $-10.55$ & $-5.59$ & $\bar{s} + s_{\uparrow} \rightarrow s_{\uparrow}$ \\
\hline
B   & $10$ & $-0.6$ & $-0.6$ & $0.48$& $-11.50$ & $-10.55$ & -- & $s_{\downarrow} + s_{\downarrow} \rightarrow \bar{s}_{\uparrow}$ \\
B2  & $10$ & $0.4$ & $-0.6$ & $0.64$ & $-12.0$ & $-10.55$ & $-21.50$ & $\bar{s}_{\uparrow} + s_{\downarrow} \rightarrow \bar{s}_{\uparrow}$  \\
\hline
\hline
\end{tabular}
\caption{List of our complete series of simulations.
We denote the initial separation $d/M$ with $M$ being the total mass,
$\chi_{1}$ and $\chi_{2} $ are the initial dimensionless spin parameters of each \bh, and
$\chi_{f} $ is the final dimensionless spin parameter of the remnant.
We use $\bar{s}$ and $s$ to denote unscalarized and scalarized states, respectively, and
the subscript $\uparrow$ ($\downarrow$) indicates spin aligned (anti-aligned) with the
orbital  angular momentum.
The coupling chosen for each simulation is given by $\beta$, whereas $\beta_{{\rm c}, 1}$ and $\beta_{{\rm c}, f}$ denote the critical couplings for the component/remnant \bh{s} respectively.
}
\label{tab:prod_list}
\end{table*}

We ran a larger series of simulations, listed in Table~\ref{tab:prod_list}, of equal-mass \bh binaries with varying initial spin that show a qualitatively same behaviour as the runs presented in the main text.
In particular, we simulated a series of initially spinning, unscalarized black holes that formed a scalarized remnant with larger spin. We also list example simulations in which one or both initial \bh{s} are scalarized and they merge into an unscalarized remnant.

\section{Validation tests}~\label{app:CodeVerification}
To validate our code, we performed a suite of convergence tests.
We ran \ref{item:setup_B}, our numerically most demanding setup, at a lower resolution of $\dif x_{\rm low} = 0.8 M$ and a higher resolution of $\dif x_{\rm high} = 0.625 M$.
The runs in the main text use a medium resolution of $\dif x_{\rm med} = 0.7 M$.
The grid setup is the same across all simulations, see Sec.~\ref{sec:Numerical_Method}.
We estimated the order of convergence $n$ and its associated convergence factor $Q_{n}$,
\begin{equation}
    Q_{n}=\frac{\left(\dif x_{\rm low}\right)^{n}-\left(\dif x_{\rm med}\right)^{n}}{\left(\dif x_{\rm med}\right)^{n}-\left(\dif x_{\rm high}\right)^{n}}
    \,.
    \label{eqn:conve_order}
\end{equation}

\begin{figure}[t]
\includegraphics[width=\columnwidth]{"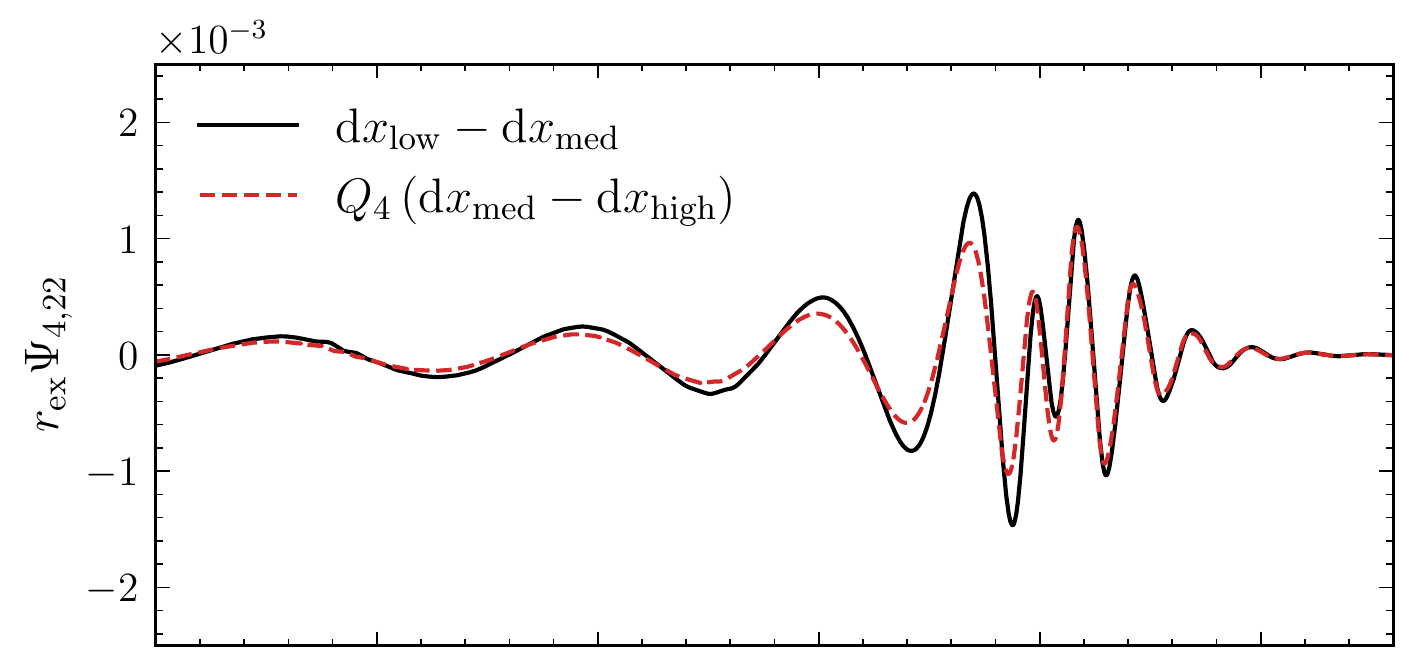"}
\includegraphics[width=\columnwidth]{"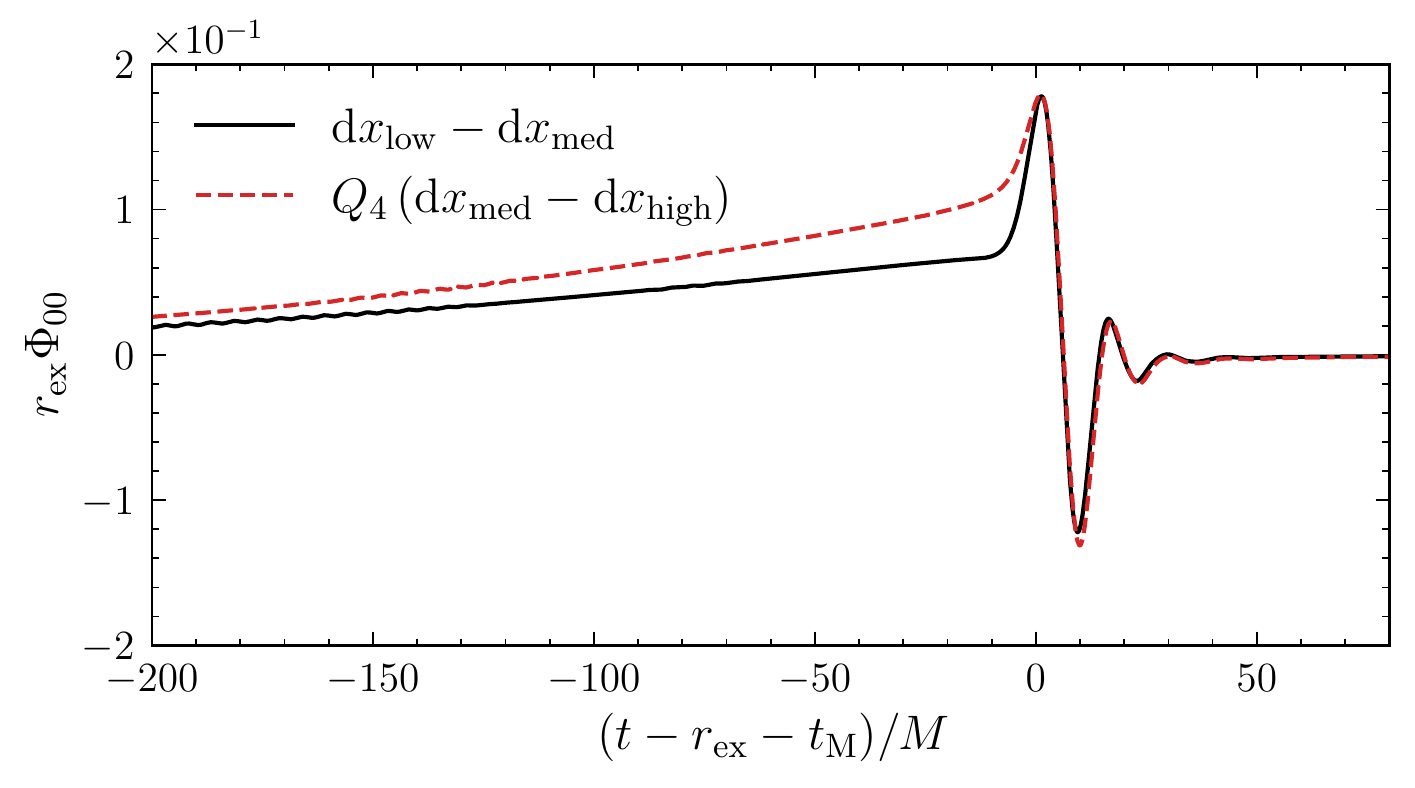"}
\caption{Convergence plots for the $\ell=m=2$ mode of the  gravitational waveform (left panel) and
the $\ell=m=0$ mode of the scalar field (right panel). In both panels,
we show the difference between the low and medium resolution run (solid line) and the medium and high resolution run (dashed line).
The latter is rescaled by $Q_{4} = 1.94$, indicating fourth order convergence.
The lines are shifted in time such that $(t-\rex-\tmg)/M=0$ indicates the time of merger and
they are rescaled by the extraction radius $\rex = 100 M$.
}
\label{fig:convergence}
\end{figure}

We computed the $n$ and $Q_{n}$
for the gravitational waveform, $\Psi_{4,22}$, of the background spacetime and for the scalar charge.
We show the corresponding convergence plots in
Fig.~\ref{fig:convergence}.
For $\Psi_{4,22}$ we find fourth order convergence, and we estimate the numerical (truncation) error to be $\Psi_{4,22}/ \Psi_{4,22} \leqslant 0.8 \%$.
For the scalar field charge, $\Phi_{00}$, we
also find fourth order convergence.
performed a convergence test on its $\ell=m=0$ multipole. We show our result in the right panel of Fig.~\ref{fig:convergence}.

We find a cumulative error $\Delta \Phi_{00}/ \Phi_{00} \leqslant 30 \%$ in the late inspiral.
The numerical error in the merger and ringdown is $\Delta \Phi_{00} /\Phi_{00} \leqslant 15\%$.
As we restrict this work to a qualitative analysis, this error does not affect the main results of the paper. Further quantitative work, such as forecasting constraints on the theory
would require this issue to be addressed.

\begin{figure}[b]
\includegraphics[width=\columnwidth]{"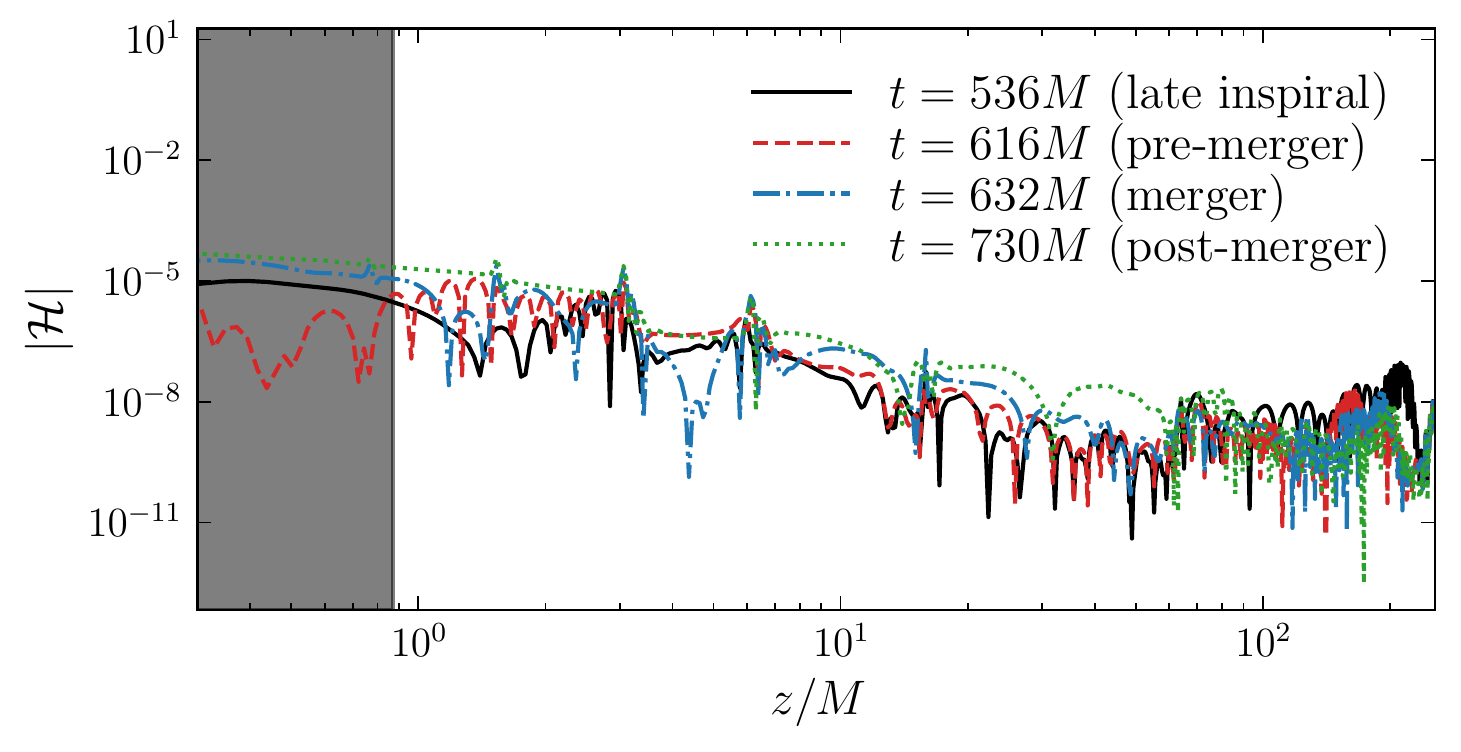"}
\caption{Hamiltonian constraint along the z-axis during the late-inspiral (solid black), half an orbit before merger (dashed red), at the time of merger from the peak of the gravitational waveform (dash-dot blue) and $100 M$ after merger (dotted green). The shaded region indicates the CAH, shown $100 M$ after merger.}
\label{fig:Hamiltonian}
\end{figure}

Finally, in Fig.~\ref{fig:Hamiltonian}, we show the Hamiltonian constraint $\H$ along the $z$-axis for \ref{item:setup_B} at different time instants. The constraint violation remains below $10^{-5}$ outside the \bh horizon through the simulation.

\bibliography{master_biblio.bib}

%apsrev4-2.bst 2019-01-14 (MD) hand-edited version of apsrev4-1.bst
%Control: key (0)
%Control: author (8) initials jnrlst
%Control: editor formatted (1) identically to author
%Control: production of article title (0) allowed
%Control: page (0) single
%Control: year (1) truncated
%Control: production of eprint (0) enabled
\begin{thebibliography}{132}%
\makeatletter
\providecommand \@ifxundefined [1]{%
 \@ifx{#1\undefined}
}%
\providecommand \@ifnum [1]{%
 \ifnum #1\expandafter \@firstoftwo
 \else \expandafter \@secondoftwo
 \fi
}%
\providecommand \@ifx [1]{%
 \ifx #1\expandafter \@firstoftwo
 \else \expandafter \@secondoftwo
 \fi
}%
\providecommand \natexlab [1]{#1}%
\providecommand \enquote  [1]{``#1''}%
\providecommand \bibnamefont  [1]{#1}%
\providecommand \bibfnamefont [1]{#1}%
\providecommand \citenamefont [1]{#1}%
\providecommand \href@noop [0]{\@secondoftwo}%
\providecommand \href [0]{\begingroup \@sanitize@url \@href}%
\providecommand \@href[1]{\@@startlink{#1}\@@href}%
\providecommand \@@href[1]{\endgroup#1\@@endlink}%
\providecommand \@sanitize@url [0]{\catcode `\\12\catcode `\$12\catcode
  `\&12\catcode `\#12\catcode `\^12\catcode `\_12\catcode `\%12\relax}%
\providecommand \@@startlink[1]{}%
\providecommand \@@endlink[0]{}%
\providecommand \url  [0]{\begingroup\@sanitize@url \@url }%
\providecommand \@url [1]{\endgroup\@href {#1}{\urlprefix }}%
\providecommand \urlprefix  [0]{URL }%
\providecommand \Eprint [0]{\href }%
\providecommand \doibase [0]{https://doi.org/}%
\providecommand \selectlanguage [0]{\@gobble}%
\providecommand \bibinfo  [0]{\@secondoftwo}%
\providecommand \bibfield  [0]{\@secondoftwo}%
\providecommand \translation [1]{[#1]}%
\providecommand \BibitemOpen [0]{}%
\providecommand \bibitemStop [0]{}%
\providecommand \bibitemNoStop [0]{.\EOS\space}%
\providecommand \EOS [0]{\spacefactor3000\relax}%
\providecommand \BibitemShut  [1]{\csname bibitem#1\endcsname}%
\let\auto@bib@innerbib\@empty
%</preamble>
\bibitem [{\citenamefont {Abbott}\ \emph
  {et~al.}(2019{\natexlab{a}})\citenamefont {Abbott} \emph
  {et~al.}}]{LIGOScientific:2018mvr}%
  \BibitemOpen
  \bibfield  {author} {\bibinfo {author} {\bibfnamefont {B.}~\bibnamefont
  {Abbott}} \emph {et~al.} (\bibinfo {collaboration} {LIGO Scientific,
  Virgo}),\ }\bibfield  {title} {\bibinfo {title} {{GWTC-1: A
  Gravitational-Wave Transient Catalog of Compact Binary Mergers Observed by
  LIGO and Virgo during the First and Second Observing Runs}},\ }\href
  {https://doi.org/10.1103/PhysRevX.9.031040} {\bibfield  {journal} {\bibinfo
  {journal} {Phys. Rev. X}\ }\textbf {\bibinfo {volume} {9}},\ \bibinfo {pages}
  {031040} (\bibinfo {year} {2019}{\natexlab{a}})},\ \Eprint
  {https://arxiv.org/abs/1811.12907} {arXiv:1811.12907 [astro-ph.HE]}
  \BibitemShut {NoStop}%
\bibitem [{\citenamefont {Abbott}\ \emph
  {et~al.}(2021{\natexlab{a}})\citenamefont {Abbott} \emph
  {et~al.}}]{LIGOScientific:2020ibl}%
  \BibitemOpen
  \bibfield  {author} {\bibinfo {author} {\bibfnamefont {R.}~\bibnamefont
  {Abbott}} \emph {et~al.} (\bibinfo {collaboration} {LIGO Scientific,
  Virgo}),\ }\bibfield  {title} {\bibinfo {title} {{GWTC-2: Compact Binary
  Coalescences Observed by LIGO and Virgo During the First Half of the Third
  Observing Run}},\ }\href {https://doi.org/10.1103/PhysRevX.11.021053}
  {\bibfield  {journal} {\bibinfo  {journal} {Phys. Rev. X}\ }\textbf {\bibinfo
  {volume} {11}},\ \bibinfo {pages} {021053} (\bibinfo {year}
  {2021}{\natexlab{a}})},\ \Eprint {https://arxiv.org/abs/2010.14527}
  {arXiv:2010.14527 [gr-qc]} \BibitemShut {NoStop}%
\bibitem [{\citenamefont {Abbott}\ \emph
  {et~al.}(2021{\natexlab{b}})\citenamefont {Abbott} \emph
  {et~al.}}]{LIGOScientific:2021djp}%
  \BibitemOpen
  \bibfield  {author} {\bibinfo {author} {\bibfnamefont {R.}~\bibnamefont
  {Abbott}} \emph {et~al.} (\bibinfo {collaboration} {LIGO Scientific, VIRGO,
  KAGRA}),\ }\href@noop {} {\bibinfo {title} {{GWTC-3: Compact Binary
  Coalescences Observed by LIGO and Virgo During the Second Part of the Third
  Observing Run}}} (\bibinfo {year} {2021}{\natexlab{b}}),\ \Eprint
  {https://arxiv.org/abs/2111.03606} {arXiv:2111.03606 [gr-qc]} \BibitemShut
  {NoStop}%
\bibitem [{\citenamefont {Yunes}\ and\ \citenamefont
  {Siemens}(2013)}]{Yunes:2013dva}%
  \BibitemOpen
  \bibfield  {author} {\bibinfo {author} {\bibfnamefont {N.}~\bibnamefont
  {Yunes}}\ and\ \bibinfo {author} {\bibfnamefont {X.}~\bibnamefont
  {Siemens}},\ }\bibfield  {title} {\bibinfo {title} {{Gravitational-Wave Tests
  of General Relativity with Ground-Based Detectors and Pulsar
  Timing-Arrays}},\ }\href {https://doi.org/10.12942/lrr-2013-9} {\bibfield
  {journal} {\bibinfo  {journal} {Living Rev. Rel.}\ }\textbf {\bibinfo
  {volume} {16}},\ \bibinfo {pages} {9} (\bibinfo {year} {2013})},\ \Eprint
  {https://arxiv.org/abs/1304.3473} {arXiv:1304.3473 [gr-qc]} \BibitemShut
  {NoStop}%
%%CITATION = ARXIV:1304.3473;%%
\bibitem [{\citenamefont {Berti}\ \emph {et~al.}(2015)\citenamefont {Berti}
  \emph {et~al.}}]{Berti:2015itd}%
  \BibitemOpen
  \bibfield  {author} {\bibinfo {author} {\bibfnamefont {E.}~\bibnamefont
  {Berti}} \emph {et~al.},\ }\bibfield  {title} {\bibinfo {title} {{Testing
  General Relativity with Present and Future Astrophysical Observations}},\
  }\href {https://doi.org/10.1088/0264-9381/32/24/243001} {\bibfield  {journal}
  {\bibinfo  {journal} {Class. Quant. Grav.}\ }\textbf {\bibinfo {volume}
  {32}},\ \bibinfo {pages} {243001} (\bibinfo {year} {2015})},\ \Eprint
  {https://arxiv.org/abs/1501.07274} {arXiv:1501.07274 [gr-qc]} \BibitemShut
  {NoStop}%
%%CITATION = ARXIV:1501.07274;%%
\bibitem [{\citenamefont {Berti}\ \emph
  {et~al.}(2018{\natexlab{a}})\citenamefont {Berti}, \citenamefont {Yagi},\
  and\ \citenamefont {Yunes}}]{Berti:2018cxi}%
  \BibitemOpen
  \bibfield  {author} {\bibinfo {author} {\bibfnamefont {E.}~\bibnamefont
  {Berti}}, \bibinfo {author} {\bibfnamefont {K.}~\bibnamefont {Yagi}},\ and\
  \bibinfo {author} {\bibfnamefont {N.}~\bibnamefont {Yunes}},\ }\bibfield
  {title} {\bibinfo {title} {{Extreme Gravity Tests with Gravitational Waves
  from Compact Binary Coalescences: (I) Inspiral-Merger}},\ }\href
  {https://doi.org/10.1007/s10714-018-2362-8} {\bibfield  {journal} {\bibinfo
  {journal} {Gen. Rel. Grav.}\ }\textbf {\bibinfo {volume} {50}},\ \bibinfo
  {pages} {46} (\bibinfo {year} {2018}{\natexlab{a}})},\ \Eprint
  {https://arxiv.org/abs/1801.03208} {arXiv:1801.03208 [gr-qc]} \BibitemShut
  {NoStop}%
\bibitem [{\citenamefont {Berti}\ \emph
  {et~al.}(2018{\natexlab{b}})\citenamefont {Berti}, \citenamefont {Yagi},
  \citenamefont {Yang},\ and\ \citenamefont {Yunes}}]{Berti:2018vdi}%
  \BibitemOpen
  \bibfield  {author} {\bibinfo {author} {\bibfnamefont {E.}~\bibnamefont
  {Berti}}, \bibinfo {author} {\bibfnamefont {K.}~\bibnamefont {Yagi}},
  \bibinfo {author} {\bibfnamefont {H.}~\bibnamefont {Yang}},\ and\ \bibinfo
  {author} {\bibfnamefont {N.}~\bibnamefont {Yunes}},\ }\bibfield  {title}
  {\bibinfo {title} {{Extreme Gravity Tests with Gravitational Waves from
  Compact Binary Coalescences: (II) Ringdown}},\ }\href
  {https://doi.org/10.1007/s10714-018-2372-6} {\bibfield  {journal} {\bibinfo
  {journal} {Gen. Rel. Grav.}\ }\textbf {\bibinfo {volume} {50}},\ \bibinfo
  {pages} {49} (\bibinfo {year} {2018}{\natexlab{b}})},\ \Eprint
  {https://arxiv.org/abs/1801.03587} {arXiv:1801.03587 [gr-qc]} \BibitemShut
  {NoStop}%
\bibitem [{\citenamefont {Yunes}\ \emph {et~al.}(2016)\citenamefont {Yunes},
  \citenamefont {Yagi},\ and\ \citenamefont {Pretorius}}]{Yunes:2016jcc}%
  \BibitemOpen
  \bibfield  {author} {\bibinfo {author} {\bibfnamefont {N.}~\bibnamefont
  {Yunes}}, \bibinfo {author} {\bibfnamefont {K.}~\bibnamefont {Yagi}},\ and\
  \bibinfo {author} {\bibfnamefont {F.}~\bibnamefont {Pretorius}},\ }\bibfield
  {title} {\bibinfo {title} {{Theoretical Physics Implications of the Binary
  Black-Hole Mergers GW150914 and GW151226}},\ }\href
  {https://doi.org/10.1103/PhysRevD.94.084002} {\bibfield  {journal} {\bibinfo
  {journal} {Phys. Rev. D}\ }\textbf {\bibinfo {volume} {94}},\ \bibinfo
  {pages} {084002} (\bibinfo {year} {2016})},\ \Eprint
  {https://arxiv.org/abs/1603.08955} {arXiv:1603.08955 [gr-qc]} \BibitemShut
  {NoStop}%
\bibitem [{\citenamefont {Abbott}\ \emph {et~al.}(2016)\citenamefont {Abbott}
  \emph {et~al.}}]{LIGOScientific:2016lio}%
  \BibitemOpen
  \bibfield  {author} {\bibinfo {author} {\bibfnamefont {B.~P.}\ \bibnamefont
  {Abbott}} \emph {et~al.} (\bibinfo {collaboration} {LIGO Scientific,
  Virgo}),\ }\bibfield  {title} {\bibinfo {title} {{Tests of general relativity
  with GW150914}},\ }\href {https://doi.org/10.1103/PhysRevLett.116.221101}
  {\bibfield  {journal} {\bibinfo  {journal} {Phys. Rev. Lett.}\ }\textbf
  {\bibinfo {volume} {116}},\ \bibinfo {pages} {221101} (\bibinfo {year}
  {2016})},\ \bibinfo {note} {[Erratum: Phys.Rev.Lett. 121, 129902 (2018)]},\
  \Eprint {https://arxiv.org/abs/1602.03841} {arXiv:1602.03841 [gr-qc]}
  \BibitemShut {NoStop}%
\bibitem [{\citenamefont {Abbott}\ \emph
  {et~al.}(2019{\natexlab{b}})\citenamefont {Abbott} \emph
  {et~al.}}]{LIGOScientific:2018dkp}%
  \BibitemOpen
  \bibfield  {author} {\bibinfo {author} {\bibfnamefont {B.~P.}\ \bibnamefont
  {Abbott}} \emph {et~al.} (\bibinfo {collaboration} {LIGO Scientific,
  Virgo}),\ }\bibfield  {title} {\bibinfo {title} {{Tests of General Relativity
  with GW170817}},\ }\href {https://doi.org/10.1103/PhysRevLett.123.011102}
  {\bibfield  {journal} {\bibinfo  {journal} {Phys. Rev. Lett.}\ }\textbf
  {\bibinfo {volume} {123}},\ \bibinfo {pages} {011102} (\bibinfo {year}
  {2019}{\natexlab{b}})},\ \Eprint {https://arxiv.org/abs/1811.00364}
  {arXiv:1811.00364 [gr-qc]} \BibitemShut {NoStop}%
\bibitem [{\citenamefont {Abbott}\ \emph
  {et~al.}(2019{\natexlab{c}})\citenamefont {Abbott} \emph
  {et~al.}}]{LIGOScientific:2019fpa}%
  \BibitemOpen
  \bibfield  {author} {\bibinfo {author} {\bibfnamefont {B.}~\bibnamefont
  {Abbott}} \emph {et~al.} (\bibinfo {collaboration} {LIGO Scientific,
  Virgo}),\ }\bibfield  {title} {\bibinfo {title} {{Tests of General Relativity
  with the Binary Black Hole Signals from the LIGO-Virgo Catalog GWTC-1}},\
  }\href {https://doi.org/10.1103/PhysRevD.100.104036} {\bibfield  {journal}
  {\bibinfo  {journal} {Phys. Rev. D}\ }\textbf {\bibinfo {volume} {100}},\
  \bibinfo {pages} {104036} (\bibinfo {year} {2019}{\natexlab{c}})},\ \Eprint
  {https://arxiv.org/abs/1903.04467} {arXiv:1903.04467 [gr-qc]} \BibitemShut
  {NoStop}%
\bibitem [{\citenamefont {Cardenas-Avendano}\ \emph {et~al.}(2020)\citenamefont
  {Cardenas-Avendano}, \citenamefont {Nampalliwar},\ and\ \citenamefont
  {Yunes}}]{Cardenas-Avendano:2019zxd}%
  \BibitemOpen
  \bibfield  {author} {\bibinfo {author} {\bibfnamefont {A.}~\bibnamefont
  {Cardenas-Avendano}}, \bibinfo {author} {\bibfnamefont {S.}~\bibnamefont
  {Nampalliwar}},\ and\ \bibinfo {author} {\bibfnamefont {N.}~\bibnamefont
  {Yunes}},\ }\bibfield  {title} {\bibinfo {title} {{Gravitational-wave versus
  X-ray tests of strong-field gravity}},\ }\href
  {https://doi.org/10.1088/1361-6382/ab8f64} {\bibfield  {journal} {\bibinfo
  {journal} {Class. Quant. Grav.}\ }\textbf {\bibinfo {volume} {37}},\ \bibinfo
  {pages} {135008} (\bibinfo {year} {2020})},\ \Eprint
  {https://arxiv.org/abs/1912.08062} {arXiv:1912.08062 [gr-qc]} \BibitemShut
  {NoStop}%
\bibitem [{\citenamefont {Abbott}\ \emph
  {et~al.}(2021{\natexlab{c}})\citenamefont {Abbott} \emph
  {et~al.}}]{LIGOScientific:2020tif}%
  \BibitemOpen
  \bibfield  {author} {\bibinfo {author} {\bibfnamefont {R.}~\bibnamefont
  {Abbott}} \emph {et~al.} (\bibinfo {collaboration} {LIGO Scientific,
  Virgo}),\ }\bibfield  {title} {\bibinfo {title} {{Tests of general relativity
  with binary black holes from the second LIGO-Virgo gravitational-wave
  transient catalog}},\ }\href {https://doi.org/10.1103/PhysRevD.103.122002}
  {\bibfield  {journal} {\bibinfo  {journal} {Phys. Rev. D}\ }\textbf {\bibinfo
  {volume} {103}},\ \bibinfo {pages} {122002} (\bibinfo {year}
  {2021}{\natexlab{c}})},\ \Eprint {https://arxiv.org/abs/2010.14529}
  {arXiv:2010.14529 [gr-qc]} \BibitemShut {NoStop}%
\bibitem [{\citenamefont {Silva}\ \emph
  {et~al.}(2021{\natexlab{a}})\citenamefont {Silva}, \citenamefont {Holgado},
  \citenamefont {C\'ardenas-Avenda\~no},\ and\ \citenamefont
  {Yunes}}]{Silva:2020acr}%
  \BibitemOpen
  \bibfield  {author} {\bibinfo {author} {\bibfnamefont {H.~O.}\ \bibnamefont
  {Silva}}, \bibinfo {author} {\bibfnamefont {A.~M.}\ \bibnamefont {Holgado}},
  \bibinfo {author} {\bibfnamefont {A.}~\bibnamefont {C\'ardenas-Avenda\~no}},\
  and\ \bibinfo {author} {\bibfnamefont {N.}~\bibnamefont {Yunes}},\ }\bibfield
   {title} {\bibinfo {title} {{Astrophysical and theoretical physics
  implications from multimessenger neutron star observations}},\ }\href
  {https://doi.org/10.1103/PhysRevLett.126.181101} {\bibfield  {journal}
  {\bibinfo  {journal} {Phys. Rev. Lett.}\ }\textbf {\bibinfo {volume} {126}},\
  \bibinfo {pages} {181101} (\bibinfo {year} {2021}{\natexlab{a}})},\ \Eprint
  {https://arxiv.org/abs/2004.01253} {arXiv:2004.01253 [gr-qc]} \BibitemShut
  {NoStop}%
\bibitem [{\citenamefont {Abbott}\ \emph
  {et~al.}(2021{\natexlab{d}})\citenamefont {Abbott} \emph
  {et~al.}}]{LIGOScientific:2021sio}%
  \BibitemOpen
  \bibfield  {author} {\bibinfo {author} {\bibfnamefont {R.}~\bibnamefont
  {Abbott}} \emph {et~al.} (\bibinfo {collaboration} {LIGO Scientific, VIRGO,
  KAGRA}),\ }\href@noop {} {\bibinfo {title} {{Tests of General Relativity with
  GWTC-3}}} (\bibinfo {year} {2021}{\natexlab{d}}),\ \Eprint
  {https://arxiv.org/abs/2112.06861} {arXiv:2112.06861 [gr-qc]} \BibitemShut
  {NoStop}%
\bibitem [{\citenamefont {Ghosh}\ \emph {et~al.}(2021)\citenamefont {Ghosh},
  \citenamefont {Brito},\ and\ \citenamefont {Buonanno}}]{Ghosh:2021mrv}%
  \BibitemOpen
  \bibfield  {author} {\bibinfo {author} {\bibfnamefont {A.}~\bibnamefont
  {Ghosh}}, \bibinfo {author} {\bibfnamefont {R.}~\bibnamefont {Brito}},\ and\
  \bibinfo {author} {\bibfnamefont {A.}~\bibnamefont {Buonanno}},\ }\bibfield
  {title} {\bibinfo {title} {{Constraints on quasinormal-mode frequencies with
  LIGO-Virgo binary\textendash{}black-hole observations}},\ }\href
  {https://doi.org/10.1103/PhysRevD.103.124041} {\bibfield  {journal} {\bibinfo
   {journal} {Phys. Rev. D}\ }\textbf {\bibinfo {volume} {103}},\ \bibinfo
  {pages} {124041} (\bibinfo {year} {2021})},\ \Eprint
  {https://arxiv.org/abs/2104.01906} {arXiv:2104.01906 [gr-qc]} \BibitemShut
  {NoStop}%
\bibitem [{\citenamefont {Carullo}(2021)}]{Carullo:2021dui}%
  \BibitemOpen
  \bibfield  {author} {\bibinfo {author} {\bibfnamefont {G.}~\bibnamefont
  {Carullo}},\ }\bibfield  {title} {\bibinfo {title} {{Enhancing modified
  gravity detection from gravitational-wave observations using the parametrized
  ringdown spin expansion coeffcients formalism}},\ }\href
  {https://doi.org/10.1103/PhysRevD.103.124043} {\bibfield  {journal} {\bibinfo
   {journal} {Phys. Rev. D}\ }\textbf {\bibinfo {volume} {103}},\ \bibinfo
  {pages} {124043} (\bibinfo {year} {2021})},\ \Eprint
  {https://arxiv.org/abs/2102.05939} {arXiv:2102.05939 [gr-qc]} \BibitemShut
  {NoStop}%
\bibitem [{\citenamefont {Sennett}\ \emph {et~al.}(2020)\citenamefont
  {Sennett}, \citenamefont {Brito}, \citenamefont {Buonanno}, \citenamefont
  {Gorbenko},\ and\ \citenamefont {Senatore}}]{Sennett:2019bpc}%
  \BibitemOpen
  \bibfield  {author} {\bibinfo {author} {\bibfnamefont {N.}~\bibnamefont
  {Sennett}}, \bibinfo {author} {\bibfnamefont {R.}~\bibnamefont {Brito}},
  \bibinfo {author} {\bibfnamefont {A.}~\bibnamefont {Buonanno}}, \bibinfo
  {author} {\bibfnamefont {V.}~\bibnamefont {Gorbenko}},\ and\ \bibinfo
  {author} {\bibfnamefont {L.}~\bibnamefont {Senatore}},\ }\bibfield  {title}
  {\bibinfo {title} {{Gravitational-Wave Constraints on an Effective
  Field-Theory Extension of General Relativity}},\ }\href
  {https://doi.org/10.1103/PhysRevD.102.044056} {\bibfield  {journal} {\bibinfo
   {journal} {Phys. Rev. D}\ }\textbf {\bibinfo {volume} {102}},\ \bibinfo
  {pages} {044056} (\bibinfo {year} {2020})},\ \Eprint
  {https://arxiv.org/abs/1912.09917} {arXiv:1912.09917 [gr-qc]} \BibitemShut
  {NoStop}%
\bibitem [{\citenamefont {Mehta}\ \emph {et~al.}(2022)\citenamefont {Mehta},
  \citenamefont {Buonanno}, \citenamefont {Cotesta}, \citenamefont {Ghosh},
  \citenamefont {Sennett},\ and\ \citenamefont {Steinhoff}}]{Mehta:2022pcn}%
  \BibitemOpen
  \bibfield  {author} {\bibinfo {author} {\bibfnamefont {A.~K.}\ \bibnamefont
  {Mehta}}, \bibinfo {author} {\bibfnamefont {A.}~\bibnamefont {Buonanno}},
  \bibinfo {author} {\bibfnamefont {R.}~\bibnamefont {Cotesta}}, \bibinfo
  {author} {\bibfnamefont {A.}~\bibnamefont {Ghosh}}, \bibinfo {author}
  {\bibfnamefont {N.}~\bibnamefont {Sennett}},\ and\ \bibinfo {author}
  {\bibfnamefont {J.}~\bibnamefont {Steinhoff}},\ }\href@noop {} {\bibinfo
  {title} {{Tests of General Relativity with Gravitational-Wave Observations
  using a Flexible-Theory-Independent Method}}} (\bibinfo {year} {2022}),\
  \Eprint {https://arxiv.org/abs/2203.13937} {arXiv:2203.13937 [gr-qc]}
  \BibitemShut {NoStop}%
\bibitem [{\citenamefont {Zhao}\ \emph {et~al.}(2019)\citenamefont {Zhao},
  \citenamefont {Shao}, \citenamefont {Cao},\ and\ \citenamefont
  {Ma}}]{Zhao:2019suc}%
  \BibitemOpen
  \bibfield  {author} {\bibinfo {author} {\bibfnamefont {J.}~\bibnamefont
  {Zhao}}, \bibinfo {author} {\bibfnamefont {L.}~\bibnamefont {Shao}}, \bibinfo
  {author} {\bibfnamefont {Z.}~\bibnamefont {Cao}},\ and\ \bibinfo {author}
  {\bibfnamefont {B.-Q.}\ \bibnamefont {Ma}},\ }\bibfield  {title} {\bibinfo
  {title} {{Reduced-order surrogate models for scalar-tensor gravity in the
  strong field regime and applications to binary pulsars and GW170817}},\
  }\href {https://doi.org/10.1103/PhysRevD.100.064034} {\bibfield  {journal}
  {\bibinfo  {journal} {Phys. Rev. D}\ }\textbf {\bibinfo {volume} {100}},\
  \bibinfo {pages} {064034} (\bibinfo {year} {2019})},\ \Eprint
  {https://arxiv.org/abs/1907.00780} {arXiv:1907.00780 [gr-qc]} \BibitemShut
  {NoStop}%
\bibitem [{\citenamefont {Wong}\ \emph {et~al.}(2022)\citenamefont {Wong},
  \citenamefont {Herdeiro},\ and\ \citenamefont {Radu}}]{Wong:2022wni}%
  \BibitemOpen
  \bibfield  {author} {\bibinfo {author} {\bibfnamefont {L.~K.}\ \bibnamefont
  {Wong}}, \bibinfo {author} {\bibfnamefont {C.~A.~R.}\ \bibnamefont
  {Herdeiro}},\ and\ \bibinfo {author} {\bibfnamefont {E.}~\bibnamefont
  {Radu}},\ }\href@noop {} {\bibinfo {title} {{Constraining spontaneous black
  hole scalarization in scalar-tensor-Gauss-Bonnet theories with current
  gravitational-wave data}}} (\bibinfo {year} {2022}),\ \Eprint
  {https://arxiv.org/abs/2204.09038} {arXiv:2204.09038 [gr-qc]} \BibitemShut
  {NoStop}%
\bibitem [{\citenamefont {Nair}\ \emph {et~al.}(2019)\citenamefont {Nair},
  \citenamefont {Perkins}, \citenamefont {Silva},\ and\ \citenamefont
  {Yunes}}]{Nair:2019iur}%
  \BibitemOpen
  \bibfield  {author} {\bibinfo {author} {\bibfnamefont {R.}~\bibnamefont
  {Nair}}, \bibinfo {author} {\bibfnamefont {S.}~\bibnamefont {Perkins}},
  \bibinfo {author} {\bibfnamefont {H.~O.}\ \bibnamefont {Silva}},\ and\
  \bibinfo {author} {\bibfnamefont {N.}~\bibnamefont {Yunes}},\ }\bibfield
  {title} {\bibinfo {title} {{Fundamental Physics Implications for
  Higher-Curvature Theories from Binary Black Hole Signals in the LIGO-Virgo
  Catalog GWTC-1}},\ }\href {https://doi.org/10.1103/PhysRevLett.123.191101}
  {\bibfield  {journal} {\bibinfo  {journal} {Phys. Rev. Lett.}\ }\textbf
  {\bibinfo {volume} {123}},\ \bibinfo {pages} {191101} (\bibinfo {year}
  {2019})},\ \Eprint {https://arxiv.org/abs/1905.00870} {arXiv:1905.00870
  [gr-qc]} \BibitemShut {NoStop}%
\bibitem [{\citenamefont {Yamada}\ \emph {et~al.}(2019)\citenamefont {Yamada},
  \citenamefont {Narikawa},\ and\ \citenamefont {Tanaka}}]{Yamada:2019zrb}%
  \BibitemOpen
  \bibfield  {author} {\bibinfo {author} {\bibfnamefont {K.}~\bibnamefont
  {Yamada}}, \bibinfo {author} {\bibfnamefont {T.}~\bibnamefont {Narikawa}},\
  and\ \bibinfo {author} {\bibfnamefont {T.}~\bibnamefont {Tanaka}},\
  }\bibfield  {title} {\bibinfo {title} {{Testing massive-field modifications
  of gravity via gravitational waves}},\ }\href
  {https://doi.org/10.1093/ptep/ptz103} {\bibfield  {journal} {\bibinfo
  {journal} {PTEP}\ }\textbf {\bibinfo {volume} {2019}},\ \bibinfo {pages}
  {103E01} (\bibinfo {year} {2019})},\ \Eprint
  {https://arxiv.org/abs/1905.11859} {arXiv:1905.11859 [gr-qc]} \BibitemShut
  {NoStop}%
\bibitem [{\citenamefont {Perkins}\ \emph
  {et~al.}(2021{\natexlab{a}})\citenamefont {Perkins}, \citenamefont {Nair},
  \citenamefont {Silva},\ and\ \citenamefont {Yunes}}]{Perkins:2021mhb}%
  \BibitemOpen
  \bibfield  {author} {\bibinfo {author} {\bibfnamefont {S.~E.}\ \bibnamefont
  {Perkins}}, \bibinfo {author} {\bibfnamefont {R.}~\bibnamefont {Nair}},
  \bibinfo {author} {\bibfnamefont {H.~O.}\ \bibnamefont {Silva}},\ and\
  \bibinfo {author} {\bibfnamefont {N.}~\bibnamefont {Yunes}},\ }\bibfield
  {title} {\bibinfo {title} {{Improved gravitational-wave constraints on
  higher-order curvature theories of gravity}},\ }\href
  {https://doi.org/10.1103/PhysRevD.104.024060} {\bibfield  {journal} {\bibinfo
   {journal} {Phys. Rev. D}\ }\textbf {\bibinfo {volume} {104}},\ \bibinfo
  {pages} {024060} (\bibinfo {year} {2021}{\natexlab{a}})},\ \Eprint
  {https://arxiv.org/abs/2104.11189} {arXiv:2104.11189 [gr-qc]} \BibitemShut
  {NoStop}%
\bibitem [{\citenamefont {Lyu}\ \emph {et~al.}(2022)\citenamefont {Lyu},
  \citenamefont {Jiang},\ and\ \citenamefont {Yagi}}]{Lyu:2022gdr}%
  \BibitemOpen
  \bibfield  {author} {\bibinfo {author} {\bibfnamefont {Z.}~\bibnamefont
  {Lyu}}, \bibinfo {author} {\bibfnamefont {N.}~\bibnamefont {Jiang}},\ and\
  \bibinfo {author} {\bibfnamefont {K.}~\bibnamefont {Yagi}},\ }\bibfield
  {title} {\bibinfo {title} {{Constraints on Einstein-dilation-Gauss-Bonnet
  gravity from black hole-neutron star gravitational wave events}},\ }\href
  {https://doi.org/10.1103/PhysRevD.105.064001} {\bibfield  {journal} {\bibinfo
   {journal} {Phys. Rev. D}\ }\textbf {\bibinfo {volume} {105}},\ \bibinfo
  {pages} {064001} (\bibinfo {year} {2022})},\ \Eprint
  {https://arxiv.org/abs/2201.02543} {arXiv:2201.02543 [gr-qc]} \BibitemShut
  {NoStop}%
\bibitem [{\citenamefont {Silva}\ \emph {et~al.}(2022)\citenamefont {Silva},
  \citenamefont {Ghosh},\ and\ \citenamefont {Buonanno}}]{Silva:2022srr}%
  \BibitemOpen
  \bibfield  {author} {\bibinfo {author} {\bibfnamefont {H.~O.}\ \bibnamefont
  {Silva}}, \bibinfo {author} {\bibfnamefont {A.}~\bibnamefont {Ghosh}},\ and\
  \bibinfo {author} {\bibfnamefont {A.}~\bibnamefont {Buonanno}},\ }\href@noop
  {} {\bibinfo {title} {{Black-hole ringdown as a probe of higher-curvature
  gravity theories}}} (\bibinfo {year} {2022}),\ \Eprint
  {https://arxiv.org/abs/2205.05132} {arXiv:2205.05132 [gr-qc]} \BibitemShut
  {NoStop}%
\bibitem [{\citenamefont {Yagi}\ \emph {et~al.}(2016)\citenamefont {Yagi},
  \citenamefont {Stein},\ and\ \citenamefont {Yunes}}]{Yagi:2015oca}%
  \BibitemOpen
  \bibfield  {author} {\bibinfo {author} {\bibfnamefont {K.}~\bibnamefont
  {Yagi}}, \bibinfo {author} {\bibfnamefont {L.~C.}\ \bibnamefont {Stein}},\
  and\ \bibinfo {author} {\bibfnamefont {N.}~\bibnamefont {Yunes}},\ }\bibfield
   {title} {\bibinfo {title} {{Challenging the Presence of Scalar Charge and
  Dipolar Radiation in Binary Pulsars}},\ }\href
  {https://doi.org/10.1103/PhysRevD.93.024010} {\bibfield  {journal} {\bibinfo
  {journal} {Phys. Rev. D}\ }\textbf {\bibinfo {volume} {93}},\ \bibinfo
  {pages} {024010} (\bibinfo {year} {2016})},\ \Eprint
  {https://arxiv.org/abs/1510.02152} {arXiv:1510.02152 [gr-qc]} \BibitemShut
  {NoStop}%
\bibitem [{\citenamefont {Metsaev}\ and\ \citenamefont
  {Tseytlin}(1987)}]{Metsaev:1987zx}%
  \BibitemOpen
  \bibfield  {author} {\bibinfo {author} {\bibfnamefont {R.}~\bibnamefont
  {Metsaev}}\ and\ \bibinfo {author} {\bibfnamefont {A.~A.}\ \bibnamefont
  {Tseytlin}},\ }\bibfield  {title} {\bibinfo {title} {{Order $\alpha'$ (Two
  Loop) Equivalence of the String Equations of Motion and the Sigma Model Weyl
  Invariance Conditions: Dependence on the Dilaton and the Antisymmetric
  Tensor}},\ }\href {https://doi.org/10.1016/0550-3213(87)90077-0} {\bibfield
  {journal} {\bibinfo  {journal} {Nucl. Phys. B}\ }\textbf {\bibinfo {volume}
  {293}},\ \bibinfo {pages} {385} (\bibinfo {year} {1987})}\BibitemShut
  {NoStop}%
\bibitem [{\citenamefont {Kanti}\ and\ \citenamefont
  {Tamvakis}(1995)}]{Kanti:1995cp}%
  \BibitemOpen
  \bibfield  {author} {\bibinfo {author} {\bibfnamefont {P.}~\bibnamefont
  {Kanti}}\ and\ \bibinfo {author} {\bibfnamefont {K.}~\bibnamefont
  {Tamvakis}},\ }\bibfield  {title} {\bibinfo {title} {{Classical moduli
  O($\alpha'$) hair}},\ }\href {https://doi.org/10.1103/PhysRevD.52.3506}
  {\bibfield  {journal} {\bibinfo  {journal} {Phys. Rev. D}\ }\textbf {\bibinfo
  {volume} {52}},\ \bibinfo {pages} {3506} (\bibinfo {year} {1995})},\ \Eprint
  {https://arxiv.org/abs/hep-th/9504031} {arXiv:hep-th/9504031} \BibitemShut
  {NoStop}%
\bibitem [{\citenamefont {Cano}\ and\ \citenamefont
  {Ruip\'erez}(2022)}]{Cano:2021rey}%
  \BibitemOpen
  \bibfield  {author} {\bibinfo {author} {\bibfnamefont {P.~A.}\ \bibnamefont
  {Cano}}\ and\ \bibinfo {author} {\bibfnamefont {A.}~\bibnamefont
  {Ruip\'erez}},\ }\bibfield  {title} {\bibinfo {title} {{String gravity in
  $D=4$}},\ }\href {https://doi.org/10.1103/PhysRevD.105.044022} {\bibfield
  {journal} {\bibinfo  {journal} {Phys. Rev. D}\ }\textbf {\bibinfo {volume}
  {105}},\ \bibinfo {pages} {044022} (\bibinfo {year} {2022})},\ \Eprint
  {https://arxiv.org/abs/2111.04750} {arXiv:2111.04750 [hep-th]} \BibitemShut
  {NoStop}%
\bibitem [{\citenamefont {Charmousis}(2015)}]{Charmousis:2014mia}%
  \BibitemOpen
  \bibfield  {author} {\bibinfo {author} {\bibfnamefont {C.}~\bibnamefont
  {Charmousis}},\ }\bibfield  {title} {\bibinfo {title} {{From Lovelock to
  Horndeski`s Generalized Scalar Tensor Theory}},\ }\href
  {https://doi.org/10.1007/978-3-319-10070-8_2} {\bibfield  {journal} {\bibinfo
   {journal} {Lect. Notes Phys.}\ }\textbf {\bibinfo {volume} {892}},\ \bibinfo
  {pages} {25} (\bibinfo {year} {2015})},\ \Eprint
  {https://arxiv.org/abs/1405.1612} {arXiv:1405.1612 [gr-qc]} \BibitemShut
  {NoStop}%
\bibitem [{\citenamefont {Kobayashi}\ \emph {et~al.}(2011)\citenamefont
  {Kobayashi}, \citenamefont {Yamaguchi},\ and\ \citenamefont
  {Yokoyama}}]{Kobayashi:2011nu}%
  \BibitemOpen
  \bibfield  {author} {\bibinfo {author} {\bibfnamefont {T.}~\bibnamefont
  {Kobayashi}}, \bibinfo {author} {\bibfnamefont {M.}~\bibnamefont
  {Yamaguchi}},\ and\ \bibinfo {author} {\bibfnamefont {J.}~\bibnamefont
  {Yokoyama}},\ }\bibfield  {title} {\bibinfo {title} {{Generalized
  G-inflation: Inflation with the most general second-order field equations}},\
  }\href {https://doi.org/10.1143/PTP.126.511} {\bibfield  {journal} {\bibinfo
  {journal} {Prog. Theor. Phys.}\ }\textbf {\bibinfo {volume} {126}},\ \bibinfo
  {pages} {511} (\bibinfo {year} {2011})},\ \Eprint
  {https://arxiv.org/abs/1105.5723} {arXiv:1105.5723 [hep-th]} \BibitemShut
  {NoStop}%
\bibitem [{\citenamefont {Kobayashi}(2019)}]{Kobayashi:2019hrl}%
  \BibitemOpen
  \bibfield  {author} {\bibinfo {author} {\bibfnamefont {T.}~\bibnamefont
  {Kobayashi}},\ }\bibfield  {title} {\bibinfo {title} {{Horndeski theory and
  beyond: a review}},\ }\href {https://doi.org/10.1088/1361-6633/ab2429}
  {\bibfield  {journal} {\bibinfo  {journal} {Rept. Prog. Phys.}\ }\textbf
  {\bibinfo {volume} {82}},\ \bibinfo {pages} {086901} (\bibinfo {year}
  {2019})},\ \Eprint {https://arxiv.org/abs/1901.07183} {arXiv:1901.07183
  [gr-qc]} \BibitemShut {NoStop}%
\bibitem [{\citenamefont {Yagi}\ \emph
  {et~al.}(2012{\natexlab{a}})\citenamefont {Yagi}, \citenamefont {Stein},
  \citenamefont {Yunes},\ and\ \citenamefont {Tanaka}}]{Yagi:2011xp}%
  \BibitemOpen
  \bibfield  {author} {\bibinfo {author} {\bibfnamefont {K.}~\bibnamefont
  {Yagi}}, \bibinfo {author} {\bibfnamefont {L.~C.}\ \bibnamefont {Stein}},
  \bibinfo {author} {\bibfnamefont {N.}~\bibnamefont {Yunes}},\ and\ \bibinfo
  {author} {\bibfnamefont {T.}~\bibnamefont {Tanaka}},\ }\bibfield  {title}
  {\bibinfo {title} {{Post-Newtonian, Quasi-Circular Binary Inspirals in
  Quadratic Modified Gravity}},\ }\href
  {https://doi.org/10.1103/PhysRevD.85.064022} {\bibfield  {journal} {\bibinfo
  {journal} {Phys. Rev. D}\ }\textbf {\bibinfo {volume} {85}},\ \bibinfo
  {pages} {064022} (\bibinfo {year} {2012}{\natexlab{a}})},\ \bibinfo {note}
  {[Erratum: Phys.Rev.D 93, 029902 (2016)]},\ \Eprint
  {https://arxiv.org/abs/1110.5950} {arXiv:1110.5950 [gr-qc]} \BibitemShut
  {NoStop}%
\bibitem [{\citenamefont {Yagi}\ \emph
  {et~al.}(2012{\natexlab{b}})\citenamefont {Yagi}, \citenamefont {Yunes},\
  and\ \citenamefont {Tanaka}}]{Yagi:2012vf}%
  \BibitemOpen
  \bibfield  {author} {\bibinfo {author} {\bibfnamefont {K.}~\bibnamefont
  {Yagi}}, \bibinfo {author} {\bibfnamefont {N.}~\bibnamefont {Yunes}},\ and\
  \bibinfo {author} {\bibfnamefont {T.}~\bibnamefont {Tanaka}},\ }\bibfield
  {title} {\bibinfo {title} {{Gravitational Waves from Quasi-Circular Black
  Hole Binaries in Dynamical Chern-Simons Gravity}},\ }\href
  {https://doi.org/10.1103/PhysRevLett.116.169902} {\bibfield  {journal}
  {\bibinfo  {journal} {Phys. Rev. Lett.}\ }\textbf {\bibinfo {volume} {109}},\
  \bibinfo {pages} {251105} (\bibinfo {year} {2012}{\natexlab{b}})},\ \bibinfo
  {note} {[Erratum: Phys.Rev.Lett. 116, 169902 (2016), Erratum: Phys.Rev.Lett.
  124, 029901 (2020)]},\ \Eprint {https://arxiv.org/abs/1208.5102}
  {arXiv:1208.5102 [gr-qc]} \BibitemShut {NoStop}%
\bibitem [{\citenamefont {Yagi}\ \emph {et~al.}(2013)\citenamefont {Yagi},
  \citenamefont {Stein}, \citenamefont {Yunes},\ and\ \citenamefont
  {Tanaka}}]{Yagi:2013mbt}%
  \BibitemOpen
  \bibfield  {author} {\bibinfo {author} {\bibfnamefont {K.}~\bibnamefont
  {Yagi}}, \bibinfo {author} {\bibfnamefont {L.~C.}\ \bibnamefont {Stein}},
  \bibinfo {author} {\bibfnamefont {N.}~\bibnamefont {Yunes}},\ and\ \bibinfo
  {author} {\bibfnamefont {T.}~\bibnamefont {Tanaka}},\ }\bibfield  {title}
  {\bibinfo {title} {{Isolated and Binary Neutron Stars in Dynamical
  Chern-Simons Gravity}},\ }\href {https://doi.org/10.1103/PhysRevD.87.084058}
  {\bibfield  {journal} {\bibinfo  {journal} {Phys. Rev. D}\ }\textbf {\bibinfo
  {volume} {87}},\ \bibinfo {pages} {084058} (\bibinfo {year} {2013})},\
  \bibinfo {note} {[Erratum: Phys.Rev.D 93, 089909 (2016)]},\ \Eprint
  {https://arxiv.org/abs/1302.1918} {arXiv:1302.1918 [gr-qc]} \BibitemShut
  {NoStop}%
\bibitem [{\citenamefont {Shiralilou}\ \emph {et~al.}(2021)\citenamefont
  {Shiralilou}, \citenamefont {Hinderer}, \citenamefont {Nissanke},
  \citenamefont {Ortiz},\ and\ \citenamefont {Witek}}]{Shiralilou:2020gah}%
  \BibitemOpen
  \bibfield  {author} {\bibinfo {author} {\bibfnamefont {B.}~\bibnamefont
  {Shiralilou}}, \bibinfo {author} {\bibfnamefont {T.}~\bibnamefont
  {Hinderer}}, \bibinfo {author} {\bibfnamefont {S.}~\bibnamefont {Nissanke}},
  \bibinfo {author} {\bibfnamefont {N.}~\bibnamefont {Ortiz}},\ and\ \bibinfo
  {author} {\bibfnamefont {H.}~\bibnamefont {Witek}},\ }\bibfield  {title}
  {\bibinfo {title} {{Nonlinear curvature effects in gravitational waves from
  inspiralling black hole binaries}},\ }\href
  {https://doi.org/10.1103/PhysRevD.103.L121503} {\bibfield  {journal}
  {\bibinfo  {journal} {Phys. Rev. D}\ }\textbf {\bibinfo {volume} {103}},\
  \bibinfo {pages} {L121503} (\bibinfo {year} {2021})},\ \Eprint
  {https://arxiv.org/abs/2012.09162} {arXiv:2012.09162 [gr-qc]} \BibitemShut
  {NoStop}%
\bibitem [{\citenamefont {Shiralilou}\ \emph {et~al.}(2022)\citenamefont
  {Shiralilou}, \citenamefont {Hinderer}, \citenamefont {Nissanke},
  \citenamefont {Ortiz},\ and\ \citenamefont {Witek}}]{Shiralilou:2021mfl}%
  \BibitemOpen
  \bibfield  {author} {\bibinfo {author} {\bibfnamefont {B.}~\bibnamefont
  {Shiralilou}}, \bibinfo {author} {\bibfnamefont {T.}~\bibnamefont
  {Hinderer}}, \bibinfo {author} {\bibfnamefont {S.~M.}\ \bibnamefont
  {Nissanke}}, \bibinfo {author} {\bibfnamefont {N.}~\bibnamefont {Ortiz}},\
  and\ \bibinfo {author} {\bibfnamefont {H.}~\bibnamefont {Witek}},\ }\bibfield
   {title} {\bibinfo {title} {{Post-Newtonian gravitational and scalar waves in
  scalar-Gauss\textendash{}Bonnet gravity}},\ }\href
  {https://doi.org/10.1088/1361-6382/ac4196} {\bibfield  {journal} {\bibinfo
  {journal} {Class. Quant. Grav.}\ }\textbf {\bibinfo {volume} {39}},\ \bibinfo
  {pages} {035002} (\bibinfo {year} {2022})},\ \Eprint
  {https://arxiv.org/abs/2105.13972} {arXiv:2105.13972 [gr-qc]} \BibitemShut
  {NoStop}%
\bibitem [{\citenamefont {Juli\'e}\ and\ \citenamefont
  {Berti}(2019)}]{Julie:2019sab}%
  \BibitemOpen
  \bibfield  {author} {\bibinfo {author} {\bibfnamefont {F.-L.}\ \bibnamefont
  {Juli\'e}}\ and\ \bibinfo {author} {\bibfnamefont {E.}~\bibnamefont
  {Berti}},\ }\bibfield  {title} {\bibinfo {title} {{Post-Newtonian dynamics
  and black hole thermodynamics in Einstein-scalar-Gauss-Bonnet gravity}},\
  }\href {https://doi.org/10.1103/PhysRevD.100.104061} {\bibfield  {journal}
  {\bibinfo  {journal} {Phys. Rev. D}\ }\textbf {\bibinfo {volume} {100}},\
  \bibinfo {pages} {104061} (\bibinfo {year} {2019})},\ \Eprint
  {https://arxiv.org/abs/1909.05258} {arXiv:1909.05258 [gr-qc]} \BibitemShut
  {NoStop}%
\bibitem [{\citenamefont {Juli\'e}\ \emph {et~al.}(2022)\citenamefont
  {Juli\'e}, \citenamefont {Silva}, \citenamefont {Berti},\ and\ \citenamefont
  {Yunes}}]{Julie:2022huo}%
  \BibitemOpen
  \bibfield  {author} {\bibinfo {author} {\bibfnamefont {F.-L.}\ \bibnamefont
  {Juli\'e}}, \bibinfo {author} {\bibfnamefont {H.~O.}\ \bibnamefont {Silva}},
  \bibinfo {author} {\bibfnamefont {E.}~\bibnamefont {Berti}},\ and\ \bibinfo
  {author} {\bibfnamefont {N.}~\bibnamefont {Yunes}},\ }\href@noop {} {\bibinfo
  {title} {{Black hole sensitivities in Einstein-scalar-Gauss-Bonnet gravity}}}
  (\bibinfo {year} {2022}),\ \Eprint {https://arxiv.org/abs/2202.01329}
  {arXiv:2202.01329 [gr-qc]} \BibitemShut {NoStop}%
\bibitem [{\citenamefont {Witek}\ \emph {et~al.}(2019)\citenamefont {Witek},
  \citenamefont {Gualtieri}, \citenamefont {Pani},\ and\ \citenamefont
  {Sotiriou}}]{Witek:2018dmd}%
  \BibitemOpen
  \bibfield  {author} {\bibinfo {author} {\bibfnamefont {H.}~\bibnamefont
  {Witek}}, \bibinfo {author} {\bibfnamefont {L.}~\bibnamefont {Gualtieri}},
  \bibinfo {author} {\bibfnamefont {P.}~\bibnamefont {Pani}},\ and\ \bibinfo
  {author} {\bibfnamefont {T.~P.}\ \bibnamefont {Sotiriou}},\ }\bibfield
  {title} {\bibinfo {title} {{Black holes and binary mergers in scalar
  Gauss-Bonnet gravity: scalar field dynamics}},\ }\href
  {https://doi.org/10.1103/PhysRevD.99.064035} {\bibfield  {journal} {\bibinfo
  {journal} {Phys. Rev. D}\ }\textbf {\bibinfo {volume} {99}},\ \bibinfo
  {pages} {064035} (\bibinfo {year} {2019})},\ \Eprint
  {https://arxiv.org/abs/1810.05177} {arXiv:1810.05177 [gr-qc]} \BibitemShut
  {NoStop}%
\bibitem [{\citenamefont {Okounkova}(2020)}]{Okounkova:2020rqw}%
  \BibitemOpen
  \bibfield  {author} {\bibinfo {author} {\bibfnamefont {M.}~\bibnamefont
  {Okounkova}},\ }\bibfield  {title} {\bibinfo {title} {{Numerical relativity
  simulation of GW150914 in Einstein dilaton Gauss-Bonnet gravity}},\ }\href
  {https://doi.org/10.1103/PhysRevD.102.084046} {\bibfield  {journal} {\bibinfo
   {journal} {Phys. Rev. D}\ }\textbf {\bibinfo {volume} {102}},\ \bibinfo
  {pages} {084046} (\bibinfo {year} {2020})},\ \Eprint
  {https://arxiv.org/abs/2001.03571} {arXiv:2001.03571 [gr-qc]} \BibitemShut
  {NoStop}%
\bibitem [{\citenamefont {East}\ and\ \citenamefont
  {Ripley}(2021{\natexlab{a}})}]{East:2020hgw}%
  \BibitemOpen
  \bibfield  {author} {\bibinfo {author} {\bibfnamefont {W.~E.}\ \bibnamefont
  {East}}\ and\ \bibinfo {author} {\bibfnamefont {J.~L.}\ \bibnamefont
  {Ripley}},\ }\bibfield  {title} {\bibinfo {title} {{Evolution of
  Einstein-scalar-Gauss-Bonnet gravity using a modified harmonic
  formulation}},\ }\href {https://doi.org/10.1103/PhysRevD.103.044040}
  {\bibfield  {journal} {\bibinfo  {journal} {Phys. Rev. D}\ }\textbf {\bibinfo
  {volume} {103}},\ \bibinfo {pages} {044040} (\bibinfo {year}
  {2021}{\natexlab{a}})},\ \Eprint {https://arxiv.org/abs/2011.03547}
  {arXiv:2011.03547 [gr-qc]} \BibitemShut {NoStop}%
\bibitem [{\citenamefont {East}\ and\ \citenamefont
  {Ripley}(2021{\natexlab{b}})}]{East:2021bqk}%
  \BibitemOpen
  \bibfield  {author} {\bibinfo {author} {\bibfnamefont {W.~E.}\ \bibnamefont
  {East}}\ and\ \bibinfo {author} {\bibfnamefont {J.~L.}\ \bibnamefont
  {Ripley}},\ }\bibfield  {title} {\bibinfo {title} {{Dynamics of Spontaneous
  Black Hole Scalarization and Mergers in Einstein-Scalar-Gauss-Bonnet
  Gravity}},\ }\href {https://doi.org/10.1103/PhysRevLett.127.101102}
  {\bibfield  {journal} {\bibinfo  {journal} {Phys. Rev. Lett.}\ }\textbf
  {\bibinfo {volume} {127}},\ \bibinfo {pages} {101102} (\bibinfo {year}
  {2021}{\natexlab{b}})},\ \Eprint {https://arxiv.org/abs/2105.08571}
  {arXiv:2105.08571 [gr-qc]} \BibitemShut {NoStop}%
\bibitem [{\citenamefont {Silva}\ \emph
  {et~al.}(2021{\natexlab{b}})\citenamefont {Silva}, \citenamefont {Witek},
  \citenamefont {Elley},\ and\ \citenamefont {Yunes}}]{Silva:2020omi}%
  \BibitemOpen
  \bibfield  {author} {\bibinfo {author} {\bibfnamefont {H.~O.}\ \bibnamefont
  {Silva}}, \bibinfo {author} {\bibfnamefont {H.}~\bibnamefont {Witek}},
  \bibinfo {author} {\bibfnamefont {M.}~\bibnamefont {Elley}},\ and\ \bibinfo
  {author} {\bibfnamefont {N.}~\bibnamefont {Yunes}},\ }\bibfield  {title}
  {\bibinfo {title} {{Dynamical Descalarization in Binary Black Hole
  Mergers}},\ }\href {https://doi.org/10.1103/PhysRevLett.127.031101}
  {\bibfield  {journal} {\bibinfo  {journal} {Phys. Rev. Lett.}\ }\textbf
  {\bibinfo {volume} {127}},\ \bibinfo {pages} {031101} (\bibinfo {year}
  {2021}{\natexlab{b}})},\ \Eprint {https://arxiv.org/abs/2012.10436}
  {arXiv:2012.10436 [gr-qc]} \BibitemShut {NoStop}%
\bibitem [{\citenamefont {Doneva}\ \emph {et~al.}(2022)\citenamefont {Doneva},
  \citenamefont {Va\~n\'o Vi\~nuales},\ and\ \citenamefont
  {Yazadjiev}}]{Doneva:2022byd}%
  \BibitemOpen
  \bibfield  {author} {\bibinfo {author} {\bibfnamefont {D.~D.}\ \bibnamefont
  {Doneva}}, \bibinfo {author} {\bibfnamefont {A.}~\bibnamefont {Va\~n\'o
  Vi\~nuales}},\ and\ \bibinfo {author} {\bibfnamefont {S.~S.}\ \bibnamefont
  {Yazadjiev}},\ }\href@noop {} {\bibinfo {title} {{Dynamical descalarization
  with a jump during black hole merger}}} (\bibinfo {year} {2022}),\ \Eprint
  {https://arxiv.org/abs/2204.05333} {arXiv:2204.05333 [gr-qc]} \BibitemShut
  {NoStop}%
\bibitem [{\citenamefont {Pani}\ and\ \citenamefont
  {Cardoso}(2009)}]{Pani:2009wy}%
  \BibitemOpen
  \bibfield  {author} {\bibinfo {author} {\bibfnamefont {P.}~\bibnamefont
  {Pani}}\ and\ \bibinfo {author} {\bibfnamefont {V.}~\bibnamefont {Cardoso}},\
  }\bibfield  {title} {\bibinfo {title} {{Are black holes in alternative
  theories serious astrophysical candidates? The Case for
  Einstein-Dilaton-Gauss-Bonnet black holes}},\ }\href
  {https://doi.org/10.1103/PhysRevD.79.084031} {\bibfield  {journal} {\bibinfo
  {journal} {Phys. Rev. D}\ }\textbf {\bibinfo {volume} {79}},\ \bibinfo
  {pages} {084031} (\bibinfo {year} {2009})},\ \Eprint
  {https://arxiv.org/abs/0902.1569} {arXiv:0902.1569 [gr-qc]} \BibitemShut
  {NoStop}%
\bibitem [{\citenamefont {Bl\'azquez-Salcedo}\ \emph
  {et~al.}(2016)\citenamefont {Bl\'azquez-Salcedo}, \citenamefont {Macedo},
  \citenamefont {Cardoso}, \citenamefont {Ferrari}, \citenamefont {Gualtieri},
  \citenamefont {Khoo}, \citenamefont {Kunz},\ and\ \citenamefont
  {Pani}}]{Blazquez-Salcedo:2016enn}%
  \BibitemOpen
  \bibfield  {author} {\bibinfo {author} {\bibfnamefont {J.~L.}\ \bibnamefont
  {Bl\'azquez-Salcedo}}, \bibinfo {author} {\bibfnamefont {C.~F.~B.}\
  \bibnamefont {Macedo}}, \bibinfo {author} {\bibfnamefont {V.}~\bibnamefont
  {Cardoso}}, \bibinfo {author} {\bibfnamefont {V.}~\bibnamefont {Ferrari}},
  \bibinfo {author} {\bibfnamefont {L.}~\bibnamefont {Gualtieri}}, \bibinfo
  {author} {\bibfnamefont {F.~S.}\ \bibnamefont {Khoo}}, \bibinfo {author}
  {\bibfnamefont {J.}~\bibnamefont {Kunz}},\ and\ \bibinfo {author}
  {\bibfnamefont {P.}~\bibnamefont {Pani}},\ }\bibfield  {title} {\bibinfo
  {title} {{Perturbed black holes in Einstein-dilaton-Gauss-Bonnet gravity:
  Stability, ringdown, and gravitational-wave emission}},\ }\href
  {https://doi.org/10.1103/PhysRevD.94.104024} {\bibfield  {journal} {\bibinfo
  {journal} {Phys. Rev. D}\ }\textbf {\bibinfo {volume} {94}},\ \bibinfo
  {pages} {104024} (\bibinfo {year} {2016})},\ \Eprint
  {https://arxiv.org/abs/1609.01286} {arXiv:1609.01286 [gr-qc]} \BibitemShut
  {NoStop}%
\bibitem [{\citenamefont {Bl\'azquez-Salcedo}\ \emph
  {et~al.}(2020{\natexlab{a}})\citenamefont {Bl\'azquez-Salcedo}, \citenamefont
  {Doneva}, \citenamefont {Kahlen}, \citenamefont {Kunz}, \citenamefont
  {Nedkova},\ and\ \citenamefont {Yazadjiev}}]{Blazquez-Salcedo:2020rhf}%
  \BibitemOpen
  \bibfield  {author} {\bibinfo {author} {\bibfnamefont {J.~L.}\ \bibnamefont
  {Bl\'azquez-Salcedo}}, \bibinfo {author} {\bibfnamefont {D.~D.}\ \bibnamefont
  {Doneva}}, \bibinfo {author} {\bibfnamefont {S.}~\bibnamefont {Kahlen}},
  \bibinfo {author} {\bibfnamefont {J.}~\bibnamefont {Kunz}}, \bibinfo {author}
  {\bibfnamefont {P.}~\bibnamefont {Nedkova}},\ and\ \bibinfo {author}
  {\bibfnamefont {S.~S.}\ \bibnamefont {Yazadjiev}},\ }\bibfield  {title}
  {\bibinfo {title} {{Axial perturbations of the scalarized
  Einstein-Gauss-Bonnet black holes}},\ }\href
  {https://doi.org/10.1103/PhysRevD.101.104006} {\bibfield  {journal} {\bibinfo
   {journal} {Phys. Rev. D}\ }\textbf {\bibinfo {volume} {101}},\ \bibinfo
  {pages} {104006} (\bibinfo {year} {2020}{\natexlab{a}})},\ \Eprint
  {https://arxiv.org/abs/2003.02862} {arXiv:2003.02862 [gr-qc]} \BibitemShut
  {NoStop}%
\bibitem [{\citenamefont {Bl\'azquez-Salcedo}\ \emph
  {et~al.}(2020{\natexlab{b}})\citenamefont {Bl\'azquez-Salcedo}, \citenamefont
  {Doneva}, \citenamefont {Kahlen}, \citenamefont {Kunz}, \citenamefont
  {Nedkova},\ and\ \citenamefont {Yazadjiev}}]{Blazquez-Salcedo:2020caw}%
  \BibitemOpen
  \bibfield  {author} {\bibinfo {author} {\bibfnamefont {J.~L.}\ \bibnamefont
  {Bl\'azquez-Salcedo}}, \bibinfo {author} {\bibfnamefont {D.~D.}\ \bibnamefont
  {Doneva}}, \bibinfo {author} {\bibfnamefont {S.}~\bibnamefont {Kahlen}},
  \bibinfo {author} {\bibfnamefont {J.}~\bibnamefont {Kunz}}, \bibinfo {author}
  {\bibfnamefont {P.}~\bibnamefont {Nedkova}},\ and\ \bibinfo {author}
  {\bibfnamefont {S.~S.}\ \bibnamefont {Yazadjiev}},\ }\bibfield  {title}
  {\bibinfo {title} {{Polar quasinormal modes of the scalarized
  Einstein-Gauss-Bonnet black holes}},\ }\href
  {https://doi.org/10.1103/PhysRevD.102.024086} {\bibfield  {journal} {\bibinfo
   {journal} {Phys. Rev. D}\ }\textbf {\bibinfo {volume} {102}},\ \bibinfo
  {pages} {024086} (\bibinfo {year} {2020}{\natexlab{b}})},\ \Eprint
  {https://arxiv.org/abs/2006.06006} {arXiv:2006.06006 [gr-qc]} \BibitemShut
  {NoStop}%
\bibitem [{\citenamefont {Pierini}\ and\ \citenamefont
  {Gualtieri}(2021)}]{Pierini:2021jxd}%
  \BibitemOpen
  \bibfield  {author} {\bibinfo {author} {\bibfnamefont {L.}~\bibnamefont
  {Pierini}}\ and\ \bibinfo {author} {\bibfnamefont {L.}~\bibnamefont
  {Gualtieri}},\ }\bibfield  {title} {\bibinfo {title} {{Quasi-normal modes of
  rotating black holes in Einstein-dilaton Gauss-Bonnet gravity: the first
  order in rotation}},\ }\href {https://doi.org/10.1103/PhysRevD.103.124017}
  {\bibfield  {journal} {\bibinfo  {journal} {Phys. Rev. D}\ }\textbf {\bibinfo
  {volume} {103}},\ \bibinfo {pages} {124017} (\bibinfo {year} {2021})},\
  \Eprint {https://arxiv.org/abs/2103.09870} {arXiv:2103.09870 [gr-qc]}
  \BibitemShut {NoStop}%
\bibitem [{\citenamefont {Bryant}\ \emph {et~al.}(2021)\citenamefont {Bryant},
  \citenamefont {Silva}, \citenamefont {Yagi},\ and\ \citenamefont
  {Glampedakis}}]{Bryant:2021xdh}%
  \BibitemOpen
  \bibfield  {author} {\bibinfo {author} {\bibfnamefont {A.}~\bibnamefont
  {Bryant}}, \bibinfo {author} {\bibfnamefont {H.~O.}\ \bibnamefont {Silva}},
  \bibinfo {author} {\bibfnamefont {K.}~\bibnamefont {Yagi}},\ and\ \bibinfo
  {author} {\bibfnamefont {K.}~\bibnamefont {Glampedakis}},\ }\bibfield
  {title} {\bibinfo {title} {{Eikonal quasinormal modes of black holes beyond
  general relativity. III. Scalar Gauss-Bonnet gravity}},\ }\href
  {https://doi.org/10.1103/PhysRevD.104.044051} {\bibfield  {journal} {\bibinfo
   {journal} {Phys. Rev. D}\ }\textbf {\bibinfo {volume} {104}},\ \bibinfo
  {pages} {044051} (\bibinfo {year} {2021})},\ \Eprint
  {https://arxiv.org/abs/2106.09657} {arXiv:2106.09657 [gr-qc]} \BibitemShut
  {NoStop}%
\bibitem [{\citenamefont {Campbell}\ \emph {et~al.}(1992)\citenamefont
  {Campbell}, \citenamefont {Kaloper},\ and\ \citenamefont
  {Olive}}]{Campbell:1991kz}%
  \BibitemOpen
  \bibfield  {author} {\bibinfo {author} {\bibfnamefont {B.~A.}\ \bibnamefont
  {Campbell}}, \bibinfo {author} {\bibfnamefont {N.}~\bibnamefont {Kaloper}},\
  and\ \bibinfo {author} {\bibfnamefont {K.~A.}\ \bibnamefont {Olive}},\
  }\bibfield  {title} {\bibinfo {title} {{Classical hair for Kerr-Newman black
  holes in string gravity}},\ }\href
  {https://doi.org/10.1016/0370-2693(92)91452-F} {\bibfield  {journal}
  {\bibinfo  {journal} {Phys. Lett. B}\ }\textbf {\bibinfo {volume} {285}},\
  \bibinfo {pages} {199} (\bibinfo {year} {1992})}\BibitemShut {NoStop}%
\bibitem [{\citenamefont {Mignemi}\ and\ \citenamefont
  {Stewart}(1993)}]{Mignemi:1992nt}%
  \BibitemOpen
  \bibfield  {author} {\bibinfo {author} {\bibfnamefont {S.}~\bibnamefont
  {Mignemi}}\ and\ \bibinfo {author} {\bibfnamefont {N.}~\bibnamefont
  {Stewart}},\ }\bibfield  {title} {\bibinfo {title} {{Charged black holes in
  effective string theory}},\ }\href {https://doi.org/10.1103/PhysRevD.47.5259}
  {\bibfield  {journal} {\bibinfo  {journal} {Phys. Rev. D}\ }\textbf {\bibinfo
  {volume} {47}},\ \bibinfo {pages} {5259} (\bibinfo {year} {1993})},\ \Eprint
  {https://arxiv.org/abs/hep-th/9212146} {arXiv:hep-th/9212146} \BibitemShut
  {NoStop}%
\bibitem [{\citenamefont {Kanti}\ \emph {et~al.}(1996)\citenamefont {Kanti},
  \citenamefont {Mavromatos}, \citenamefont {Rizos}, \citenamefont {Tamvakis},\
  and\ \citenamefont {Winstanley}}]{Kanti:1995vq}%
  \BibitemOpen
  \bibfield  {author} {\bibinfo {author} {\bibfnamefont {P.}~\bibnamefont
  {Kanti}}, \bibinfo {author} {\bibfnamefont {N.}~\bibnamefont {Mavromatos}},
  \bibinfo {author} {\bibfnamefont {J.}~\bibnamefont {Rizos}}, \bibinfo
  {author} {\bibfnamefont {K.}~\bibnamefont {Tamvakis}},\ and\ \bibinfo
  {author} {\bibfnamefont {E.}~\bibnamefont {Winstanley}},\ }\bibfield  {title}
  {\bibinfo {title} {{Dilatonic black holes in higher curvature string
  gravity}},\ }\href {https://doi.org/10.1103/PhysRevD.54.5049} {\bibfield
  {journal} {\bibinfo  {journal} {Phys. Rev. D}\ }\textbf {\bibinfo {volume}
  {54}},\ \bibinfo {pages} {5049} (\bibinfo {year} {1996})},\ \Eprint
  {https://arxiv.org/abs/hep-th/9511071} {arXiv:hep-th/9511071} \BibitemShut
  {NoStop}%
\bibitem [{\citenamefont {Torii}\ \emph {et~al.}(1997)\citenamefont {Torii},
  \citenamefont {Yajima},\ and\ \citenamefont {Maeda}}]{Torii:1996yi}%
  \BibitemOpen
  \bibfield  {author} {\bibinfo {author} {\bibfnamefont {T.}~\bibnamefont
  {Torii}}, \bibinfo {author} {\bibfnamefont {H.}~\bibnamefont {Yajima}},\ and\
  \bibinfo {author} {\bibfnamefont {K.-i.}\ \bibnamefont {Maeda}},\ }\bibfield
  {title} {\bibinfo {title} {{Dilatonic black holes with Gauss-Bonnet term}},\
  }\href {https://doi.org/10.1103/PhysRevD.55.739} {\bibfield  {journal}
  {\bibinfo  {journal} {Phys. Rev. D}\ }\textbf {\bibinfo {volume} {55}},\
  \bibinfo {pages} {739} (\bibinfo {year} {1997})},\ \Eprint
  {https://arxiv.org/abs/gr-qc/9606034} {arXiv:gr-qc/9606034} \BibitemShut
  {NoStop}%
\bibitem [{\citenamefont {Guo}\ \emph {et~al.}(2008)\citenamefont {Guo},
  \citenamefont {Ohta},\ and\ \citenamefont {Torii}}]{Guo:2008hf}%
  \BibitemOpen
  \bibfield  {author} {\bibinfo {author} {\bibfnamefont {Z.-K.}\ \bibnamefont
  {Guo}}, \bibinfo {author} {\bibfnamefont {N.}~\bibnamefont {Ohta}},\ and\
  \bibinfo {author} {\bibfnamefont {T.}~\bibnamefont {Torii}},\ }\bibfield
  {title} {\bibinfo {title} {{Black Holes in the Dilatonic
  Einstein-Gauss-Bonnet Theory in Various Dimensions. I. Asymptotically Flat
  Black Holes}},\ }\href {https://doi.org/10.1143/PTP.120.581} {\bibfield
  {journal} {\bibinfo  {journal} {Prog. Theor. Phys.}\ }\textbf {\bibinfo
  {volume} {120}},\ \bibinfo {pages} {581} (\bibinfo {year} {2008})},\ \Eprint
  {https://arxiv.org/abs/0806.2481} {arXiv:0806.2481 [gr-qc]} \BibitemShut
  {NoStop}%
\bibitem [{\citenamefont {Yunes}\ and\ \citenamefont
  {Stein}(2011)}]{Yunes:2011we}%
  \BibitemOpen
  \bibfield  {author} {\bibinfo {author} {\bibfnamefont {N.}~\bibnamefont
  {Yunes}}\ and\ \bibinfo {author} {\bibfnamefont {L.~C.}\ \bibnamefont
  {Stein}},\ }\bibfield  {title} {\bibinfo {title} {{Non-Spinning Black Holes
  in Alternative Theories of Gravity}},\ }\href
  {https://doi.org/10.1103/PhysRevD.83.104002} {\bibfield  {journal} {\bibinfo
  {journal} {Phys. Rev. D}\ }\textbf {\bibinfo {volume} {83}},\ \bibinfo
  {pages} {104002} (\bibinfo {year} {2011})},\ \Eprint
  {https://arxiv.org/abs/1101.2921} {arXiv:1101.2921 [gr-qc]} \BibitemShut
  {NoStop}%
%%CITATION = ARXIV:1101.2921;%%
\bibitem [{\citenamefont {Sotiriou}\ and\ \citenamefont
  {Zhou}(2014{\natexlab{a}})}]{Sotiriou:2013qea}%
  \BibitemOpen
  \bibfield  {author} {\bibinfo {author} {\bibfnamefont {T.~P.}\ \bibnamefont
  {Sotiriou}}\ and\ \bibinfo {author} {\bibfnamefont {S.-Y.}\ \bibnamefont
  {Zhou}},\ }\bibfield  {title} {\bibinfo {title} {{Black hole hair in
  generalized scalar-tensor gravity}},\ }\href
  {https://doi.org/10.1103/PhysRevLett.112.251102} {\bibfield  {journal}
  {\bibinfo  {journal} {Phys. Rev. Lett.}\ }\textbf {\bibinfo {volume} {112}},\
  \bibinfo {pages} {251102} (\bibinfo {year} {2014}{\natexlab{a}})},\ \Eprint
  {https://arxiv.org/abs/1312.3622} {arXiv:1312.3622 [gr-qc]} \BibitemShut
  {NoStop}%
\bibitem [{\citenamefont {Sotiriou}\ and\ \citenamefont
  {Zhou}(2014{\natexlab{b}})}]{Sotiriou:2014pfa}%
  \BibitemOpen
  \bibfield  {author} {\bibinfo {author} {\bibfnamefont {T.~P.}\ \bibnamefont
  {Sotiriou}}\ and\ \bibinfo {author} {\bibfnamefont {S.-Y.}\ \bibnamefont
  {Zhou}},\ }\bibfield  {title} {\bibinfo {title} {{Black hole hair in
  generalized scalar-tensor gravity: An explicit example}},\ }\href
  {https://doi.org/10.1103/PhysRevD.90.124063} {\bibfield  {journal} {\bibinfo
  {journal} {Phys. Rev. D}\ }\textbf {\bibinfo {volume} {90}},\ \bibinfo
  {pages} {124063} (\bibinfo {year} {2014}{\natexlab{b}})},\ \Eprint
  {https://arxiv.org/abs/1408.1698} {arXiv:1408.1698 [gr-qc]} \BibitemShut
  {NoStop}%
\bibitem [{\citenamefont {Benkel}\ \emph {et~al.}(2016)\citenamefont {Benkel},
  \citenamefont {Sotiriou},\ and\ \citenamefont {Witek}}]{Benkel:2016kcq}%
  \BibitemOpen
  \bibfield  {author} {\bibinfo {author} {\bibfnamefont {R.}~\bibnamefont
  {Benkel}}, \bibinfo {author} {\bibfnamefont {T.~P.}\ \bibnamefont
  {Sotiriou}},\ and\ \bibinfo {author} {\bibfnamefont {H.}~\bibnamefont
  {Witek}},\ }\bibfield  {title} {\bibinfo {title} {{Dynamical scalar hair
  formation around a Schwarzschild black hole}},\ }\href
  {https://doi.org/10.1103/PhysRevD.94.121503} {\bibfield  {journal} {\bibinfo
  {journal} {Phys. Rev. D}\ }\textbf {\bibinfo {volume} {94}},\ \bibinfo
  {pages} {121503} (\bibinfo {year} {2016})},\ \Eprint
  {https://arxiv.org/abs/1612.08184} {arXiv:1612.08184 [gr-qc]} \BibitemShut
  {NoStop}%
%%CITATION = ARXIV:1612.08184;%%
\bibitem [{\citenamefont {Benkel}\ \emph {et~al.}(2017)\citenamefont {Benkel},
  \citenamefont {Sotiriou},\ and\ \citenamefont {Witek}}]{Benkel:2016rlz}%
  \BibitemOpen
  \bibfield  {author} {\bibinfo {author} {\bibfnamefont {R.}~\bibnamefont
  {Benkel}}, \bibinfo {author} {\bibfnamefont {T.~P.}\ \bibnamefont
  {Sotiriou}},\ and\ \bibinfo {author} {\bibfnamefont {H.}~\bibnamefont
  {Witek}},\ }\bibfield  {title} {\bibinfo {title} {{Black hole hair formation
  in shift-symmetric generalised scalar-tensor gravity}},\ }\href
  {https://doi.org/10.1088/1361-6382/aa5ce7} {\bibfield  {journal} {\bibinfo
  {journal} {Class. Quant. Grav.}\ }\textbf {\bibinfo {volume} {34}},\ \bibinfo
  {pages} {064001} (\bibinfo {year} {2017})},\ \Eprint
  {https://arxiv.org/abs/1610.09168} {arXiv:1610.09168 [gr-qc]} \BibitemShut
  {NoStop}%
%%CITATION = ARXIV:1610.09168;%%
\bibitem [{\citenamefont {Antoniou}\ \emph
  {et~al.}(2018{\natexlab{a}})\citenamefont {Antoniou}, \citenamefont
  {Bakopoulos},\ and\ \citenamefont {Kanti}}]{Antoniou:2017acq}%
  \BibitemOpen
  \bibfield  {author} {\bibinfo {author} {\bibfnamefont {G.}~\bibnamefont
  {Antoniou}}, \bibinfo {author} {\bibfnamefont {A.}~\bibnamefont
  {Bakopoulos}},\ and\ \bibinfo {author} {\bibfnamefont {P.}~\bibnamefont
  {Kanti}},\ }\bibfield  {title} {\bibinfo {title} {{Evasion of No-Hair
  Theorems and Novel Black-Hole Solutions in Gauss-Bonnet Theories}},\ }\href
  {https://doi.org/10.1103/PhysRevLett.120.131102} {\bibfield  {journal}
  {\bibinfo  {journal} {Phys. Rev. Lett.}\ }\textbf {\bibinfo {volume} {120}},\
  \bibinfo {pages} {131102} (\bibinfo {year} {2018}{\natexlab{a}})},\ \Eprint
  {https://arxiv.org/abs/1711.03390} {arXiv:1711.03390 [hep-th]} \BibitemShut
  {NoStop}%
\bibitem [{\citenamefont {Antoniou}\ \emph
  {et~al.}(2018{\natexlab{b}})\citenamefont {Antoniou}, \citenamefont
  {Bakopoulos},\ and\ \citenamefont {Kanti}}]{Antoniou:2017hxj}%
  \BibitemOpen
  \bibfield  {author} {\bibinfo {author} {\bibfnamefont {G.}~\bibnamefont
  {Antoniou}}, \bibinfo {author} {\bibfnamefont {A.}~\bibnamefont
  {Bakopoulos}},\ and\ \bibinfo {author} {\bibfnamefont {P.}~\bibnamefont
  {Kanti}},\ }\bibfield  {title} {\bibinfo {title} {{Black-Hole Solutions with
  Scalar Hair in Einstein-Scalar-Gauss-Bonnet Theories}},\ }\href
  {https://doi.org/10.1103/PhysRevD.97.084037} {\bibfield  {journal} {\bibinfo
  {journal} {Phys. Rev. D}\ }\textbf {\bibinfo {volume} {97}},\ \bibinfo
  {pages} {084037} (\bibinfo {year} {2018}{\natexlab{b}})},\ \Eprint
  {https://arxiv.org/abs/1711.07431} {arXiv:1711.07431 [hep-th]} \BibitemShut
  {NoStop}%
\bibitem [{\citenamefont {Prabhu}\ and\ \citenamefont
  {Stein}(2018)}]{Prabhu:2018aun}%
  \BibitemOpen
  \bibfield  {author} {\bibinfo {author} {\bibfnamefont {K.}~\bibnamefont
  {Prabhu}}\ and\ \bibinfo {author} {\bibfnamefont {L.~C.}\ \bibnamefont
  {Stein}},\ }\bibfield  {title} {\bibinfo {title} {{Black hole scalar charge
  from a topological horizon integral in Einstein-dilaton-Gauss-Bonnet
  gravity}},\ }\href {https://doi.org/10.1103/PhysRevD.98.021503} {\bibfield
  {journal} {\bibinfo  {journal} {Phys. Rev. D}\ }\textbf {\bibinfo {volume}
  {98}},\ \bibinfo {pages} {021503} (\bibinfo {year} {2018})},\ \Eprint
  {https://arxiv.org/abs/1805.02668} {arXiv:1805.02668 [gr-qc]} \BibitemShut
  {NoStop}%
\bibitem [{\citenamefont {Saravani}\ and\ \citenamefont
  {Sotiriou}(2019)}]{Saravani:2019xwx}%
  \BibitemOpen
  \bibfield  {author} {\bibinfo {author} {\bibfnamefont {M.}~\bibnamefont
  {Saravani}}\ and\ \bibinfo {author} {\bibfnamefont {T.~P.}\ \bibnamefont
  {Sotiriou}},\ }\bibfield  {title} {\bibinfo {title} {{Classification of
  shift-symmetric Horndeski theories and hairy black holes}},\ }\href
  {https://doi.org/10.1103/PhysRevD.99.124004} {\bibfield  {journal} {\bibinfo
  {journal} {Phys. Rev. D}\ }\textbf {\bibinfo {volume} {99}},\ \bibinfo
  {pages} {124004} (\bibinfo {year} {2019})},\ \Eprint
  {https://arxiv.org/abs/1903.02055} {arXiv:1903.02055 [gr-qc]} \BibitemShut
  {NoStop}%
\bibitem [{\citenamefont {R.}\ \emph {et~al.}(2022)\citenamefont {R.},
  \citenamefont {Most}, \citenamefont {Noronha}, \citenamefont {Witek},\ and\
  \citenamefont {Yunes}}]{R:2022cwe}%
  \BibitemOpen
  \bibfield  {author} {\bibinfo {author} {\bibfnamefont {A.~H.~K.}\
  \bibnamefont {R.}}, \bibinfo {author} {\bibfnamefont {E.~R.}\ \bibnamefont
  {Most}}, \bibinfo {author} {\bibfnamefont {J.}~\bibnamefont {Noronha}},
  \bibinfo {author} {\bibfnamefont {H.}~\bibnamefont {Witek}},\ and\ \bibinfo
  {author} {\bibfnamefont {N.}~\bibnamefont {Yunes}},\ }\bibfield  {title}
  {\bibinfo {title} {{How do spherical black holes grow monopole hair?}},\
  }\href {https://doi.org/10.1103/PhysRevD.105.064041} {\bibfield  {journal}
  {\bibinfo  {journal} {Phys. Rev. D}\ }\textbf {\bibinfo {volume} {105}},\
  \bibinfo {pages} {064041} (\bibinfo {year} {2022})},\ \Eprint
  {https://arxiv.org/abs/2201.05178} {arXiv:2201.05178 [gr-qc]} \BibitemShut
  {NoStop}%
\bibitem [{\citenamefont {Doneva}\ and\ \citenamefont
  {Yazadjiev}(2018)}]{Doneva:2017bvd}%
  \BibitemOpen
  \bibfield  {author} {\bibinfo {author} {\bibfnamefont {D.~D.}\ \bibnamefont
  {Doneva}}\ and\ \bibinfo {author} {\bibfnamefont {S.~S.}\ \bibnamefont
  {Yazadjiev}},\ }\bibfield  {title} {\bibinfo {title} {{New Gauss-Bonnet Black
  Holes with Curvature-Induced Scalarization in Extended Scalar-Tensor
  Theories}},\ }\href {https://doi.org/10.1103/PhysRevLett.120.131103}
  {\bibfield  {journal} {\bibinfo  {journal} {Phys. Rev. Lett.}\ }\textbf
  {\bibinfo {volume} {120}},\ \bibinfo {pages} {131103} (\bibinfo {year}
  {2018})},\ \Eprint {https://arxiv.org/abs/1711.01187} {arXiv:1711.01187
  [gr-qc]} \BibitemShut {NoStop}%
\bibitem [{\citenamefont {Silva}\ \emph {et~al.}(2018)\citenamefont {Silva},
  \citenamefont {Sakstein}, \citenamefont {Gualtieri}, \citenamefont
  {Sotiriou},\ and\ \citenamefont {Berti}}]{Silva:2017uqg}%
  \BibitemOpen
  \bibfield  {author} {\bibinfo {author} {\bibfnamefont {H.~O.}\ \bibnamefont
  {Silva}}, \bibinfo {author} {\bibfnamefont {J.}~\bibnamefont {Sakstein}},
  \bibinfo {author} {\bibfnamefont {L.}~\bibnamefont {Gualtieri}}, \bibinfo
  {author} {\bibfnamefont {T.~P.}\ \bibnamefont {Sotiriou}},\ and\ \bibinfo
  {author} {\bibfnamefont {E.}~\bibnamefont {Berti}},\ }\bibfield  {title}
  {\bibinfo {title} {{Spontaneous scalarization of black holes and compact
  stars from a Gauss-Bonnet coupling}},\ }\href
  {https://doi.org/10.1103/PhysRevLett.120.131104} {\bibfield  {journal}
  {\bibinfo  {journal} {Phys. Rev. Lett.}\ }\textbf {\bibinfo {volume} {120}},\
  \bibinfo {pages} {131104} (\bibinfo {year} {2018})},\ \Eprint
  {https://arxiv.org/abs/1711.02080} {arXiv:1711.02080 [gr-qc]} \BibitemShut
  {NoStop}%
\bibitem [{\citenamefont {Macedo}\ \emph {et~al.}(2019)\citenamefont {Macedo},
  \citenamefont {Sakstein}, \citenamefont {Berti}, \citenamefont {Gualtieri},
  \citenamefont {Silva},\ and\ \citenamefont {Sotiriou}}]{Macedo:2019sem}%
  \BibitemOpen
  \bibfield  {author} {\bibinfo {author} {\bibfnamefont {C.~F.}\ \bibnamefont
  {Macedo}}, \bibinfo {author} {\bibfnamefont {J.}~\bibnamefont {Sakstein}},
  \bibinfo {author} {\bibfnamefont {E.}~\bibnamefont {Berti}}, \bibinfo
  {author} {\bibfnamefont {L.}~\bibnamefont {Gualtieri}}, \bibinfo {author}
  {\bibfnamefont {H.~O.}\ \bibnamefont {Silva}},\ and\ \bibinfo {author}
  {\bibfnamefont {T.~P.}\ \bibnamefont {Sotiriou}},\ }\bibfield  {title}
  {\bibinfo {title} {{Self-interactions and Spontaneous Black Hole
  Scalarization}},\ }\href {https://doi.org/10.1103/PhysRevD.99.104041}
  {\bibfield  {journal} {\bibinfo  {journal} {Phys. Rev. D}\ }\textbf {\bibinfo
  {volume} {99}},\ \bibinfo {pages} {104041} (\bibinfo {year} {2019})},\
  \Eprint {https://arxiv.org/abs/1903.06784} {arXiv:1903.06784 [gr-qc]}
  \BibitemShut {NoStop}%
\bibitem [{\citenamefont {Cunha}\ \emph {et~al.}(2019)\citenamefont {Cunha},
  \citenamefont {Herdeiro},\ and\ \citenamefont {Radu}}]{Cunha:2019dwb}%
  \BibitemOpen
  \bibfield  {author} {\bibinfo {author} {\bibfnamefont {P.~V.}\ \bibnamefont
  {Cunha}}, \bibinfo {author} {\bibfnamefont {C.~A.}\ \bibnamefont
  {Herdeiro}},\ and\ \bibinfo {author} {\bibfnamefont {E.}~\bibnamefont
  {Radu}},\ }\bibfield  {title} {\bibinfo {title} {{Spontaneously Scalarized
  Kerr Black Holes in Extended Scalar-Tensor\textendash{}Gauss-Bonnet
  Gravity}},\ }\href {https://doi.org/10.1103/PhysRevLett.123.011101}
  {\bibfield  {journal} {\bibinfo  {journal} {Phys. Rev. Lett.}\ }\textbf
  {\bibinfo {volume} {123}},\ \bibinfo {pages} {011101} (\bibinfo {year}
  {2019})},\ \Eprint {https://arxiv.org/abs/1904.09997} {arXiv:1904.09997
  [gr-qc]} \BibitemShut {NoStop}%
\bibitem [{\citenamefont {Collodel}\ \emph {et~al.}(2020)\citenamefont
  {Collodel}, \citenamefont {Kleihaus}, \citenamefont {Kunz},\ and\
  \citenamefont {Berti}}]{Collodel:2019kkx}%
  \BibitemOpen
  \bibfield  {author} {\bibinfo {author} {\bibfnamefont {L.~G.}\ \bibnamefont
  {Collodel}}, \bibinfo {author} {\bibfnamefont {B.}~\bibnamefont {Kleihaus}},
  \bibinfo {author} {\bibfnamefont {J.}~\bibnamefont {Kunz}},\ and\ \bibinfo
  {author} {\bibfnamefont {E.}~\bibnamefont {Berti}},\ }\bibfield  {title}
  {\bibinfo {title} {{Spinning and excited black holes in
  Einstein-scalar-Gauss\textendash{}Bonnet theory}},\ }\href
  {https://doi.org/10.1088/1361-6382/ab74f9} {\bibfield  {journal} {\bibinfo
  {journal} {Class. Quant. Grav.}\ }\textbf {\bibinfo {volume} {37}},\ \bibinfo
  {pages} {075018} (\bibinfo {year} {2020})},\ \Eprint
  {https://arxiv.org/abs/1912.05382} {arXiv:1912.05382 [gr-qc]} \BibitemShut
  {NoStop}%
\bibitem [{\citenamefont {Dima}\ \emph {et~al.}(2020)\citenamefont {Dima},
  \citenamefont {Barausse}, \citenamefont {Franchini},\ and\ \citenamefont
  {Sotiriou}}]{Dima:2020yac}%
  \BibitemOpen
  \bibfield  {author} {\bibinfo {author} {\bibfnamefont {A.}~\bibnamefont
  {Dima}}, \bibinfo {author} {\bibfnamefont {E.}~\bibnamefont {Barausse}},
  \bibinfo {author} {\bibfnamefont {N.}~\bibnamefont {Franchini}},\ and\
  \bibinfo {author} {\bibfnamefont {T.~P.}\ \bibnamefont {Sotiriou}},\
  }\bibfield  {title} {\bibinfo {title} {{Spin-induced black hole spontaneous
  scalarization}},\ }\href {https://doi.org/10.1103/PhysRevLett.125.231101}
  {\bibfield  {journal} {\bibinfo  {journal} {Phys. Rev. Lett.}\ }\textbf
  {\bibinfo {volume} {125}},\ \bibinfo {pages} {231101} (\bibinfo {year}
  {2020})},\ \Eprint {https://arxiv.org/abs/2006.03095} {arXiv:2006.03095
  [gr-qc]} \BibitemShut {NoStop}%
\bibitem [{\citenamefont {Herdeiro}\ \emph {et~al.}(2021)\citenamefont
  {Herdeiro}, \citenamefont {Radu}, \citenamefont {Silva}, \citenamefont
  {Sotiriou},\ and\ \citenamefont {Yunes}}]{Herdeiro:2020wei}%
  \BibitemOpen
  \bibfield  {author} {\bibinfo {author} {\bibfnamefont {C.~A.~R.}\
  \bibnamefont {Herdeiro}}, \bibinfo {author} {\bibfnamefont {E.}~\bibnamefont
  {Radu}}, \bibinfo {author} {\bibfnamefont {H.~O.}\ \bibnamefont {Silva}},
  \bibinfo {author} {\bibfnamefont {T.~P.}\ \bibnamefont {Sotiriou}},\ and\
  \bibinfo {author} {\bibfnamefont {N.}~\bibnamefont {Yunes}},\ }\bibfield
  {title} {\bibinfo {title} {{Spin-induced scalarized black holes}},\ }\href
  {https://doi.org/10.1103/PhysRevLett.126.011103} {\bibfield  {journal}
  {\bibinfo  {journal} {Phys. Rev. Lett.}\ }\textbf {\bibinfo {volume} {126}},\
  \bibinfo {pages} {011103} (\bibinfo {year} {2021})},\ \Eprint
  {https://arxiv.org/abs/2009.03904} {arXiv:2009.03904 [gr-qc]} \BibitemShut
  {NoStop}%
\bibitem [{\citenamefont {Berti}\ \emph {et~al.}(2021)\citenamefont {Berti},
  \citenamefont {Collodel}, \citenamefont {Kleihaus},\ and\ \citenamefont
  {Kunz}}]{Berti:2020kgk}%
  \BibitemOpen
  \bibfield  {author} {\bibinfo {author} {\bibfnamefont {E.}~\bibnamefont
  {Berti}}, \bibinfo {author} {\bibfnamefont {L.~G.}\ \bibnamefont {Collodel}},
  \bibinfo {author} {\bibfnamefont {B.}~\bibnamefont {Kleihaus}},\ and\
  \bibinfo {author} {\bibfnamefont {J.}~\bibnamefont {Kunz}},\ }\bibfield
  {title} {\bibinfo {title} {{Spin-induced black-hole scalarization in
  Einstein-scalar-Gauss-Bonnet theory}},\ }\href
  {https://doi.org/10.1103/PhysRevLett.126.011104} {\bibfield  {journal}
  {\bibinfo  {journal} {Phys. Rev. Lett.}\ }\textbf {\bibinfo {volume} {126}},\
  \bibinfo {pages} {011104} (\bibinfo {year} {2021})},\ \Eprint
  {https://arxiv.org/abs/2009.03905} {arXiv:2009.03905 [gr-qc]} \BibitemShut
  {NoStop}%
\bibitem [{\citenamefont {Hod}(2020)}]{Hod:2020jjy}%
  \BibitemOpen
  \bibfield  {author} {\bibinfo {author} {\bibfnamefont {S.}~\bibnamefont
  {Hod}},\ }\bibfield  {title} {\bibinfo {title} {{Onset of spontaneous
  scalarization in spinning Gauss-Bonnet black holes}},\ }\href
  {https://doi.org/10.1103/PhysRevD.102.084060} {\bibfield  {journal} {\bibinfo
   {journal} {Phys. Rev. D}\ }\textbf {\bibinfo {volume} {102}},\ \bibinfo
  {pages} {084060} (\bibinfo {year} {2020})},\ \Eprint
  {https://arxiv.org/abs/2006.09399} {arXiv:2006.09399 [gr-qc]} \BibitemShut
  {NoStop}%
\bibitem [{\citenamefont {Doneva}\ \emph {et~al.}(2020)\citenamefont {Doneva},
  \citenamefont {Collodel}, \citenamefont {Kr\"uger},\ and\ \citenamefont
  {Yazadjiev}}]{Doneva:2020nbb}%
  \BibitemOpen
  \bibfield  {author} {\bibinfo {author} {\bibfnamefont {D.~D.}\ \bibnamefont
  {Doneva}}, \bibinfo {author} {\bibfnamefont {L.~G.}\ \bibnamefont
  {Collodel}}, \bibinfo {author} {\bibfnamefont {C.~J.}\ \bibnamefont
  {Kr\"uger}},\ and\ \bibinfo {author} {\bibfnamefont {S.~S.}\ \bibnamefont
  {Yazadjiev}},\ }\bibfield  {title} {\bibinfo {title} {{Black hole
  scalarization induced by the spin: 2+1 time evolution}},\ }\href
  {https://doi.org/10.1103/PhysRevD.102.104027} {\bibfield  {journal} {\bibinfo
   {journal} {Phys. Rev. D}\ }\textbf {\bibinfo {volume} {102}},\ \bibinfo
  {pages} {104027} (\bibinfo {year} {2020})},\ \Eprint
  {https://arxiv.org/abs/2008.07391} {arXiv:2008.07391 [gr-qc]} \BibitemShut
  {NoStop}%
\bibitem [{\citenamefont {Hod}(2022)}]{Hod:2022hfm}%
  \BibitemOpen
  \bibfield  {author} {\bibinfo {author} {\bibfnamefont {S.}~\bibnamefont
  {Hod}},\ }\bibfield  {title} {\bibinfo {title} {{Spin-induced black hole
  spontaneous scalarization: Analytic treatment in the large-coupling
  regime}},\ }\href {https://doi.org/10.1103/PhysRevD.105.024074} {\bibfield
  {journal} {\bibinfo  {journal} {Phys. Rev. D}\ }\textbf {\bibinfo {volume}
  {105}},\ \bibinfo {pages} {024074} (\bibinfo {year} {2022})}\BibitemShut
  {NoStop}%
\bibitem [{\citenamefont {Ripley}\ and\ \citenamefont
  {Pretorius}(2020)}]{Ripley:2020vpk}%
  \BibitemOpen
  \bibfield  {author} {\bibinfo {author} {\bibfnamefont {J.~L.}\ \bibnamefont
  {Ripley}}\ and\ \bibinfo {author} {\bibfnamefont {F.}~\bibnamefont
  {Pretorius}},\ }\bibfield  {title} {\bibinfo {title} {{Dynamics of a $\mathbb
  Z_2$ symmetric EdGB gravity in spherical symmetry}},\ }\href
  {https://doi.org/10.1088/1361-6382/ab9bbb} {\bibfield  {journal} {\bibinfo
  {journal} {Class. Quant. Grav.}\ }\textbf {\bibinfo {volume} {37}},\ \bibinfo
  {pages} {155003} (\bibinfo {year} {2020})},\ \Eprint
  {https://arxiv.org/abs/2005.05417} {arXiv:2005.05417 [gr-qc]} \BibitemShut
  {NoStop}%
\bibitem [{\citenamefont {Damour}\ and\ \citenamefont
  {Esposito-Far\`ese}(1993)}]{Damour:1993hw}%
  \BibitemOpen
  \bibfield  {author} {\bibinfo {author} {\bibfnamefont {T.}~\bibnamefont
  {Damour}}\ and\ \bibinfo {author} {\bibfnamefont {G.}~\bibnamefont
  {Esposito-Far\`ese}},\ }\bibfield  {title} {\bibinfo {title}
  {{Nonperturbative strong field effects in tensor-scalar theories of
  gravitation}},\ }\href {https://doi.org/10.1103/PhysRevLett.70.2220}
  {\bibfield  {journal} {\bibinfo  {journal} {Phys. Rev. Lett.}\ }\textbf
  {\bibinfo {volume} {70}},\ \bibinfo {pages} {2220} (\bibinfo {year}
  {1993})}\BibitemShut {NoStop}%
\bibitem [{\citenamefont {Damour}\ and\ \citenamefont
  {Esposito-Far\`ese}(1996)}]{Damour:1996ke}%
  \BibitemOpen
  \bibfield  {author} {\bibinfo {author} {\bibfnamefont {T.}~\bibnamefont
  {Damour}}\ and\ \bibinfo {author} {\bibfnamefont {G.}~\bibnamefont
  {Esposito-Far\`ese}},\ }\bibfield  {title} {\bibinfo {title} {{Tensor-scalar
  gravity and binary pulsar experiments}},\ }\href
  {https://doi.org/10.1103/PhysRevD.54.1474} {\bibfield  {journal} {\bibinfo
  {journal} {Phys. Rev. D}\ }\textbf {\bibinfo {volume} {54}},\ \bibinfo
  {pages} {1474} (\bibinfo {year} {1996})},\ \Eprint
  {https://arxiv.org/abs/gr-qc/9602056} {arXiv:gr-qc/9602056} \BibitemShut
  {NoStop}%
\bibitem [{\citenamefont {Silva}\ \emph {et~al.}(2015)\citenamefont {Silva},
  \citenamefont {Macedo}, \citenamefont {Berti},\ and\ \citenamefont
  {Crispino}}]{Silva:2014fca}%
  \BibitemOpen
  \bibfield  {author} {\bibinfo {author} {\bibfnamefont {H.~O.}\ \bibnamefont
  {Silva}}, \bibinfo {author} {\bibfnamefont {C.~F.~B.}\ \bibnamefont
  {Macedo}}, \bibinfo {author} {\bibfnamefont {E.}~\bibnamefont {Berti}},\ and\
  \bibinfo {author} {\bibfnamefont {L.~C.~B.}\ \bibnamefont {Crispino}},\
  }\bibfield  {title} {\bibinfo {title} {{Slowly rotating anisotropic neutron
  stars in general relativity and scalar\textendash{}tensor theory}},\ }\href
  {https://doi.org/10.1088/0264-9381/32/14/145008} {\bibfield  {journal}
  {\bibinfo  {journal} {Class. Quant. Grav.}\ }\textbf {\bibinfo {volume}
  {32}},\ \bibinfo {pages} {145008} (\bibinfo {year} {2015})},\ \Eprint
  {https://arxiv.org/abs/1411.6286} {arXiv:1411.6286 [gr-qc]} \BibitemShut
  {NoStop}%
\bibitem [{\citenamefont {Cherubini}\ \emph {et~al.}(2002)\citenamefont
  {Cherubini}, \citenamefont {Bini}, \citenamefont {Capozziello},\ and\
  \citenamefont {Ruffini}}]{Cherubini:2002gen}%
  \BibitemOpen
  \bibfield  {author} {\bibinfo {author} {\bibfnamefont {C.}~\bibnamefont
  {Cherubini}}, \bibinfo {author} {\bibfnamefont {D.}~\bibnamefont {Bini}},
  \bibinfo {author} {\bibfnamefont {S.}~\bibnamefont {Capozziello}},\ and\
  \bibinfo {author} {\bibfnamefont {R.}~\bibnamefont {Ruffini}},\ }\bibfield
  {title} {\bibinfo {title} {{Second order scalar invariants of the Riemann
  tensor: Applications to black hole space-times}},\ }\href
  {https://doi.org/10.1142/S0218271802002037} {\bibfield  {journal} {\bibinfo
  {journal} {Int. J. Mod. Phys. D}\ }\textbf {\bibinfo {volume} {11}},\
  \bibinfo {pages} {827} (\bibinfo {year} {2002})},\ \Eprint
  {https://arxiv.org/abs/gr-qc/0302095} {arXiv:gr-qc/0302095} \BibitemShut
  {NoStop}%
\bibitem [{\citenamefont {Bl\'azquez-Salcedo}\ \emph
  {et~al.}(2018)\citenamefont {Bl\'azquez-Salcedo}, \citenamefont {Doneva},
  \citenamefont {Kunz},\ and\ \citenamefont
  {Yazadjiev}}]{Blazquez-Salcedo:2018jnn}%
  \BibitemOpen
  \bibfield  {author} {\bibinfo {author} {\bibfnamefont {J.~L.}\ \bibnamefont
  {Bl\'azquez-Salcedo}}, \bibinfo {author} {\bibfnamefont {D.~D.}\ \bibnamefont
  {Doneva}}, \bibinfo {author} {\bibfnamefont {J.}~\bibnamefont {Kunz}},\ and\
  \bibinfo {author} {\bibfnamefont {S.~S.}\ \bibnamefont {Yazadjiev}},\
  }\bibfield  {title} {\bibinfo {title} {{Radial perturbations of the
  scalarized Einstein-Gauss-Bonnet black holes}},\ }\href
  {https://doi.org/10.1103/PhysRevD.98.084011} {\bibfield  {journal} {\bibinfo
  {journal} {Phys. Rev. D}\ }\textbf {\bibinfo {volume} {98}},\ \bibinfo
  {pages} {084011} (\bibinfo {year} {2018})},\ \Eprint
  {https://arxiv.org/abs/1805.05755} {arXiv:1805.05755 [gr-qc]} \BibitemShut
  {NoStop}%
\bibitem [{\citenamefont {Minamitsuji}\ and\ \citenamefont
  {Ikeda}(2019)}]{Minamitsuji:2018xde}%
  \BibitemOpen
  \bibfield  {author} {\bibinfo {author} {\bibfnamefont {M.}~\bibnamefont
  {Minamitsuji}}\ and\ \bibinfo {author} {\bibfnamefont {T.}~\bibnamefont
  {Ikeda}},\ }\bibfield  {title} {\bibinfo {title} {{Scalarized black holes in
  the presence of the coupling to Gauss-Bonnet gravity}},\ }\href
  {https://doi.org/10.1103/PhysRevD.99.044017} {\bibfield  {journal} {\bibinfo
  {journal} {Phys. Rev. D}\ }\textbf {\bibinfo {volume} {99}},\ \bibinfo
  {pages} {044017} (\bibinfo {year} {2019})},\ \Eprint
  {https://arxiv.org/abs/1812.03551} {arXiv:1812.03551 [gr-qc]} \BibitemShut
  {NoStop}%
\bibitem [{\citenamefont {Silva}\ \emph {et~al.}(2019)\citenamefont {Silva},
  \citenamefont {Macedo}, \citenamefont {Sotiriou}, \citenamefont {Gualtieri},
  \citenamefont {Sakstein},\ and\ \citenamefont {Berti}}]{Silva:2018qhn}%
  \BibitemOpen
  \bibfield  {author} {\bibinfo {author} {\bibfnamefont {H.~O.}\ \bibnamefont
  {Silva}}, \bibinfo {author} {\bibfnamefont {C.~F.}\ \bibnamefont {Macedo}},
  \bibinfo {author} {\bibfnamefont {T.~P.}\ \bibnamefont {Sotiriou}}, \bibinfo
  {author} {\bibfnamefont {L.}~\bibnamefont {Gualtieri}}, \bibinfo {author}
  {\bibfnamefont {J.}~\bibnamefont {Sakstein}},\ and\ \bibinfo {author}
  {\bibfnamefont {E.}~\bibnamefont {Berti}},\ }\bibfield  {title} {\bibinfo
  {title} {{Stability of scalarized black hole solutions in scalar-Gauss-Bonnet
  gravity}},\ }\href {https://doi.org/10.1103/PhysRevD.99.064011} {\bibfield
  {journal} {\bibinfo  {journal} {Phys. Rev. D}\ }\textbf {\bibinfo {volume}
  {99}},\ \bibinfo {pages} {064011} (\bibinfo {year} {2019})},\ \Eprint
  {https://arxiv.org/abs/1812.05590} {arXiv:1812.05590 [gr-qc]} \BibitemShut
  {NoStop}%
\bibitem [{\citenamefont {Antoniou}\ \emph {et~al.}(2021)\citenamefont
  {Antoniou}, \citenamefont {Leh\'ebel}, \citenamefont {Ventagli},\ and\
  \citenamefont {Sotiriou}}]{Antoniou:2021zoy}%
  \BibitemOpen
  \bibfield  {author} {\bibinfo {author} {\bibfnamefont {G.}~\bibnamefont
  {Antoniou}}, \bibinfo {author} {\bibfnamefont {A.}~\bibnamefont {Leh\'ebel}},
  \bibinfo {author} {\bibfnamefont {G.}~\bibnamefont {Ventagli}},\ and\
  \bibinfo {author} {\bibfnamefont {T.~P.}\ \bibnamefont {Sotiriou}},\
  }\bibfield  {title} {\bibinfo {title} {{Black hole scalarization with
  Gauss-Bonnet and Ricci scalar couplings}},\ }\href
  {https://doi.org/10.1103/PhysRevD.104.044002} {\bibfield  {journal} {\bibinfo
   {journal} {Phys. Rev. D}\ }\textbf {\bibinfo {volume} {104}},\ \bibinfo
  {pages} {044002} (\bibinfo {year} {2021})},\ \Eprint
  {https://arxiv.org/abs/2105.04479} {arXiv:2105.04479 [gr-qc]} \BibitemShut
  {NoStop}%
\bibitem [{\citenamefont {Antoniou}\ \emph {et~al.}(2022)\citenamefont
  {Antoniou}, \citenamefont {Macedo}, \citenamefont {McManus},\ and\
  \citenamefont {Sotiriou}}]{Antoniou:2022agj}%
  \BibitemOpen
  \bibfield  {author} {\bibinfo {author} {\bibfnamefont {G.}~\bibnamefont
  {Antoniou}}, \bibinfo {author} {\bibfnamefont {C.~F.~B.}\ \bibnamefont
  {Macedo}}, \bibinfo {author} {\bibfnamefont {R.}~\bibnamefont {McManus}},\
  and\ \bibinfo {author} {\bibfnamefont {T.~P.}\ \bibnamefont {Sotiriou}},\
  }\href@noop {} {\bibinfo {title} {{Stable spontaneously-scalarized black
  holes in generalized scalar-tensor theories}}} (\bibinfo {year} {2022}),\
  \Eprint {https://arxiv.org/abs/2204.01684} {arXiv:2204.01684 [gr-qc]}
  \BibitemShut {NoStop}%
\bibitem [{\citenamefont {Alcubierre}(2008)}]{Alcubierre:2008}%
  \BibitemOpen
  \bibfield  {author} {\bibinfo {author} {\bibfnamefont {M.}~\bibnamefont
  {Alcubierre}},\ }\href
  {https://doi.org/10.1093/acprof:oso/9780199205677.001.0001} {\emph {\bibinfo
  {title} {{Introduction to 3+1 numerical relativity}}}},\ International Series
  of Monographs on Physics\ (\bibinfo  {publisher} {Oxford Univ. Press},\
  \bibinfo {address} {Oxford},\ \bibinfo {year} {2008})\BibitemShut {NoStop}%
\bibitem [{\citenamefont {Shibata}\ and\ \citenamefont
  {Nakamura}(1995)}]{Shibata:1995we}%
  \BibitemOpen
  \bibfield  {author} {\bibinfo {author} {\bibfnamefont {M.}~\bibnamefont
  {Shibata}}\ and\ \bibinfo {author} {\bibfnamefont {T.}~\bibnamefont
  {Nakamura}},\ }\bibfield  {title} {\bibinfo {title} {{Evolution of
  three-dimensional gravitational waves: Harmonic slicing case}},\ }\href
  {https://doi.org/10.1103/PhysRevD.52.5428} {\bibfield  {journal} {\bibinfo
  {journal} {Phys. Rev. D}\ }\textbf {\bibinfo {volume} {52}},\ \bibinfo
  {pages} {5428} (\bibinfo {year} {1995})}\BibitemShut {NoStop}%
\bibitem [{\citenamefont {Baumgarte}\ and\ \citenamefont
  {Shapiro}(1999)}]{Baumgarte:1998te}%
  \BibitemOpen
  \bibfield  {author} {\bibinfo {author} {\bibfnamefont {T.~W.}\ \bibnamefont
  {Baumgarte}}\ and\ \bibinfo {author} {\bibfnamefont {S.~L.}\ \bibnamefont
  {Shapiro}},\ }\bibfield  {title} {\bibinfo {title} {{On the numerical
  integration of Einstein's field equations}},\ }\href
  {https://doi.org/10.1103/PhysRevD.59.024007} {\bibfield  {journal} {\bibinfo
  {journal} {Phys. Rev. D}\ }\textbf {\bibinfo {volume} {59}},\ \bibinfo
  {pages} {024007} (\bibinfo {year} {1999})},\ \Eprint
  {https://arxiv.org/abs/gr-qc/9810065} {arXiv:gr-qc/9810065} \BibitemShut
  {NoStop}%
\bibitem [{\citenamefont {Campanelli}\ \emph {et~al.}(2006)\citenamefont
  {Campanelli}, \citenamefont {Lousto}, \citenamefont {Marronetti},\ and\
  \citenamefont {Zlochower}}]{Campanelli:2005dd}%
  \BibitemOpen
  \bibfield  {author} {\bibinfo {author} {\bibfnamefont {M.}~\bibnamefont
  {Campanelli}}, \bibinfo {author} {\bibfnamefont {C.}~\bibnamefont {Lousto}},
  \bibinfo {author} {\bibfnamefont {P.}~\bibnamefont {Marronetti}},\ and\
  \bibinfo {author} {\bibfnamefont {Y.}~\bibnamefont {Zlochower}},\ }\bibfield
  {title} {\bibinfo {title} {{Accurate evolutions of orbiting black-hole
  binaries without excision}},\ }\href
  {https://doi.org/10.1103/PhysRevLett.96.111101} {\bibfield  {journal}
  {\bibinfo  {journal} {Phys. Rev. Lett.}\ }\textbf {\bibinfo {volume} {96}},\
  \bibinfo {pages} {111101} (\bibinfo {year} {2006})},\ \Eprint
  {https://arxiv.org/abs/gr-qc/0511048} {arXiv:gr-qc/0511048} \BibitemShut
  {NoStop}%
\bibitem [{\citenamefont {Baker}\ \emph {et~al.}(2006)\citenamefont {Baker},
  \citenamefont {Centrella}, \citenamefont {Choi}, \citenamefont {Koppitz},\
  and\ \citenamefont {van Meter}}]{Baker:2005vv}%
  \BibitemOpen
  \bibfield  {author} {\bibinfo {author} {\bibfnamefont {J.~G.}\ \bibnamefont
  {Baker}}, \bibinfo {author} {\bibfnamefont {J.}~\bibnamefont {Centrella}},
  \bibinfo {author} {\bibfnamefont {D.-I.}\ \bibnamefont {Choi}}, \bibinfo
  {author} {\bibfnamefont {M.}~\bibnamefont {Koppitz}},\ and\ \bibinfo {author}
  {\bibfnamefont {J.}~\bibnamefont {van Meter}},\ }\bibfield  {title} {\bibinfo
  {title} {{Gravitational wave extraction from an inspiraling configuration of
  merging black holes}},\ }\href
  {https://doi.org/10.1103/PhysRevLett.96.111102} {\bibfield  {journal}
  {\bibinfo  {journal} {Phys. Rev. Lett.}\ }\textbf {\bibinfo {volume} {96}},\
  \bibinfo {pages} {111102} (\bibinfo {year} {2006})},\ \Eprint
  {https://arxiv.org/abs/gr-qc/0511103} {arXiv:gr-qc/0511103} \BibitemShut
  {NoStop}%
\bibitem [{\citenamefont {Bowen}\ and\ \citenamefont
  {York}(1980)}]{Bowen:1980yu}%
  \BibitemOpen
  \bibfield  {author} {\bibinfo {author} {\bibfnamefont {J.~M.}\ \bibnamefont
  {Bowen}}\ and\ \bibinfo {author} {\bibfnamefont {J.}~\bibnamefont {York},
  \bibfnamefont {James~W.}},\ }\bibfield  {title} {\bibinfo {title} {{Time
  asymmetric initial data for black holes and black hole collisions}},\ }\href
  {https://doi.org/10.1103/PhysRevD.21.2047} {\bibfield  {journal} {\bibinfo
  {journal} {Phys. Rev. D}\ }\textbf {\bibinfo {volume} {21}},\ \bibinfo
  {pages} {2047} (\bibinfo {year} {1980})}\BibitemShut {NoStop}%
\bibitem [{\citenamefont {Brandt}\ and\ \citenamefont
  {Bruegmann}(1997)}]{Brandt:1997tf}%
  \BibitemOpen
  \bibfield  {author} {\bibinfo {author} {\bibfnamefont {S.}~\bibnamefont
  {Brandt}}\ and\ \bibinfo {author} {\bibfnamefont {B.}~\bibnamefont
  {Bruegmann}},\ }\bibfield  {title} {\bibinfo {title} {{A Simple construction
  of initial data for multiple black holes}},\ }\href
  {https://doi.org/10.1103/PhysRevLett.78.3606} {\bibfield  {journal} {\bibinfo
   {journal} {Phys. Rev. Lett.}\ }\textbf {\bibinfo {volume} {78}},\ \bibinfo
  {pages} {3606} (\bibinfo {year} {1997})},\ \Eprint
  {https://arxiv.org/abs/gr-qc/9703066} {arXiv:gr-qc/9703066} \BibitemShut
  {NoStop}%
\bibitem [{\citenamefont {Witek}\ \emph {et~al.}(2020)\citenamefont {Witek},
  \citenamefont {Gualtieri},\ and\ \citenamefont {Pani}}]{Witek:2020uzz}%
  \BibitemOpen
  \bibfield  {author} {\bibinfo {author} {\bibfnamefont {H.}~\bibnamefont
  {Witek}}, \bibinfo {author} {\bibfnamefont {L.}~\bibnamefont {Gualtieri}},\
  and\ \bibinfo {author} {\bibfnamefont {P.}~\bibnamefont {Pani}},\ }\bibfield
  {title} {\bibinfo {title} {{Towards numerical relativity in scalar
  Gauss-Bonnet gravity: $3+1$ decomposition beyond the small-coupling limit}},\
  }\href {https://doi.org/10.1103/PhysRevD.101.124055} {\bibfield  {journal}
  {\bibinfo  {journal} {Phys. Rev. D}\ }\textbf {\bibinfo {volume} {101}},\
  \bibinfo {pages} {124055} (\bibinfo {year} {2020})},\ \Eprint
  {https://arxiv.org/abs/2004.00009} {arXiv:2004.00009 [gr-qc]} \BibitemShut
  {NoStop}%
\bibitem [{\citenamefont {Witek}\ \emph {et~al.}(2021)\citenamefont {Witek},
  \citenamefont {Zilhao}, \citenamefont {Bozzola}, \citenamefont {Elley},
  \citenamefont {Ficarra}, \citenamefont {Ikeda}, \citenamefont
  {Sanchis-Gual},\ and\ \citenamefont {Silva}}]{witek_helvi_2021_5520862}%
  \BibitemOpen
  \bibfield  {author} {\bibinfo {author} {\bibfnamefont {H.}~\bibnamefont
  {Witek}}, \bibinfo {author} {\bibfnamefont {M.}~\bibnamefont {Zilhao}},
  \bibinfo {author} {\bibfnamefont {G.}~\bibnamefont {Bozzola}}, \bibinfo
  {author} {\bibfnamefont {M.}~\bibnamefont {Elley}}, \bibinfo {author}
  {\bibfnamefont {G.}~\bibnamefont {Ficarra}}, \bibinfo {author} {\bibfnamefont
  {T.}~\bibnamefont {Ikeda}}, \bibinfo {author} {\bibfnamefont
  {N.}~\bibnamefont {Sanchis-Gual}},\ and\ \bibinfo {author} {\bibfnamefont
  {H.}~\bibnamefont {Silva}},\ }\href {https://doi.org/10.5281/zenodo.5520862}
  {\bibinfo {title} {{Canuda: a public numerical relativity library to probe
  fundamental physics}}} (\bibinfo {year} {2021})\BibitemShut {NoStop}%
\bibitem [{\citenamefont {Okawa}\ \emph {et~al.}(2014)\citenamefont {Okawa},
  \citenamefont {Witek},\ and\ \citenamefont {Cardoso}}]{Okawa:2014nda}%
  \BibitemOpen
  \bibfield  {author} {\bibinfo {author} {\bibfnamefont {H.}~\bibnamefont
  {Okawa}}, \bibinfo {author} {\bibfnamefont {H.}~\bibnamefont {Witek}},\ and\
  \bibinfo {author} {\bibfnamefont {V.}~\bibnamefont {Cardoso}},\ }\bibfield
  {title} {\bibinfo {title} {{Black holes and fundamental fields in Numerical
  Relativity: initial data construction and evolution of bound states}},\
  }\href {https://doi.org/10.1103/PhysRevD.89.104032} {\bibfield  {journal}
  {\bibinfo  {journal} {Phys. Rev. D}\ }\textbf {\bibinfo {volume} {89}},\
  \bibinfo {pages} {104032} (\bibinfo {year} {2014})},\ \Eprint
  {https://arxiv.org/abs/1401.1548} {arXiv:1401.1548 [gr-qc]} \BibitemShut
  {NoStop}%
\bibitem [{\citenamefont {Zilh\~ao}\ \emph {et~al.}(2015)\citenamefont
  {Zilh\~ao}, \citenamefont {Witek},\ and\ \citenamefont
  {Cardoso}}]{Zilhao:2015tya}%
  \BibitemOpen
  \bibfield  {author} {\bibinfo {author} {\bibfnamefont {M.}~\bibnamefont
  {Zilh\~ao}}, \bibinfo {author} {\bibfnamefont {H.}~\bibnamefont {Witek}},\
  and\ \bibinfo {author} {\bibfnamefont {V.}~\bibnamefont {Cardoso}},\
  }\bibfield  {title} {\bibinfo {title} {{Nonlinear interactions between black
  holes and Proca fields}},\ }\href
  {https://doi.org/10.1088/0264-9381/32/23/234003} {\bibfield  {journal}
  {\bibinfo  {journal} {Class. Quant. Grav.}\ }\textbf {\bibinfo {volume}
  {32}},\ \bibinfo {pages} {234003} (\bibinfo {year} {2015})},\ \Eprint
  {https://arxiv.org/abs/1505.00797} {arXiv:1505.00797 [gr-qc]} \BibitemShut
  {NoStop}%
\bibitem [{\citenamefont {Brandt}\ \emph {et~al.}(2021)\citenamefont {Brandt},
  \citenamefont {Bozzola}, \citenamefont {Cheng}, \citenamefont {Diener},
  \citenamefont {Dima}, \citenamefont {Gabella}, \citenamefont
  {Gracia-Linares}, \citenamefont {Haas}, \citenamefont {Zlochower},
  \citenamefont {Alcubierre}, \citenamefont {Alic}, \citenamefont {Allen},
  \citenamefont {Ansorg}, \citenamefont {Babiuc-Hamilton}, \citenamefont
  {Baiotti}, \citenamefont {Benger}, \citenamefont {Bentivegna}, \citenamefont
  {Bernuzzi}, \citenamefont {Bode}, \citenamefont {Brendal}, \citenamefont
  {Bruegmann}, \citenamefont {Campanelli}, \citenamefont {Cipolletta},
  \citenamefont {Corvino}, \citenamefont {Cupp}, \citenamefont {Pietri},
  \citenamefont {Dimmelmeier}, \citenamefont {Dooley}, \citenamefont {Dorband},
  \citenamefont {Elley}, \citenamefont {Khamra}, \citenamefont {Etienne},
  \citenamefont {Faber}, \citenamefont {Font}, \citenamefont {Frieben},
  \citenamefont {Giacomazzo}, \citenamefont {Goodale}, \citenamefont
  {Gundlach}, \citenamefont {Hawke}, \citenamefont {Hawley}, \citenamefont
  {Hinder}, \citenamefont {Huerta}, \citenamefont {Husa}, \citenamefont {Iyer},
  \citenamefont {Johnson}, \citenamefont {Joshi}, \citenamefont {Kastaun},
  \citenamefont {Kellermann}, \citenamefont {Knapp}, \citenamefont {Koppitz},
  \citenamefont {Laguna}, \citenamefont {Lanferman}, \citenamefont {Löffler},
  \citenamefont {Masso}, \citenamefont {Menger}, \citenamefont {Merzky},
  \citenamefont {Miller}, \citenamefont {Miller}, \citenamefont {Moesta},
  \citenamefont {Montero}, \citenamefont {Mundim}, \citenamefont {Nerozzi},
  \citenamefont {Noble}, \citenamefont {Ott}, \citenamefont {Paruchuri},
  \citenamefont {Pollney}, \citenamefont {Radice}, \citenamefont {Radke},
  \citenamefont {Reisswig}, \citenamefont {Rezzolla}, \citenamefont {Rideout},
  \citenamefont {Ripeanu}, \citenamefont {Sala}, \citenamefont {Schewtschenko},
  \citenamefont {Schnetter}, \citenamefont {Schutz}, \citenamefont {Seidel},
  \citenamefont {Seidel}, \citenamefont {Shalf}, \citenamefont {Sible},
  \citenamefont {Sperhake}, \citenamefont {Stergioulas}, \citenamefont {Suen},
  \citenamefont {Szilagyi}, \citenamefont {Takahashi}, \citenamefont {Thomas},
  \citenamefont {Thornburg}, \citenamefont {Tobias}, \citenamefont {Tonita},
  \citenamefont {Walker}, \citenamefont {Wan}, \citenamefont {Wardell},
  \citenamefont {Werneck}, \citenamefont {Witek}, \citenamefont {Zilhão},\
  and\ \citenamefont {Zink}}]{steven_r_brandt_2021_5770803}%
  \BibitemOpen
  \bibfield  {author} {\bibinfo {author} {\bibfnamefont {S.~R.}\ \bibnamefont
  {Brandt}}, \bibinfo {author} {\bibfnamefont {G.}~\bibnamefont {Bozzola}},
  \bibinfo {author} {\bibfnamefont {C.-H.}\ \bibnamefont {Cheng}}, \bibinfo
  {author} {\bibfnamefont {P.}~\bibnamefont {Diener}}, \bibinfo {author}
  {\bibfnamefont {A.}~\bibnamefont {Dima}}, \bibinfo {author} {\bibfnamefont
  {W.~E.}\ \bibnamefont {Gabella}}, \bibinfo {author} {\bibfnamefont
  {M.}~\bibnamefont {Gracia-Linares}}, \bibinfo {author} {\bibfnamefont
  {R.}~\bibnamefont {Haas}}, \bibinfo {author} {\bibfnamefont {Y.}~\bibnamefont
  {Zlochower}}, \bibinfo {author} {\bibfnamefont {M.}~\bibnamefont
  {Alcubierre}}, \bibinfo {author} {\bibfnamefont {D.}~\bibnamefont {Alic}},
  \bibinfo {author} {\bibfnamefont {G.}~\bibnamefont {Allen}}, \bibinfo
  {author} {\bibfnamefont {M.}~\bibnamefont {Ansorg}}, \bibinfo {author}
  {\bibfnamefont {M.}~\bibnamefont {Babiuc-Hamilton}}, \bibinfo {author}
  {\bibfnamefont {L.}~\bibnamefont {Baiotti}}, \bibinfo {author} {\bibfnamefont
  {W.}~\bibnamefont {Benger}}, \bibinfo {author} {\bibfnamefont
  {E.}~\bibnamefont {Bentivegna}}, \bibinfo {author} {\bibfnamefont
  {S.}~\bibnamefont {Bernuzzi}}, \bibinfo {author} {\bibfnamefont
  {T.}~\bibnamefont {Bode}}, \bibinfo {author} {\bibfnamefont {B.}~\bibnamefont
  {Brendal}}, \bibinfo {author} {\bibfnamefont {B.}~\bibnamefont {Bruegmann}},
  \bibinfo {author} {\bibfnamefont {M.}~\bibnamefont {Campanelli}}, \bibinfo
  {author} {\bibfnamefont {F.}~\bibnamefont {Cipolletta}}, \bibinfo {author}
  {\bibfnamefont {G.}~\bibnamefont {Corvino}}, \bibinfo {author} {\bibfnamefont
  {S.}~\bibnamefont {Cupp}}, \bibinfo {author} {\bibfnamefont {R.~D.}\
  \bibnamefont {Pietri}}, \bibinfo {author} {\bibfnamefont {H.}~\bibnamefont
  {Dimmelmeier}}, \bibinfo {author} {\bibfnamefont {R.}~\bibnamefont {Dooley}},
  \bibinfo {author} {\bibfnamefont {N.}~\bibnamefont {Dorband}}, \bibinfo
  {author} {\bibfnamefont {M.}~\bibnamefont {Elley}}, \bibinfo {author}
  {\bibfnamefont {Y.~E.}\ \bibnamefont {Khamra}}, \bibinfo {author}
  {\bibfnamefont {Z.}~\bibnamefont {Etienne}}, \bibinfo {author} {\bibfnamefont
  {J.}~\bibnamefont {Faber}}, \bibinfo {author} {\bibfnamefont
  {T.}~\bibnamefont {Font}}, \bibinfo {author} {\bibfnamefont {J.}~\bibnamefont
  {Frieben}}, \bibinfo {author} {\bibfnamefont {B.}~\bibnamefont {Giacomazzo}},
  \bibinfo {author} {\bibfnamefont {T.}~\bibnamefont {Goodale}}, \bibinfo
  {author} {\bibfnamefont {C.}~\bibnamefont {Gundlach}}, \bibinfo {author}
  {\bibfnamefont {I.}~\bibnamefont {Hawke}}, \bibinfo {author} {\bibfnamefont
  {S.}~\bibnamefont {Hawley}}, \bibinfo {author} {\bibfnamefont
  {I.}~\bibnamefont {Hinder}}, \bibinfo {author} {\bibfnamefont {E.~A.}\
  \bibnamefont {Huerta}}, \bibinfo {author} {\bibfnamefont {S.}~\bibnamefont
  {Husa}}, \bibinfo {author} {\bibfnamefont {S.}~\bibnamefont {Iyer}}, \bibinfo
  {author} {\bibfnamefont {D.}~\bibnamefont {Johnson}}, \bibinfo {author}
  {\bibfnamefont {A.~V.}\ \bibnamefont {Joshi}}, \bibinfo {author}
  {\bibfnamefont {W.}~\bibnamefont {Kastaun}}, \bibinfo {author} {\bibfnamefont
  {T.}~\bibnamefont {Kellermann}}, \bibinfo {author} {\bibfnamefont
  {A.}~\bibnamefont {Knapp}}, \bibinfo {author} {\bibfnamefont
  {M.}~\bibnamefont {Koppitz}}, \bibinfo {author} {\bibfnamefont
  {P.}~\bibnamefont {Laguna}}, \bibinfo {author} {\bibfnamefont
  {G.}~\bibnamefont {Lanferman}}, \bibinfo {author} {\bibfnamefont
  {F.}~\bibnamefont {Löffler}}, \bibinfo {author} {\bibfnamefont
  {J.}~\bibnamefont {Masso}}, \bibinfo {author} {\bibfnamefont
  {L.}~\bibnamefont {Menger}}, \bibinfo {author} {\bibfnamefont
  {A.}~\bibnamefont {Merzky}}, \bibinfo {author} {\bibfnamefont {J.~M.}\
  \bibnamefont {Miller}}, \bibinfo {author} {\bibfnamefont {M.}~\bibnamefont
  {Miller}}, \bibinfo {author} {\bibfnamefont {P.}~\bibnamefont {Moesta}},
  \bibinfo {author} {\bibfnamefont {P.}~\bibnamefont {Montero}}, \bibinfo
  {author} {\bibfnamefont {B.}~\bibnamefont {Mundim}}, \bibinfo {author}
  {\bibfnamefont {A.}~\bibnamefont {Nerozzi}}, \bibinfo {author} {\bibfnamefont
  {S.~C.}\ \bibnamefont {Noble}}, \bibinfo {author} {\bibfnamefont
  {C.}~\bibnamefont {Ott}}, \bibinfo {author} {\bibfnamefont {R.}~\bibnamefont
  {Paruchuri}}, \bibinfo {author} {\bibfnamefont {D.}~\bibnamefont {Pollney}},
  \bibinfo {author} {\bibfnamefont {D.}~\bibnamefont {Radice}}, \bibinfo
  {author} {\bibfnamefont {T.}~\bibnamefont {Radke}}, \bibinfo {author}
  {\bibfnamefont {C.}~\bibnamefont {Reisswig}}, \bibinfo {author}
  {\bibfnamefont {L.}~\bibnamefont {Rezzolla}}, \bibinfo {author}
  {\bibfnamefont {D.}~\bibnamefont {Rideout}}, \bibinfo {author} {\bibfnamefont
  {M.}~\bibnamefont {Ripeanu}}, \bibinfo {author} {\bibfnamefont
  {L.}~\bibnamefont {Sala}}, \bibinfo {author} {\bibfnamefont {J.~A.}\
  \bibnamefont {Schewtschenko}}, \bibinfo {author} {\bibfnamefont
  {E.}~\bibnamefont {Schnetter}}, \bibinfo {author} {\bibfnamefont
  {B.}~\bibnamefont {Schutz}}, \bibinfo {author} {\bibfnamefont
  {E.}~\bibnamefont {Seidel}}, \bibinfo {author} {\bibfnamefont
  {E.}~\bibnamefont {Seidel}}, \bibinfo {author} {\bibfnamefont
  {J.}~\bibnamefont {Shalf}}, \bibinfo {author} {\bibfnamefont
  {K.}~\bibnamefont {Sible}}, \bibinfo {author} {\bibfnamefont
  {U.}~\bibnamefont {Sperhake}}, \bibinfo {author} {\bibfnamefont
  {N.}~\bibnamefont {Stergioulas}}, \bibinfo {author} {\bibfnamefont {W.-M.}\
  \bibnamefont {Suen}}, \bibinfo {author} {\bibfnamefont {B.}~\bibnamefont
  {Szilagyi}}, \bibinfo {author} {\bibfnamefont {R.}~\bibnamefont {Takahashi}},
  \bibinfo {author} {\bibfnamefont {M.}~\bibnamefont {Thomas}}, \bibinfo
  {author} {\bibfnamefont {J.}~\bibnamefont {Thornburg}}, \bibinfo {author}
  {\bibfnamefont {M.}~\bibnamefont {Tobias}}, \bibinfo {author} {\bibfnamefont
  {A.}~\bibnamefont {Tonita}}, \bibinfo {author} {\bibfnamefont
  {P.}~\bibnamefont {Walker}}, \bibinfo {author} {\bibfnamefont {M.-B.}\
  \bibnamefont {Wan}}, \bibinfo {author} {\bibfnamefont {B.}~\bibnamefont
  {Wardell}}, \bibinfo {author} {\bibfnamefont {L.}~\bibnamefont {Werneck}},
  \bibinfo {author} {\bibfnamefont {H.}~\bibnamefont {Witek}}, \bibinfo
  {author} {\bibfnamefont {M.}~\bibnamefont {Zilhão}},\ and\ \bibinfo {author}
  {\bibfnamefont {B.}~\bibnamefont {Zink}},\ }\href
  {https://doi.org/10.5281/zenodo.5770803} {\bibinfo {title} {The {E}instein
  {T}oolkit}} (\bibinfo {year} {2021}),\ \bibinfo {note} {to find out more,
  visit \url{http://einsteintoolkit.org}}\BibitemShut {NoStop}%
\bibitem [{\citenamefont {L{\"o}ffler}\ \emph {et~al.}(2012)\citenamefont
  {L{\"o}ffler} \emph {et~al.}}]{Loffler:2011ay}%
  \BibitemOpen
  \bibfield  {author} {\bibinfo {author} {\bibfnamefont {F.}~\bibnamefont
  {L{\"o}ffler}} \emph {et~al.},\ }\bibfield  {title} {\bibinfo {title} {{The
  Einstein Toolkit: A Community Computational Infrastructure for Relativistic
  Astrophysics}},\ }\href {https://doi.org/10.1088/0264-9381/29/11/115001}
  {\bibfield  {journal} {\bibinfo  {journal} {Class. Quant. Grav.}\ }\textbf
  {\bibinfo {volume} {29}},\ \bibinfo {pages} {115001} (\bibinfo {year}
  {2012})},\ \Eprint {https://arxiv.org/abs/1111.3344} {arXiv:1111.3344
  [gr-qc]} \BibitemShut {NoStop}%
\bibitem [{\citenamefont {Zilh\~ao}\ and\ \citenamefont
  {L\"offler}(2013)}]{Zilhao:2013hia}%
  \BibitemOpen
  \bibfield  {author} {\bibinfo {author} {\bibfnamefont {M.}~\bibnamefont
  {Zilh\~ao}}\ and\ \bibinfo {author} {\bibfnamefont {F.}~\bibnamefont
  {L\"offler}},\ }\bibfield  {title} {\bibinfo {title} {{An Introduction to the
  Einstein Toolkit}},\ }\href {https://doi.org/10.1142/S0217751X13400149}
  {\bibfield  {journal} {\bibinfo  {journal} {Int. J. Mod. Phys. A}\ }\textbf
  {\bibinfo {volume} {28}},\ \bibinfo {pages} {1340014} (\bibinfo {year}
  {2013})},\ \Eprint {https://arxiv.org/abs/1305.5299} {arXiv:1305.5299
  [gr-qc]} \BibitemShut {NoStop}%
\bibitem [{\citenamefont {Goodale}\ \emph {et~al.}(2003)\citenamefont
  {Goodale}, \citenamefont {Allen}, \citenamefont {Lanfermann}, \citenamefont
  {Mass{\'o}}, \citenamefont {Radke}, \citenamefont {Seidel},\ and\
  \citenamefont {Shalf}}]{Goodale:2002a}%
  \BibitemOpen
  \bibfield  {author} {\bibinfo {author} {\bibfnamefont {T.}~\bibnamefont
  {Goodale}}, \bibinfo {author} {\bibfnamefont {G.}~\bibnamefont {Allen}},
  \bibinfo {author} {\bibfnamefont {G.}~\bibnamefont {Lanfermann}}, \bibinfo
  {author} {\bibfnamefont {J.}~\bibnamefont {Mass{\'o}}}, \bibinfo {author}
  {\bibfnamefont {T.}~\bibnamefont {Radke}}, \bibinfo {author} {\bibfnamefont
  {E.}~\bibnamefont {Seidel}},\ and\ \bibinfo {author} {\bibfnamefont
  {J.}~\bibnamefont {Shalf}},\ }\bibfield  {title} {\bibinfo {title} {The
  {Cactus} framework and toolkit: Design and applications},\ }in\ \href
  {http://edoc.mpg.de/3341} {\emph {\bibinfo {booktitle} {Vector and Parallel
  Processing -- VECPAR'2002, 5th International Conference, Lecture Notes in
  Computer Science}}}\ (\bibinfo  {publisher} {Springer},\ \bibinfo {address}
  {Berlin},\ \bibinfo {year} {2003})\BibitemShut {NoStop}%
\bibitem [{Cactus developers()}]{Cactuscode:web}%
  \BibitemOpen
  Cactus developers,\ \href {http://www.cactuscode.org/} {\bibinfo {title}
  {{Cactus Computational Toolkit}}}\BibitemShut {NoStop}%
\bibitem [{\citenamefont {Schnetter}\ \emph {et~al.}(2004)\citenamefont
  {Schnetter}, \citenamefont {Hawley},\ and\ \citenamefont
  {Hawke}}]{Schnetter:2003rb}%
  \BibitemOpen
  \bibfield  {author} {\bibinfo {author} {\bibfnamefont {E.}~\bibnamefont
  {Schnetter}}, \bibinfo {author} {\bibfnamefont {S.~H.}\ \bibnamefont
  {Hawley}},\ and\ \bibinfo {author} {\bibfnamefont {I.}~\bibnamefont
  {Hawke}},\ }\bibfield  {title} {\bibinfo {title} {{Evolutions in 3-D
  numerical relativity using fixed mesh refinement}},\ }\href
  {https://doi.org/10.1088/0264-9381/21/6/014} {\bibfield  {journal} {\bibinfo
  {journal} {Class. Quant. Grav.}\ }\textbf {\bibinfo {volume} {21}},\ \bibinfo
  {pages} {1465} (\bibinfo {year} {2004})},\ \Eprint
  {https://arxiv.org/abs/gr-qc/0310042} {arXiv:gr-qc/0310042} \BibitemShut
  {NoStop}%
\bibitem [{Carpet()}]{CarpetCode:web}%
  \BibitemOpen
  Carpet,\ \href {https://bitbucket.org/eschnett/carpet.git} {}\bibinfo {note}
  {{Carpet}: Adaptive Mesh Refinement for the {Cactus} Framework}\BibitemShut
  {NoStop}%
\bibitem [{\citenamefont {Ansorg}\ \emph {et~al.}(2004)\citenamefont {Ansorg},
  \citenamefont {Br{\"u}gmann},\ and\ \citenamefont {Tichy}}]{Ansorg:2004ds}%
  \BibitemOpen
  \bibfield  {author} {\bibinfo {author} {\bibfnamefont {M.}~\bibnamefont
  {Ansorg}}, \bibinfo {author} {\bibfnamefont {B.}~\bibnamefont
  {Br{\"u}gmann}},\ and\ \bibinfo {author} {\bibfnamefont {W.}~\bibnamefont
  {Tichy}},\ }\bibfield  {title} {\bibinfo {title} {{A Single-domain spectral
  method for black hole puncture data}},\ }\href
  {https://doi.org/10.1103/PhysRevD.70.064011} {\bibfield  {journal} {\bibinfo
  {journal} {Phys. Rev. D}\ }\textbf {\bibinfo {volume} {70}},\ \bibinfo
  {pages} {064011} (\bibinfo {year} {2004})},\ \Eprint
  {https://arxiv.org/abs/gr-qc/0404056} {arXiv:gr-qc/0404056} \BibitemShut
  {NoStop}%
\bibitem [{\citenamefont {Sperhake}(2007)}]{Sperhake:2006cy}%
  \BibitemOpen
  \bibfield  {author} {\bibinfo {author} {\bibfnamefont {U.}~\bibnamefont
  {Sperhake}},\ }\bibfield  {title} {\bibinfo {title} {{Binary black-hole
  evolutions of excision and puncture data}},\ }\href
  {https://doi.org/10.1103/PhysRevD.76.104015} {\bibfield  {journal} {\bibinfo
  {journal} {Phys. Rev. D}\ }\textbf {\bibinfo {volume} {76}},\ \bibinfo
  {pages} {104015} (\bibinfo {year} {2007})},\ \Eprint
  {https://arxiv.org/abs/gr-qc/0606079} {arXiv:gr-qc/0606079} \BibitemShut
  {NoStop}%
\bibitem [{\citenamefont {Dreyer}\ \emph {et~al.}(2003)\citenamefont {Dreyer},
  \citenamefont {Krishnan}, \citenamefont {Shoemaker},\ and\ \citenamefont
  {Schnetter}}]{Dreyer:2002mx}%
  \BibitemOpen
  \bibfield  {author} {\bibinfo {author} {\bibfnamefont {O.}~\bibnamefont
  {Dreyer}}, \bibinfo {author} {\bibfnamefont {B.}~\bibnamefont {Krishnan}},
  \bibinfo {author} {\bibfnamefont {D.}~\bibnamefont {Shoemaker}},\ and\
  \bibinfo {author} {\bibfnamefont {E.}~\bibnamefont {Schnetter}},\ }\bibfield
  {title} {\bibinfo {title} {{Introduction to isolated horizons in numerical
  relativity}},\ }\href {https://doi.org/10.1103/PhysRevD.67.024018} {\bibfield
   {journal} {\bibinfo  {journal} {Phys. Rev. D}\ }\textbf {\bibinfo {volume}
  {67}},\ \bibinfo {pages} {024018} (\bibinfo {year} {2003})},\ \Eprint
  {https://arxiv.org/abs/gr-qc/0206008} {arXiv:gr-qc/0206008} \BibitemShut
  {NoStop}%
\bibitem [{\citenamefont {Thornburg}(1996)}]{Thornburg:1995cp}%
  \BibitemOpen
  \bibfield  {author} {\bibinfo {author} {\bibfnamefont {J.}~\bibnamefont
  {Thornburg}},\ }\bibfield  {title} {\bibinfo {title} {{Finding apparent
  horizons in numerical relativity}},\ }\href
  {https://doi.org/10.1103/PhysRevD.54.4899} {\bibfield  {journal} {\bibinfo
  {journal} {Phys. Rev. D}\ }\textbf {\bibinfo {volume} {54}},\ \bibinfo
  {pages} {4899} (\bibinfo {year} {1996})},\ \Eprint
  {https://arxiv.org/abs/gr-qc/9508014} {arXiv:gr-qc/9508014} \BibitemShut
  {NoStop}%
\bibitem [{\citenamefont {Thornburg}(2004)}]{Thornburg:2003sf}%
  \BibitemOpen
  \bibfield  {author} {\bibinfo {author} {\bibfnamefont {J.}~\bibnamefont
  {Thornburg}},\ }\bibfield  {title} {\bibinfo {title} {{A Fast apparent
  horizon finder for three-dimensional Cartesian grids in numerical
  relativity}},\ }\href {https://doi.org/10.1088/0264-9381/21/2/026} {\bibfield
   {journal} {\bibinfo  {journal} {Class. Quant. Grav.}\ }\textbf {\bibinfo
  {volume} {21}},\ \bibinfo {pages} {743} (\bibinfo {year} {2004})},\ \Eprint
  {https://arxiv.org/abs/gr-qc/0306056} {arXiv:gr-qc/0306056} \BibitemShut
  {NoStop}%
\bibitem [{\citenamefont
  {Shlapentokh-Rothman}(2014)}]{Shlapentokh-Rothman:2013ysa}%
  \BibitemOpen
  \bibfield  {author} {\bibinfo {author} {\bibfnamefont {Y.}~\bibnamefont
  {Shlapentokh-Rothman}},\ }\bibfield  {title} {\bibinfo {title}
  {{Exponentially growing finite energy solutions for the Klein-Gordon equation
  on sub-extremal Kerr spacetimes}},\ }\href
  {https://doi.org/10.1007/s00220-014-2033-x} {\bibfield  {journal} {\bibinfo
  {journal} {Commun. Math. Phys.}\ }\textbf {\bibinfo {volume} {329}},\
  \bibinfo {pages} {859} (\bibinfo {year} {2014})},\ \Eprint
  {https://arxiv.org/abs/1302.3448} {arXiv:1302.3448 [gr-qc]} \BibitemShut
  {NoStop}%
\bibitem [{\citenamefont {Brito}\ \emph {et~al.}(2015)\citenamefont {Brito},
  \citenamefont {Cardoso},\ and\ \citenamefont {Pani}}]{Brito:2015oca}%
  \BibitemOpen
  \bibfield  {author} {\bibinfo {author} {\bibfnamefont {R.}~\bibnamefont
  {Brito}}, \bibinfo {author} {\bibfnamefont {V.}~\bibnamefont {Cardoso}},\
  and\ \bibinfo {author} {\bibfnamefont {P.}~\bibnamefont {Pani}},\ }\bibfield
  {title} {\bibinfo {title} {{Superradiance}: {New Frontiers in Black Hole
  Physics}},\ }\href {https://doi.org/10.1007/978-3-319-19000-6} {\bibfield
  {journal} {\bibinfo  {journal} {Lect. Notes Phys.}\ }\textbf {\bibinfo
  {volume} {906}},\ \bibinfo {pages} {pp.1} (\bibinfo {year} {2015})},\ \Eprint
  {https://arxiv.org/abs/1501.06570} {arXiv:1501.06570 [gr-qc]} \BibitemShut
  {NoStop}%
\bibitem [{\citenamefont {Moschidis}(2016)}]{Moschidis:2016wew}%
  \BibitemOpen
  \bibfield  {author} {\bibinfo {author} {\bibfnamefont {G.}~\bibnamefont
  {Moschidis}},\ }\href@noop {} {\bibinfo {title} {{Superradiant instabilities
  for short-range non-negative potentials on Kerr spacetimes and
  applications}}} (\bibinfo {year} {2016}),\ \Eprint
  {https://arxiv.org/abs/1608.02041} {arXiv:1608.02041 [math.AP]} \BibitemShut
  {NoStop}%
\bibitem [{\citenamefont {Dolan}(2007)}]{Dolan:2007mj}%
  \BibitemOpen
  \bibfield  {author} {\bibinfo {author} {\bibfnamefont {S.~R.}\ \bibnamefont
  {Dolan}},\ }\bibfield  {title} {\bibinfo {title} {{Instability of the massive
  Klein-Gordon field on the Kerr spacetime}},\ }\href
  {https://doi.org/10.1103/PhysRevD.76.084001} {\bibfield  {journal} {\bibinfo
  {journal} {Phys. Rev. D}\ }\textbf {\bibinfo {volume} {76}},\ \bibinfo
  {pages} {084001} (\bibinfo {year} {2007})},\ \Eprint
  {https://arxiv.org/abs/0705.2880} {arXiv:0705.2880 [gr-qc]} \BibitemShut
  {NoStop}%
\bibitem [{\citenamefont {Dolan}(2013)}]{Dolan:2012yt}%
  \BibitemOpen
  \bibfield  {author} {\bibinfo {author} {\bibfnamefont {S.~R.}\ \bibnamefont
  {Dolan}},\ }\bibfield  {title} {\bibinfo {title} {{Superradiant instabilities
  of rotating black holes in the time domain}},\ }\href
  {https://doi.org/10.1103/PhysRevD.87.124026} {\bibfield  {journal} {\bibinfo
  {journal} {Phys. Rev. D}\ }\textbf {\bibinfo {volume} {87}},\ \bibinfo
  {pages} {124026} (\bibinfo {year} {2013})},\ \Eprint
  {https://arxiv.org/abs/1212.1477} {arXiv:1212.1477 [gr-qc]} \BibitemShut
  {NoStop}%
\bibitem [{\citenamefont {Berti}\ \emph {et~al.}(2009)\citenamefont {Berti},
  \citenamefont {Cardoso},\ and\ \citenamefont {Starinets}}]{Berti:2009kk}%
  \BibitemOpen
  \bibfield  {author} {\bibinfo {author} {\bibfnamefont {E.}~\bibnamefont
  {Berti}}, \bibinfo {author} {\bibfnamefont {V.}~\bibnamefont {Cardoso}},\
  and\ \bibinfo {author} {\bibfnamefont {A.~O.}\ \bibnamefont {Starinets}},\
  }\bibfield  {title} {\bibinfo {title} {{Quasinormal modes of black holes and
  black branes}},\ }\href {https://doi.org/10.1088/0264-9381/26/16/163001}
  {\bibfield  {journal} {\bibinfo  {journal} {Class. Quant. Grav.}\ }\textbf
  {\bibinfo {volume} {26}},\ \bibinfo {pages} {163001} (\bibinfo {year}
  {2009})},\ \Eprint {https://arxiv.org/abs/0905.2975} {arXiv:0905.2975
  [gr-qc]} \BibitemShut {NoStop}%
\bibitem [{\citenamefont {Carson}\ \emph {et~al.}(2020)\citenamefont {Carson},
  \citenamefont {Seymour},\ and\ \citenamefont {Yagi}}]{Carson:2019fxr}%
  \BibitemOpen
  \bibfield  {author} {\bibinfo {author} {\bibfnamefont {Z.}~\bibnamefont
  {Carson}}, \bibinfo {author} {\bibfnamefont {B.~C.}\ \bibnamefont
  {Seymour}},\ and\ \bibinfo {author} {\bibfnamefont {K.}~\bibnamefont
  {Yagi}},\ }\bibfield  {title} {\bibinfo {title} {{Future prospects for
  probing scalar\textendash{}tensor theories with gravitational waves from
  mixed binaries}},\ }\href {https://doi.org/10.1088/1361-6382/ab6a1f}
  {\bibfield  {journal} {\bibinfo  {journal} {Class. Quant. Grav.}\ }\textbf
  {\bibinfo {volume} {37}},\ \bibinfo {pages} {065008} (\bibinfo {year}
  {2020})},\ \Eprint {https://arxiv.org/abs/1907.03897} {arXiv:1907.03897
  [gr-qc]} \BibitemShut {NoStop}%
\bibitem [{\citenamefont {Perkins}\ \emph
  {et~al.}(2021{\natexlab{b}})\citenamefont {Perkins}, \citenamefont {Yunes},\
  and\ \citenamefont {Berti}}]{Perkins:2020tra}%
  \BibitemOpen
  \bibfield  {author} {\bibinfo {author} {\bibfnamefont {S.~E.}\ \bibnamefont
  {Perkins}}, \bibinfo {author} {\bibfnamefont {N.}~\bibnamefont {Yunes}},\
  and\ \bibinfo {author} {\bibfnamefont {E.}~\bibnamefont {Berti}},\ }\bibfield
   {title} {\bibinfo {title} {{Probing Fundamental Physics with Gravitational
  Waves: The Next Generation}},\ }\href
  {https://doi.org/10.1103/PhysRevD.103.044024} {\bibfield  {journal} {\bibinfo
   {journal} {Phys. Rev. D}\ }\textbf {\bibinfo {volume} {103}},\ \bibinfo
  {pages} {044024} (\bibinfo {year} {2021}{\natexlab{b}})},\ \Eprint
  {https://arxiv.org/abs/2010.09010} {arXiv:2010.09010 [gr-qc]} \BibitemShut
  {NoStop}%
\bibitem [{\citenamefont {Yunes}\ and\ \citenamefont
  {Pretorius}(2009)}]{Yunes:2009ke}%
  \BibitemOpen
  \bibfield  {author} {\bibinfo {author} {\bibfnamefont {N.}~\bibnamefont
  {Yunes}}\ and\ \bibinfo {author} {\bibfnamefont {F.}~\bibnamefont
  {Pretorius}},\ }\bibfield  {title} {\bibinfo {title} {{Fundamental
  Theoretical Bias in Gravitational Wave Astrophysics and the Parameterized
  Post-Einsteinian Framework}},\ }\href
  {https://doi.org/10.1103/PhysRevD.80.122003} {\bibfield  {journal} {\bibinfo
  {journal} {Phys. Rev. D}\ }\textbf {\bibinfo {volume} {80}},\ \bibinfo
  {pages} {122003} (\bibinfo {year} {2009})},\ \Eprint
  {https://arxiv.org/abs/0909.3328} {arXiv:0909.3328 [gr-qc]} \BibitemShut
  {NoStop}%
\bibitem [{\citenamefont {Cornish}\ \emph {et~al.}(2011)\citenamefont
  {Cornish}, \citenamefont {Sampson}, \citenamefont {Yunes},\ and\
  \citenamefont {Pretorius}}]{Cornish:2011ys}%
  \BibitemOpen
  \bibfield  {author} {\bibinfo {author} {\bibfnamefont {N.}~\bibnamefont
  {Cornish}}, \bibinfo {author} {\bibfnamefont {L.}~\bibnamefont {Sampson}},
  \bibinfo {author} {\bibfnamefont {N.}~\bibnamefont {Yunes}},\ and\ \bibinfo
  {author} {\bibfnamefont {F.}~\bibnamefont {Pretorius}},\ }\bibfield  {title}
  {\bibinfo {title} {{Gravitational Wave Tests of General Relativity with the
  Parameterized Post-Einsteinian Framework}},\ }\href
  {https://doi.org/10.1103/PhysRevD.84.062003} {\bibfield  {journal} {\bibinfo
  {journal} {Phys. Rev. D}\ }\textbf {\bibinfo {volume} {84}},\ \bibinfo
  {pages} {062003} (\bibinfo {year} {2011})},\ \Eprint
  {https://arxiv.org/abs/1105.2088} {arXiv:1105.2088 [gr-qc]} \BibitemShut
  {NoStop}%
\bibitem [{\citenamefont {Tahura}\ and\ \citenamefont
  {Yagi}(2018)}]{Tahura:2018zuq}%
  \BibitemOpen
  \bibfield  {author} {\bibinfo {author} {\bibfnamefont {S.}~\bibnamefont
  {Tahura}}\ and\ \bibinfo {author} {\bibfnamefont {K.}~\bibnamefont {Yagi}},\
  }\bibfield  {title} {\bibinfo {title} {{Parameterized Post-Einsteinian
  Gravitational Waveforms in Various Modified Theories of Gravity}},\ }\href
  {https://doi.org/10.1103/PhysRevD.98.084042} {\bibfield  {journal} {\bibinfo
  {journal} {Phys. Rev. D}\ }\textbf {\bibinfo {volume} {98}},\ \bibinfo
  {pages} {084042} (\bibinfo {year} {2018})},\ \bibinfo {note} {[Erratum:
  Phys.Rev.D 101, 109902 (2020)]},\ \Eprint {https://arxiv.org/abs/1809.00259}
  {arXiv:1809.00259 [gr-qc]} \BibitemShut {NoStop}%
\bibitem [{\citenamefont {Perkins}\ and\ \citenamefont
  {Yunes}(2022)}]{Perkins:2022fhr}%
  \BibitemOpen
  \bibfield  {author} {\bibinfo {author} {\bibfnamefont {S.}~\bibnamefont
  {Perkins}}\ and\ \bibinfo {author} {\bibfnamefont {N.}~\bibnamefont
  {Yunes}},\ }\href@noop {} {\bibinfo {title} {{Are Parametrized Tests of
  General Relativity with Gravitational Waves Robust to Unknown Higher
  Post-Newtonian Order Effects?}}} (\bibinfo {year} {2022}),\ \Eprint
  {https://arxiv.org/abs/2201.02542} {arXiv:2201.02542 [gr-qc]} \BibitemShut
  {NoStop}%
\bibitem [{\citenamefont {Abbott}\ \emph {et~al.}(2020)\citenamefont {Abbott}
  \emph {et~al.}}]{LIGOScientific:2020zkf}%
  \BibitemOpen
  \bibfield  {author} {\bibinfo {author} {\bibfnamefont {R.}~\bibnamefont
  {Abbott}} \emph {et~al.} (\bibinfo {collaboration} {LIGO Scientific,
  Virgo}),\ }\bibfield  {title} {\bibinfo {title} {{GW190814: Gravitational
  Waves from the Coalescence of a 23 Solar Mass Black Hole with a 2.6 Solar
  Mass Compact Object}},\ }\href {https://doi.org/10.3847/2041-8213/ab960f}
  {\bibfield  {journal} {\bibinfo  {journal} {Astrophys. J. Lett.}\ }\textbf
  {\bibinfo {volume} {896}},\ \bibinfo {pages} {L44} (\bibinfo {year}
  {2020})},\ \Eprint {https://arxiv.org/abs/2006.12611} {arXiv:2006.12611
  [astro-ph.HE]} \BibitemShut {NoStop}%
\bibitem [{\citenamefont {Brizuela}\ \emph {et~al.}(2009)\citenamefont
  {Brizuela}, \citenamefont {Martín-García},\ and\ \citenamefont
  {Mena~Marugan}}]{Brizuela:2008ra}%
  \BibitemOpen
  \bibfield  {author} {\bibinfo {author} {\bibfnamefont {D.}~\bibnamefont
  {Brizuela}}, \bibinfo {author} {\bibfnamefont {J.~M.}\ \bibnamefont
  {Martín-García}},\ and\ \bibinfo {author} {\bibfnamefont {G.~A.}\
  \bibnamefont {Mena~Marugan}},\ }\bibfield  {title} {\bibinfo {title} {{xPert:
  Computer algebra for metric perturbation theory}},\ }\href
  {https://doi.org/10.1007/s10714-009-0773-2} {\bibfield  {journal} {\bibinfo
  {journal} {Gen. Rel. Grav.}\ }\textbf {\bibinfo {volume} {41}},\ \bibinfo
  {pages} {2415} (\bibinfo {year} {2009})},\ \Eprint
  {https://arxiv.org/abs/0807.0824} {arXiv:0807.0824 [gr-qc]} \BibitemShut
  {NoStop}%
\bibitem [{\citenamefont {Martín-García}\ \emph {et~al.}(2007)\citenamefont
  {Martín-García}, \citenamefont {Portugal},\ and\ \citenamefont
  {Manssur}}]{Martin-Garcia:2007bqa}%
  \BibitemOpen
  \bibfield  {author} {\bibinfo {author} {\bibfnamefont {J.~M.}\ \bibnamefont
  {Martín-García}}, \bibinfo {author} {\bibfnamefont {R.}~\bibnamefont
  {Portugal}},\ and\ \bibinfo {author} {\bibfnamefont {L.~R.~U.}\ \bibnamefont
  {Manssur}},\ }\bibfield  {title} {\bibinfo {title} {{The Invar Tensor
  Package}},\ }\href {https://doi.org/10.1016/j.cpc.2007.05.015} {\bibfield
  {journal} {\bibinfo  {journal} {Comput. Phys. Commun.}\ }\textbf {\bibinfo
  {volume} {177}},\ \bibinfo {pages} {640} (\bibinfo {year} {2007})},\ \Eprint
  {https://arxiv.org/abs/0704.1756} {arXiv:0704.1756 [cs.SC]} \BibitemShut
  {NoStop}%
\bibitem [{\citenamefont {Martín-García}\ \emph {et~al.}(2008)\citenamefont
  {Martín-García}, \citenamefont {Yllanes},\ and\ \citenamefont
  {Portugal}}]{Martin-Garcia:2008yei}%
  \BibitemOpen
  \bibfield  {author} {\bibinfo {author} {\bibfnamefont {J.~M.}\ \bibnamefont
  {Martín-García}}, \bibinfo {author} {\bibfnamefont {D.}~\bibnamefont
  {Yllanes}},\ and\ \bibinfo {author} {\bibfnamefont {R.}~\bibnamefont
  {Portugal}},\ }\bibfield  {title} {\bibinfo {title} {{The Invar tensor
  package: Differential invariants of Riemann}},\ }\href
  {https://doi.org/10.1016/j.cpc.2008.04.018} {\bibfield  {journal} {\bibinfo
  {journal} {Comput. Phys. Commun.}\ }\textbf {\bibinfo {volume} {179}},\
  \bibinfo {pages} {586} (\bibinfo {year} {2008})},\ \Eprint
  {https://arxiv.org/abs/0802.1274} {arXiv:0802.1274 [cs.SC]} \BibitemShut
  {NoStop}%
\bibitem [{\citenamefont {Martín-García}(2008)}]{Mart_n_Garc_a_2008}%
  \BibitemOpen
  \bibfield  {author} {\bibinfo {author} {\bibfnamefont {J.~M.}\ \bibnamefont
  {Martín-García}},\ }\bibfield  {title} {\bibinfo {title} {x{P}erm: fast
  index canonicalization for tensor computer algebra},\ }\href
  {https://doi.org/10.1016/j.cpc.2008.05.009} {\bibfield  {journal} {\bibinfo
  {journal} {Computer Physics Communications}\ }\textbf {\bibinfo {volume}
  {179}},\ \bibinfo {pages} {597} (\bibinfo {year} {2008})},\ \Eprint
  {https://arxiv.org/abs/0803.0862} {arXiv:0803.0862 [cs-sc]} \BibitemShut
  {NoStop}%
\bibitem [{xAc()}]{xAct}%
  \BibitemOpen
  \href@noop {} {\bibinfo {title} {\emph{``x{A}ct: Efficient tensor computer
  algebra for the Wolfram Language''}}},\ \bibinfo {howpublished}
  {\url{http://www.xact.es/ }}\BibitemShut {NoStop}%
\bibitem [{\citenamefont {Hunter}(2007)}]{Hunter:2007}%
  \BibitemOpen
  \bibfield  {author} {\bibinfo {author} {\bibfnamefont {J.~D.}\ \bibnamefont
  {Hunter}},\ }\bibfield  {title} {\bibinfo {title} {Matplotlib: A {2D}
  graphics environment},\ }\href {https://doi.org/10.1109/MCSE.2007.55}
  {\bibfield  {journal} {\bibinfo  {journal} {Computing in Science \&
  Engineering}\ }\textbf {\bibinfo {volume} {9}},\ \bibinfo {pages} {90}
  (\bibinfo {year} {2007})}\BibitemShut {NoStop}%
\bibitem [{\citenamefont {{Bozzola}}(2021)}]{kuibit}%
  \BibitemOpen
  \bibfield  {author} {\bibinfo {author} {\bibfnamefont {G.}~\bibnamefont
  {{Bozzola}}},\ }\bibfield  {title} {\bibinfo {title} {{kuibit: Analyzing
  Einstein Toolkit simulations with Python}},\ }\href
  {https://doi.org/10.21105/joss.03099} {\bibfield  {journal} {\bibinfo
  {journal} {The Journal of Open Source Software}\ }\textbf {\bibinfo {volume}
  {6}},\ \bibinfo {eid} {3099} (\bibinfo {year} {2021})},\ \Eprint
  {https://arxiv.org/abs/2104.06376} {arXiv:2104.06376 [gr-qc]} \BibitemShut
  {NoStop}%
\bibitem [{\citenamefont {Ellis}(2017)}]{Ellis:2016jkw}%
  \BibitemOpen
  \bibfield  {author} {\bibinfo {author} {\bibfnamefont {J.}~\bibnamefont
  {Ellis}},\ }\bibfield  {title} {\bibinfo {title} {{TikZ-Feynman: Feynman
  diagrams with TikZ}},\ }\href {https://doi.org/10.1016/j.cpc.2016.08.019}
  {\bibfield  {journal} {\bibinfo  {journal} {Comput. Phys. Commun.}\ }\textbf
  {\bibinfo {volume} {210}},\ \bibinfo {pages} {103} (\bibinfo {year}
  {2017})},\ \Eprint {https://arxiv.org/abs/1601.05437} {arXiv:1601.05437
  [hep-ph]} \BibitemShut {NoStop}%
\end{thebibliography}%
\end{document}